\RequirePackage{fixltx2e}

\documentclass[prd,twocolumn,showpacs,aps,nofootinbib,mediabb]{revtex4-1}

\usepackage{lastpage}
\usepackage{verbatim}
\usepackage{url,hyperref}
\usepackage[noabbrev]{cleveref}
\usepackage{balance}
\usepackage{bm}
\usepackage{microtype}
\usepackage{bbold}
\usepackage[utf8]{inputenc}
\usepackage[english]{babel}
\usepackage{latexsym,fancyref}
\usepackage{float}
\usepackage{xcolor}
\usepackage[T1]{fontenc}
\usepackage{textcomp}
\usepackage{tipa}
\usepackage{amstext,amssymb,amsthm,amscd,amsmath,amsfonts,mathtools,amsfonts}
\usepackage{esint}
\usepackage{pstricks-add}
\usepackage{braket,fixltx2e}
\usepackage{graphicx,tikz,epsfig}
\usepackage{lmodern}

\makeatletter
\def\p@subsection{}
\makeatother

\begin{document}
	\title{More on Homological Supersymmetric Quantum Mechanics}	
	\author{Alireza Behtash}
	\email{abehtas@ncsu.edu}
	\affiliation{Department of Physics, North Carolina State University, 
		Raleigh, NC 27695, USA}
	\begin{abstract}
		In this work, we first solve complex Morse flow equations for the simplest case of a bosonic harmonic oscillator to discuss localization in the context of Picard-Lefschetz theory. We briefly touch on the exact non-BPS solutions of the bosonized supersymmetric quantum mechanics on algebraic geometric grounds and report that their complex phases can be accessed through the cohomology of WKB 1-form of the underlying singular spectral curve subject to necessary cohomological corrections for non-zero genus. Motivated by Picard-Lefschetz theory, we write down a general formula for the index of $\mathcal{N} = 4$ quantum mechanics with background $R$-symmetry gauge fields. We conjecture that certain symmetries of the refined Witten index and singularities of the moduli space may be used to determine the correct intersection coefficients. A few examples, where this conjecture holds, are shown in both linear and closed quivers with rank-one quiver gauge groups. 
		The $R$-anomaly removal along the ``Morsified'' relative homology cycles also called ``Lefschetz thimbles'' is shown to lead to the appearance of Stokes lines. 
		We show that the Fayet-Iliopoulos (FI) parameters appear in the intersection coefficients for the relative homology of the quiver quantum mechanics resulting from dimensional reduction of $2d$ $\mathcal{N}=(2,2)$ gauge theory on a circle and explicitly calculate integrals along the Lefschetz thimbles in $\mathcal{N}=4$ $\mathbb{CP}^{k-1}$ model. The Stokes jumping of coefficients and its relation to wall crossing phenomena is briefly discussed. We also find that the notion of ``on-the-wall'' index is related to the invariant Lefschetz thimbles
		under Stokes phenomena. An implication of the Lefschetz thimbles in constructing knots from quiver quantum mechanics is indicated. 
	\end{abstract}
	\maketitle
	\tableofcontents
	\balance
	\section{Introduction and Summary}\label{sec:intro}
	
	Supersymmetric quantum mechanics is a very fruitful tool for studying mathematical properties of manifolds and algebraic curves. It also provides tractable prototypes to discuss (algebraic) topological invariants that arise from physical systems described by vacuum (e.g. zero-energy) states of generic supersymmetric quantum field theories. Therefore, at least from a topological standpoint, supersymmetric quantum mechanics boils down to calculating the ground states and their proper counting based on their parity under the action of $(-1)^F$ operator where $F$ is the fermion number operator. The result of this counting, considering the parity for every solution, leads to the most basic topological invariant protected by supersymmetry, namely, Witten index \cite{Witten:1981nf}:
	\begin{equation}
	\mathcal{I}=\text{tr} (-1)^F.\label{wittenindex}
	\end{equation}
	$\mathcal{I}$ also gives the Euler characteristic for the target manifold of the underlying field theory. 
	
	The study of (\ref{wittenindex}) led to a seminal work \cite{Witten:1982im} where the de Rham cohomology of compact finite-dimensional manifolds was computed using a deep relation between Morse inequalities and supersymmetric quantum mechanics. Two years later, Atiyah and Bott \cite{Atiyah:1984px} showed that de Rham version of the localization theorems in the context of
	equivariant cohomology are very closely related to Witten's result and gave a generalized exact stationary phase formula. A Morse theoretic study of Yang-Mills equations over Riemann surfaces (or algebraic curves) $M$ was done in \cite{Atiyah:1982fa} that proved useful in deriving the cohomology of the moduli spaces of stable algebraic vector bundles over $M$. These efforts paved the way for applying Morse theory to quantum field theories as well. 	

	The Picard-Lefschetz theory is a more recent attempt to deliver a crucial understanding of complexified path integrals and index technology by studying the topology of path space via the critical points of the holomorphized action. In this regard, a possible use of this theory in calculating the Witten index of $2d$ gauge theories was first addressed in $2d$ supersymmetric Landau-Ginzburg models in \cite{Cecotti:1992rm}. Not long ago, Witten introduced this theory into an analytically continued version of Chern-Simons path integral with Wilson loop operator insertions in an endeavor to recreate Jones polynomial of knots \cite{Witten:2010cx}. Another application was given in \cite{Harlow:2011ny} where it was shown that the complexified Liouville theory produces correct DOZZ formula only when its path integral is evaluated on integration cycles pieced together of  Lefschetz thimbles attached to multivalued complex saddle solutions. 
	
	More recent applications were considered in the context of $\mathcal{N}=1$ and $\mathcal{N}=2$ quantum mechanics \cite{Behtash:2015loa,Behtash:2015kva} as well as $\mathbb{CP}^{k-1}$ models \cite{Fujimori:2016ljw} from a semi-classical point of view.  
	Analytic continuation of path integrals have been shown to be a fundamental necessity of quantum theories to yield a correct semi-classical analysis consistent with supersymmetry \cite{Behtash:2015zha}. What was shown to be of crucial importance, was that no matter what real theories we are considering, the axiom of holomorphization is a first step to set the stage for understanding the path integrals. Above all else lies the fact that when one applies Picard-Lefschetz theory to a holomorphized theory, we would get so much than we initially asked for: Complex configurations that were missing in the analysis of ground state properties of physical theories, seem to crack the puzzles of semi-classics wide open. These mysterious puzzles including the vanishing of gluon condensate in $\mathcal{N}=1$ SYM theory and the positive semi-definiteness of non-perturbative contributions to the ground state energy of the bosonized $\mathcal{N}=1$ supersymmetric quantum mechanics were finally settled down using the aforementioned ideas combined with a key data that Picard-Lefschetz theory provides: Hidden topological angle (HTA) \cite{Behtash:2015kna}. Each amplitude assigned to saddle point in the stationary phase formula has to carry an extra topological phase coming from the homology of the (relative) cycle to which it is attached. This invariant phase along the cycle is usually a sign in supersymmetric theories. If one breaks supersymmtry, say takes the number of fermions to be real in quantum mechanics, ``resurgence'' would kick in because of the fact that complex phase changes the amount of the saddle amplitudes \cite{Dunne:2015eaa,Dorigoni:2014hea,Cherman:2014ofa}. There is an anomaly cancellation argument given in Subsec.~\ref{sub:signofAHTA} that could also be used. Furthermore, we already have a holomorphized action that contains most of what we need.  
	
	In supersymmetric gauge theories, localization principle provides an effective 1-loop function out of a complicated path integral that needs to be integrated on $1d$ integration cycles in a finite dimensional moduli space. The calculation of path integral is then literally boiled down to picking the right integration cycle around the poles of effective 1-loop function. Localization was formally applied to a supersymmetric quantum gauge theory in \cite{Pestun:2007rz}. Here, we are interested in the index of supersymmetric quantum mechanics arising from $2d$ Witten index refined by external gauge fields, leading to a categorification of (\ref{wittenindex}) often called {\it refined} Witten index, specifically, the $2d$ index of $\mathcal{N}=(2,2)$ theory on a torus $T^2$ known as ``elliptic genus'' which was calculated by implementing the Jeffrey-Kirwan (JK) residue operation in \cite{Benini:2013xpa,Benini:2013nda,Hori:2014tda,Benini:2016qnm} and using matrix models in \cite{Gadde:2013ftv}. Following these works, Ref. \cite{Cordova:2014oxa} computed the refined index of
	the $\mathcal{N}=4$ quantum mechanics resulting from the dimensional reduction of elliptic genus on a circle and Ref. \cite{Hori:2014tda}
	went through a thorough examination of the $1d$ Witten index of gauged linear sigma models with at least $\mathcal{N}=2$ by considering the full path integral, which is to date the most concrete account of $1d$ Witten index.
	
	We consider the same gauged quantum mechanical theories with four real supercharges with only finitely many vacua as in \cite{Cordova:2014oxa}. We partially address choosing the correct integration cycles by going through a Picard-Lefschetz analysis of the theory, and discuss whether the wall crossing phenomena are resulted from deformation of integration cycle away from the Stokes rays of the moduli space of the solutions to BPS
    equations in the localized theory e.g., $u$-space, that is suitably non-compact in $1d$. A more explicit answer will need a regularization term in the effective 1-loop determinant, as well as a Fayet-Iliopoulos (FI) term in the original theory \cite{Fayet},
	\begin{equation}
	S_{FI}=\zeta\int d^2\theta d^2\bar{\theta} V,
	\end{equation}
   where $V$ denotes the 4-supercharge vector superfield and $\zeta$
   aka FI parameter takes real values unless otherwise stated. For a complete answer, that does not require the use of regularization, we need to consider the Coulomb branch of the full $\mathcal{N}=(2,2)$ theory that will be done elsewhere where explicit appearance of FI parameters in the effective theory helps to put things into perspective in the Lefschetz thimble construction. 
	
	The organization of this paper and summary of results is in order. In Sec.~\ref{sec:Review} we use the elementary dynamics of a bosonic harmonic 
	oscillator by solving its Morse flow equations to pave the way for understanding the idea of localization to constant paths in the context of
	path integrals. We first holomorphize (complexify) the configuration space of the oscillator in question and
	then identify the imaginary direction with the momentum in an attempt to define the conserved Hamiltonian flow which 
	implies the localization once the boundary conditions are correctly set. The constant paths are downward flows attached to the critical points, which are universally referred to as \textit{Lefschetz thimbles}, form a relative homology of the non-compact ace, study of which falls under the umbrella of complex Morse and Picard-Lefschetz theories.   
	The Hamiltonian flow then associates a topological angle to 
	Lefschetz thimbles that in supersymmetric
	quantum theories would be necessary to fix their long-standing semi-classical issues in a rather systematic way. 
	
	The cohomology of singular algebraic curves corresponding to the bosonized potential of the supersymmetric quantum mechanics 
	fixed at the ground state energy level, will be considered in Sec.~\ref{sec:WKB} to provide a competing method of calculating the topological invariants of 
	the path space. We argue that for singular algebraic curves of supersymmetric quantum mechanics with non-zero genus, apart from higher order quantum corrections, the (classical) WKB 1-form (namely, that of WKB theory at zero order expansion in $g$) is not enough to capture the complex phases of its saddle configurations hidden in the topology of thimbles, so it needs to be corrected by considering the sheaf of holomorphic 1-forms. We compute these phases for the double- and triple- well systems.
	
	In Sec.~\ref{sec:Neq4QM}, the refined Witten index formula of supersymmetric quantum mechanics with four real supercharges will be given 
	by considering an integration cycle $\gamma$ that is along the non-compact directions of the moduli space. In the vast literature of 	index calculations, localization is the key feature of supersymmetric theories that produces a meromorphic top-form -1-loop determinant to be integrated over (more precisely the cover of) moduli space along a given path to be determined \cite{Pestun:2007rz}. We point out a common feature of gauge theories that is in the presence of a group action $G$, the saddle points
	will form orbits of the group, rendering them non-isolated \cite{Witten:2010cx}. Thus, it is appropriate to call such points ``saddle rims'' as they happen to be the boundaries of the path space at infinite imaginary directions. For a $U(1)$ gauge theory, the path space is made of infinitely many copies of $\mathbb{C}\setminus\{0\} \cong S^1 \times \mathbb{R}$, where the non-compact direction corresponds to the Lie algebra of the maximal torus of $U(1)$ gauge symmetry that is the Coulomb branch of the $1d$ quiver quantum mechanics \cite{Hori:2014tda}. Throughout this paper, for a reason explained below, we will refer to this non-compact direction in the path space as the ``Stokes wall'' on which the integration cycle $\gamma$ dwells. Using path homotopy constraint, we form a set of Lefschetz thimbles that flow out of the degenerate saddle rims. The elements of this set form a basis for the relative homology upon Morsification and therefore the contour will be written as $\gamma=\sum_a n_a\mathcal{J}_a$ for integer $n_a$. The index calculated on the Stokes wall will then be interpreted  as ``on-the-wall'' index that obviously receives only contributions from thimble integrals attached to saddle rims on the boundaries of the path space. Exactness of saddle point approximation will come in handy to avoid difficulties of taking these integrals in general quiver theories. 
	
	We note that by the Stokes wall we really mean a Stokes ray in the $u$-space and it happens that the integration cycle $\gamma$ sits on this ray, which is slightly different in general from the concept of the wall in the FI parameter space (e.g. $\zeta$-space). But the former should depend on $\zeta$ that is possible and controllable in the language of Picard-Lefschetz theory in the Coulomb branch of the problem wherein the two concepts of wall coincide \footnote{\label{foot1}The advantage of doing the latter is that we totally understand the structure of the contour in the JK residue operation in general $U(1)$ and higher-rank theories without the need to consider the individual poles which will be discussed elsewhere. There a Stokes wall is unambiguously identical to the wall in $\zeta$-space.}. The Stokes ray connects distinct supersymmetric vacua and upon regularization a computation of off-the-wall index yields naturally an index which is basically the difference between Witten indices in two chambers. In search of a way to connect $\zeta$ to the integration contour in Picard-Lefschetz theory, we propose that a gauge-invariant $\zeta$-dependent regularization term would give rise to unique Stokes jumps as different chambers in $\zeta$-space are probed in the process. This allows for a complex Morse function that subsequently
	helps to construct Lefschetz thimbles explicitly. This is done for the case of $\mathbb{CP}^{k-1}$ model.  We study the index of linear quiver systems of total gauge group $G=U(1)^\alpha$ in Sec.~\ref{sec:LAQM}, and closed quivers in Sec.~\ref{sec:ocq}. We observe that integration contour would be modified through this addition explicitly at the expense of losing non-compactness of the moduli space. In all the examples studied we find that the proposed index formula exactly reproduces the results of \cite{Cordova:2014oxa,Hori:2014tda}.
	
	Sec.~\ref{sec:2nodetheory} is intended to elaborate on the simple example of $2$-node quiver theory in detail. In Subsec.~\ref{subsec:thimbleCal}, we solve the alternative condition on the thimbles, namely the imaginary part of effective action is constant on the Lefschetz thimbles attached to the same saddle rims. Subsec.~\ref{subsec:FIdep} shows the connection between intersection coefficients and FI parameters starting from the $2d$ theory. Being 
	on a Stokes ray means that the effective action should satisfy a certain condition as studied in \cite{Witten:2010cx}. It is then shown that this condition imposes a constraint on the $R$-symmetry fugacity $y$ that in turn gives rise to the $R$-anomaly removal as is the case for the chiral theories where only one type of fermions (left or right) would couple to external gauge fields present in the original $\mathcal{N}=(2,2)$ theory. Sec.~\ref{sec:GenXYZ} is devoted to structuring a paradigm in which the data concerning integration cycle on the Stokes wall tells us about the value of index in all chambers as well as the jumping of Lefschetz thimbles.
	
	Finally, we discuss a possible future direction of research in Sec.~\ref{sec:knots} regarding knots. There, we make a connection between the HOMFLY polynomial of an unknot and refined Witten index of $\mathcal{N}=4$ $\mathbb{CP}^{k-1}$ model and hint at possibility
	of projecting a mirror symmetric knot $K$ onto the path space of some $\mathcal{N}=4$ quiver quantum mechanics in such a way that good regions and
	Lefschetz thimbles are identified with crossing points and line segments between crossings of $K$, respectively. 
	
	\section{Review of Picard-Lefschetz Theory, Axiom of Holomorphization and Localization}
	\label{sec:Review}
	In $d$ dimensions, Euclidean path integral of a (non)relativistic field $\phi:=\phi(\mathbf{x},t)$  propagating from point $\mathbf{x}_i$ to $\mathbf{x}_f$ reads
	\begin{equation}
	Z= \int \mathcal{D}\phi e^{-S[\phi(\mathbf{x},t)]},
	\end{equation}
	subject to the boundary conditions $\phi(\mathbf{x}_i,t_i)=\phi_i$ and $\phi(\mathbf{x}_f,t_f)=\phi_f$. Here $\mathbf{x}:=\{x_i\}_{i=1}^{d-1}$. A natural Picard-Lefschetz theory treatment of path integrals requires holomorphization of fields and coordinates as $\phi\rightarrow \hat{\phi}$ and promoting the action to a $\textit{holomorphic}$ action functional $\hat{{S}}[\hat{\phi}(z)]$ such that $\hat{S}=\int dt \hat{{L}}\equiv  h + i\theta$ where $h=\Re(\hat{S})$ is a ``Morse function'' (a real valued function with nondegenerate critical points, that here is chosen to be real part of a holomorphic action so should no dependence on $\bar{\hat{\phi}}$ exist), and $\Im(\hat{S})=\theta$ is a conserved quantity along the ``Morse flow'', also called steepest descent path. We will refer to $\hat{{S}}$ as holomorphic action 
	functional which will play a significant role in constructing flow equations \cite{Witten:2016qzs}.
	 
	The paths are governed by the following set of equations, sometimes referred to as {\it complex Morse equations}:
	\begin{equation}
	\frac{\partial \hat{\phi}}{\partial\tau} =\frac{\delta \bar{\hat{{S}}}}{\delta\bar{\hat{\phi}}},\quad \frac{\partial\bar{\hat{\phi}}}{\partial\tau} = \frac{\delta \hat{{S}}}{\delta\hat{\phi}} \label{FlowEqs}
	\end{equation}
	It is important to note that the parameter $\tau$ is a \textit{flow parameter} that may be independent of time direction. In general, we take $\tau\in \mathbb{R}$ and therefore the infinite dimensional complexified field space is defined to be $\Gamma_\mathbb{C}:=\{\hat{\phi}(\mathbf{x}(t,\tau))|\mathbf{x}(t,\tau)\in\mathbb{R}^{d-1}\}$, which are subject to the non-Cauchy boundary condition
	\begin{equation}
	\lim_{\tau\rightarrow -\infty}\hat{\phi}(\mathbf{x}(t,\tau))= \hat{\phi}_{\rm cr}(\mathbf{x}(t)),
	\end{equation}
    where $\hat{\phi}_{\rm cr}(\mathbf{x}(t))$ is the path over which the action $\hat{S}$ is stationary.
    In $0+1$-dim, the action functional in Euclidean signature is written as
    \begin{equation}
    \hat{S}[z]= \int dt (\tfrac{1}{2}\dot{z}^2+V(z))
    \end{equation}
     where $z(t)=x(t)+iy(t)$ and over-dot means derivative with respect to time $t$. We note that $z$ is a coordinate, i.e. our field $\hat{\phi}(t,\tau)$,
     that takes values in $\Gamma_{\mathbb{C}}=\mathbb{C}.$  The flow equations \eqref{FlowEqs} can be replaced by
     the heat equation \cite{Salamon:2003}  
    \begin{equation}
    \frac{\partial \bar{z}(t,\tau)}{\partial\tau} =  \ddot{z}(t,\tau) -\frac{d V(z)}{dz} \label{eq:heat}
    \end{equation}
	and its complex conjugate. 
	 
	 Let us now focus on a simple harmonic potential $V(z)=\tfrac{1}{2} \omega^2z^2$ in $0+1$-dim where the path space is simply $\mathbb{C}$.
	 Also, the motion begins at $t=0$ from the point $(x_0,y_0)$ which we set them to be the point $(0,0)$, is found to be
	 at the point $(x_T,y_T)$ at $t=T$.
	 Then \eqref{eq:heat} becomes
	 \begin{equation}
	   -\partial_\tau x(t,\tau)+i\partial_\tau y(t,\tau)=  (\ddot{x}-\omega^2 x )+i(\ddot{y}-\omega^2 y ).\label{eq:heat_HO}
	 \end{equation}
	 Since $x,y$ are both real we can decouple this into two equations 
	 \begin{equation}
	 	(-\partial_\tau-\partial_t^2 +\omega^2)x(t,\tau) =0,\quad (\partial_\tau-\partial_t^2 +\omega^2)y(t,\tau) =0.
	 \end{equation}
	 Therefore, the solutions can be simply given as
	 \begin{subequations}
	 \begin{eqnarray}
	 x(t,\tau)&=&x_T\frac{\sinh(\omega t)}{\sinh(\omega T)}-\sum_{n=0}^\infty C_n(\tau) \sinh(\omega_n t),\\
	 y(t,\tau)&=&y_T\frac{\sinh(\omega t)}{\sinh(\omega T)}-\sum_{n=0}^\infty C'_n(\tau)\sinh(\omega_n t), \label{eq:HOthimble}
	 \end{eqnarray}
 \end{subequations}
     where we have defined $C_n(\tau)=c_n\exp[{-(\omega_n^2-\omega^2)\tau}]\ge0$ and $C'_n(\tau)=c_n\exp[{-(\omega^2-\omega_n^2)\tau}]\ge0$. Here the mode frequencies are $\omega_n = \pi n/T$ where $T$ is the period of the motion. It can be seen that $\lim_{\tau\rightarrow-\infty}(x+iy)=
     (x_T+iy_T)\sinh(\omega t)/\sinh(\omega T)$  provided that there is a separate bound
     on each sum over modes for the convergence of both solutions simultaneously. Thus, there is an integer $n^*$ such that $\omega_{n^*}^{2}$ 
     is closest to $\omega^2$ from above and with which then one
     correctly reproduces the real-time harmonic motion at the bottom of the well, $x=y=0$, 
     \begin{flalign}
     z(t,\tau) &= C_n(\tau)\left(-\sum^{n^* -1}_{n=1} \sinh(\omega_n t)-i\sum_{n=n^*}^{\infty} \sinh(\omega_n t)\right)~+~\nonumber\\ 
     &\hspace{2.7cm} ~+(x_T+iy_T)\frac{\sinh(\omega t)}{\sinh(\omega T)}, \label{eq:HOthimble_vf}
     \end{flalign}
     the only saddle point configuration
     present in the harmonic oscillator. We denote the path with coordinates \eqref{eq:HOthimble_vf} 
     by $\mathcal{J}$. Of course when calculating the path integral for an oscillating system, the 
     ultimate goal is to analytically continue back to real time, $t\rightarrow i t_r$. Doing this would result in  
     a product of  $\mu_M\equiv n^*-1$ factors of $-i$ that coalesce to create an overall phase of
     \begin{equation}
     e^{-i\mu_M\pi/2}, \label{eq:Maslov}
     \end{equation}
     where $\mu_M$ is known as ``Maslov index'', which plays an important role 
     in the classification of
     Lagrangian submanifolds in symplectic topology \cite{Piccione2008} and WKB approximation. Here, its presence is explicitly justified 
     by the holomorphization of path space and flow equations where the role of symplectic structure $dx\wedge dp$ in the phase space 
     is replaced by $dxdy$ in the complexified configuration space. This has been also noticed in the real-time analysis of the same problem
     in \cite{Tanizaki:2014xba}. With this factor included, the measure boils down to the flow line attached to 
     the saddle configuration at $z=0$ and fluctuations around it, namely
     \begin{equation}
      \int_{\Gamma_\mathbb{C}}Dz\rightarrow\int_{\mathcal{J}}Dz \sim e^{-i\mu_M\pi/2} \int\prod_{n=1}^\infty dC_n.
     \end{equation}
    The rest of the computation follows easily from here. We emphasize that this equation
    shows that the path integral of the harmonic oscillator is localized to a specific path connected to a bosonic 
    saddle point, which provides a simple example of {\it localization principle}.
    
     We note the solutions along $x$ and $y$ directions are not the same and the complexified system is not 
     equivalent to a pair of coupled harmonic oscillators; they satisfy a set of {\it distinct heat equations} where along one, the eigenvalues
     of the flow Hamiltonian increase and they do the opposite along the other but still keep their positive-definiteness. This has a consequence in terms of dynamics of the motion which is throughly discussed
      in \cite{Behtash:2015loa} for more general $0+1$-dim systems.
     
     In the real-time formalism of path integrals, the Lagrangian becomes proportional to Hamiltonian and one now has to worry about the implicit $\tau$ dependence of $x$ and $y$ coordinates and ask to see if there is any conserved quantity along the flow line $\mathcal{J}$ that is preserved under time translation. The answer is in the affirmative because using (\ref{FlowEqs}), it is evident that 
     \begin{equation}
     \frac{d\theta}{d\tau} \propto \frac{d[\hat{{S}}-\bar{\hat{{S}}}]}{d\tau} =\frac{\delta\bar{\hat{{{S}}}}}{\delta \bar{\hat{\phi}}}\frac{\partial\bar{\hat{\phi}}}{\partial\tau}-\frac{\delta\hat{{{S}}}}{\delta \hat{\phi}}\frac{\partial\hat{\phi}}{\partial\tau}=0.
     \label{eq:imH}
     \end{equation} 
     In the motion of a simple harmonic oscillator, this readily is given by
     \begin{equation}
     \theta = \int dt\, (\dot{x}(t,\tau) \dot{y}(t,\tau)+\omega^2 x(t,\tau) y(t,\tau)).
     \end{equation}
     To simplify things, we notice that if the derivative is zero, it is zero even at the saddle
     solution. Hence,
     \begin{eqnarray}
     \theta&=&\frac{x_Ty_T\,\omega^2}{\sinh^2(\omega T)}\int_{0}^{T} dt\,(\cosh^2(\omega t)+\sinh^2(\omega t)) \label{eq:pdx}\nonumber\\
     &=&x_Ty_T\,\omega\,\coth(\omega T).
     \end{eqnarray}
     
	  Technically speaking, the union of flow lines reached by a critical point at $\tau\rightarrow -\infty$ is called a ``Lefschetz thimble''
	  which we already denoted by $\mathcal{J}$. Lefschetz thimbles are middle dimensional manifolds attached to a critical point and flow down to good regions $G_a$. Here, `good' means that the integral converges in these regions. There is a homological interpretation of this idea that will come in a bit.
	
	However, the opposite scenario would have led to an \emph{upward} flow - also known as $\mathcal{K}$-cycle over which $h\rightarrow-\infty$. This is not good because a $\mathcal{K}$-cycle always flows in orthogonal directions to $\mathcal{J}$-cycles and therefore the integral always diverge on them. This orthogonality means that in general, the intersection of a $\mathcal{K}_a$-cycle and a $\mathcal{J}_b$-cycle happens to give either one or zero. This  in compact notation is written as $\langle\mathcal{K}_a,\mathcal{J}_b \rangle=\delta_{ij}$.  It may be worth mentioning that a downward flow that starts at one critical point cannot end at another critical point and will always flow in the direction of $h\rightarrow\infty$ unless we are on a ``Stokes ray'' where flow lines `overlap', that is to say that the paths of steepest descent become one that connects two critical points. By crossing a Stokes ray, a cycle may or may not jump but it is ill-defined exactly on the ray  itself which requires further work to be made sensible.
	
	Given a (non-compact) manifold $X$ of arbitrary dimension, if for some very large flow time $\tau < \tau^\star$, $h>L$ with $L$ being really large, then $X_{\tau^\star}$ denotes the union of good regions $G_a$. So we can form $\mathbb{Z}$-valued relative real $k$th-homology group $H_k(X,X_{\tau^\star})$ and use a Morse function $h$ on $X$ to determine an upper bound for the rank of each group. It can be shown that if the differences between the Morse indices ($=$ number of negative eigenvalues of the Hessian matrix at a given critical point) of distinct critical points of $h$ are different from $\pm 1$, then the rank of $m$-dim homology groups is equal to the number of critical points of Morse index $m$. In our trivial example, the Morse index of $\rho=0$ is one so $H_1(X,X_{\tau^\star})$ is of rank 1 and all other relative homology groups vanish. In the context of supersymmetric theories in physics, the invariant measured by this relative homology is usually a Witten-type index, which by a clue from the bosonic harmonic theory discussed above, roots back to localization to Lefschetz thimbles attached to BPS configurations. 
	
	In general, this holomorphization procedure for any path integral over a field space $\Gamma_\mathbb{R}$ yields an equivalent formulation with a path integral over an integration cycle in the complexified field space $\Gamma_\mathbb{C}$, such that we can write the path integral as
	\begin{equation}
	Z= \int_{\Gamma_\mathbb{R}} \mathcal{D}\phi e^{-{S}[\phi]}\equiv\sum_{a\in \Sigma}n_a\int_{\mathcal{J}_a} \mathcal{D}\hat{\phi}  e^{-\hat{S}[\hat{\phi}]}\label{PLFormula}
	\end{equation}
	where $n_a\in \{\pm1,0\}$ and $\Sigma$ is the set of all critical points of $\hat{S}$. This is what one means by Picard-Lefschetz theory.
	We notice that the critical points must be non-degenerate and isolated for \eqref{PLFormula} to follow through unambiguously. 
	
	As a last remark, we point out that in case the critical points are degenerate such that their Morse index is ill-defined, or equivalently if the real part of the (effective) action is not a Morse function, the formula (\ref{PLFormula}) is not well-defined. Asking $n_a$ to take values in $\mathbb Z$, instead, we keep track of this degeneracy factor which is infinite since $1/\mathcal{H}(p)=\pm\infty$ where $\mathcal{H}(p)$ is the Hessian (determinant of Hessian matrix) at the degenerate saddle point $p$,
	\begin{equation}
	\mathcal{H} = \det \left(\frac{\partial^2 h}{\partial {\hat{\phi}_a} \partial {\hat{\phi}_b}}\right),
	\end{equation}
	where $\hat{\phi}_a$ are the set of all real and imaginary parts of every complexified field that the Morse function 
	$h$ depends on.
	
	 In supersymmetric quantum mechanics, the degeneracy can be removed topologically by path homotopy equivalence of Lefschetz thimbles. In short, any two Lefschetz thimbles attached to degenerate (and non-isolated as in gauge theories) saddle points that \emph{belong} to the same homotopy class are deemed equivalent and should be counted only once. Homologically, the $\cal J$-cycles in $H_k(X,X_{\tau^\star})$ must describe a basis for any generic cycle $\gamma$ in $X$ with its boundaries lying in $X_{\tau^\star}$. 
	
	Even though in a theory with degenerate saddle points, a Morse function is ill-defined (unless as shown in Subsec.~\ref{sec:2nodequiver} regularized in a way that degeneracy is lifted or, the {\it Morsification} process \cite{Arnold}), still we can distinguish between saddle points and corresponding 
	Lefschetz thimbles by path homotopy. This enables us to define the irreducible $\mathbf{J}$-set whose elements $\mathcal{J}_a$ are (1) generators of $H_k(\mathfrak{F},\mathfrak{F}_{\tau^\star})$ and (2) form a homotopy class $[\mathcal{J}_a]_\pi$ where $[.]_\pi$ means `modulo path homotopy'. This removes the degeneracy of saddle points in theory and renders a clear interpretation of the homological quantum mechanics in the presence of a gauge group action provided that the intersection coefficients be determined by Witten index check or other alternative checks discussed in Subsec.~\ref{subsec:intersection}. We will concentrate on the imaginary phase of this effective action at the saddle rims defining the elements of $\mathbf{J}$-set in our treatment of Picard-Lefschetz theory. 
	
	\section{Hidden Topological Angles, Spectral Curves and Non-BPS objects}\label{sec:WKB}
	In this section we explain the connection between degenerate (or singular) spectral curves and
	Non-BPS solutions to the integrable/quasi-exact-solvable systems related to them.
	\subsection{Complex and bions and singular spectral curves}
	In general, the topological invariants hidden in the complexified space of all paths described by flow equations (\ref{FlowEqs}) are related to the 
	phase shifts that occur by crossing some turning points included in the trajectory of a particle. These invariants are closely related to the action-angle
	variables in the study of symplectic manifolds that could describe the phase space of the very same particle. These invariants are often angles that arise as a result of holding action constant along a trajectory. For instance, the
	well-known periodic trajectory of a harmonic oscillator in the phase space classically gives the following action-angle variable (in units of $\hbar$)
	\begin{equation}
	H_a=i\oint_\gamma p dx=2i\pi k_a \Rightarrow \theta_a = 2\pi k_a, \label{HTA_v1}
	\end{equation} 
	for integer $k_a$ and $\gamma$ being a circle of radius $\sqrt{2E}$ at energy level $E$. Quantum mechanically, the Maslov correction
	adds a phase shift of $\pi$ that accounts for the $1/2$ energy of the ground state.
	
	Yet another method to get the phases of saddle points would be through studying algebraic geometry and (co)homology theory of the potential curves. In (non)supersymmetric theories with double-well superpotential $W=  z^3/3 - z$  \cite{Behtash:2015kna}, the only non-perturbative contribution to ground state energy comes from a non-BPS 
	exact complex instanton-antiinstanton solution which starts at the true vacuum of the system at $z^m$ and turns back 
	from either the complex turning point $z^T$ or its complex conjugate $\bar{z}^T$, completing the periodic motion with infinite periodicity. This system is part of a bigger family of integrable systems which are described by the singular spectral curves $\mathfrak{S}_a(\mathbb C)$ defined (in Euclidean signature) by
	\begin{eqnarray}
	y^2 &=&(W'(z))^2 + k_a g W''(z)+2E^m_a\label{spec_gen}\\
	&=& (z^2-1)^2 + 2 k_a g z +2E^m_a \label{spec_dw}
	\end{eqnarray} 
	where $g\in \mathbb{R}$ is a coupling constant, $k_a$ is the number of fermionic degrees of freedom, and $-E^m_a$ is energy at the global minimum of the potentials 
	$V_a(z)=\tfrac{1}{2}(W'(z))^2 + \tfrac{k_a}{2} g W''(z)$ with $W'(z)$ being of at least degree $d=2$. In general, $k_a$ may be an integer but only for $k_a=1$ (or $k_a = -1$ in certain cases)
	the theory is supersymmetric. For double-well superpotential, the maps (\ref{spec_dw}) look like $y^2\sim(12(z^{m}_{a})^2-4)(z-z^{m}_{a})^2$ near $z^m_a\in \mathbb R$, so $z^m_a$ have multiplicity 2 and since they are global minima of potential graphs, hence simple ramification points of index $\nu_{z^m_a}=2$. The first relative (co)homology group of the spectral curves (\ref{spec_dw}) with respect to the ramification divisor $\mathcal D = 2z^m_a$ is $\mathbb Z/2\mathbb Z$-graded (for the supersymmetric case $k_a=\pm1$) which is shown by computing the integral of the $1$-form $\sqrt{y^2}dz$ over the branch cut $\mathcal C$ that connects two complex conjugate pair of fixed points, e.g. turning points, $[z_a^T,\bar{z}_a^T]$ of $\mathfrak{S}_a(\mathbb C)$ at energy level $E^m_a$:
	\begin{equation}
	H_a=\frac{1}{ g}\int^{z_a^T}_{\bar{z}_a^T} dz \sqrt{2E_a^m+W'(z)^2 + k_a g W''(z)}, \label{period_gamma_dw}
	\end{equation} 
	which is literally nothing but the phase of WKB approximation at zero order in $g$-expansion. In the rest of this section,
	we take $\Im(z_a^T)>0$.
	
	It is 
	interesting to note that for the double-well system, one has
    \begin{equation}
     z_a^m\Im(z_a^T) = -g\,{\rm sign}(k_a).\label{eq:Imzm}
    \end{equation}
	 The integral \eqref{period_gamma_dw} for a genus zero singular algebraic curve is calculated to be $H_a=i\, \pi k_a$ up to a sign, giving what we had called early on a hidden topological angle (HTA): $\theta_a = \pi k_a$  \cite{Behtash:2015kna}. A motivation for this name comes from the lack of any input in flow equations (\ref{FlowEqs}) that includes a topological term to produce {\it a priori} a phase for the saddle points. 
	 The integral introduced in \eqref{period_gamma_dw}
	 may also be equally represented by the residue formula 
     \begin{eqnarray}
     H_a &=& \frac{1}{2g} \oint_{\mathcal{C}_\odot}\sqrt{(z^2-1)^2 + 2 k_a g z +2E_a^m }\,dz\nonumber\\
     &=& \frac{1}{2g}  \oint_{\mathcal{C}_\odot}\sqrt{(z-z_a^m)^2(z-z_a^T)(z-\bar{z}_a^T)}\,dz\nonumber\\
     &=& \frac{1}{2g}  \oint_{\mathcal{C}_\odot}t^2\sqrt{1+\frac{4z_a^m}{t}+\frac{(\Im(z_a^T))^2+4(z_a^m)^2}{t^2}}\,dt\nonumber\\
     &=& \frac{1}{2g}  \oint_{\mathcal{C}_\odot} t^2 \left(1+\dots-\frac{\Im(z_a^T)z_a^m}{t^3}+\mathcal{O}(t^{-4})\right)dt\nonumber\\
     &=& i\pi k_a\, z_a^m\Im(z_a^T) \nonumber\\
     &=& i\pi k_a\, {\rm sign}(k_a),\label{eq:newHi}
    \end{eqnarray}
    where $\mathcal{C}_\odot$ is a contour enclosing the global minimum $z_a^m$. Here,
    use was made of the substitution $z-z_a^m=t$ and eq.~\eqref{eq:Imzm}. 
    It is clear that away from the roots of $y^2$, the 1-form $\sqrt{y^2}dz$
    is holomorphic and the HTA is a measure of the monodromy of this differential
    around the singular (degeneration) point $z=z_a^m$ since exactly there it fails to
    be exact.
    
	We now seek a set of appropriate transformations to reduce $\mathfrak E_a(\mathbb C)$ with $W=  z^3/3 - z$ to nodal cubic curves and comment on the connection between singular points and HTA. 
	
	Starting with the generic algebraic curve 
	\begin{equation}
	y^2 + l^a_1 z y + l^a_3 y - (z^3 + l^a_2 z^2 + l^a_4 z + l^a_6)=0,  \label{cubic_form}
	\end{equation}
	where 
	\begin{eqnarray}
	l^a_1 &=&\frac{ 2k_a g}{\sqrt{1 + 2 E^m_a}}, l^a_2 = -2 - \frac{k_a^2 g^2}{
		1 + 2 E^m_a},\nonumber\\
	l^a_3 &=& 0, l^a_4 = -4 (1 + 2 E^m_a), l^a_6 = 4 k_a^2 g^2 + 8(1 + 2 E^m_a),\nonumber
	\end{eqnarray}
	we find that the set of transformations
	\begin{eqnarray}
	z&=&\frac{1}{u^2} (2 \sqrt{1 + 2 E^m_a} (v + \sqrt{1 + 2 E^m_a}) + 2 k_a g u)\nonumber\\
	y&=&\frac{1}{u^3}(4 (1 + 2 E^m_a) (v + \sqrt{1 + 2 E^m_a}) \nonumber\\
	&+& 2 \sqrt{1 + 2 E^m_a} (2 k_a g u + c u^2) - \frac{d^2 u^2}{2 \sqrt{1 + 2 E^m_a}}) \nonumber,
	\end{eqnarray}
	turn (\ref{cubic_form}) into the quartic equation $v^2=2 E^m_a + 2 k_a g u + ( u^2-1)^2$ describing $\mathfrak E_a(\mathbb C)$ in $(u,v)$ coordinates.
	Hence, the algebraic curves
	\begin{eqnarray}
	y^2 &-& z^3 + (2 + \frac{ k^2_ag^2}{1 + 2 E^m_a}) z^2 + z (4 + 8 E^m_a + \frac{ 2k_ag}{\sqrt{1 + 2 E^m_a}}) \nonumber\\
	&-& 8 - 16 E^m_a - 4 g^2 k_a^2=0,  \label{cubic_form_vf}
	\end{eqnarray}
	are the reduced spectral curves associated with degree 4 orbits at energy $E^m_a$ in the inverted double-well potential.
	Introducing 
	\begin{eqnarray}
	m^a_1 &=& (l^a_1)^2 + 4 l^a_2,\nonumber\\
	m^a_4 &=& l^a_1 l^a_3 + 2l^a_4, m^a_6 = l^a_3 + 4l^a_6,\nonumber
	\end{eqnarray}
	and a change of variables $y\rightarrow 2y$ and $ z\rightarrow z - \frac{l^a_2}{12}$, equation (\ref{cubic_form_vf}) turns into
	\begin{equation}
	y^2 - z^3 +(\tfrac{16}{3} + 8 E^m_a) z  - 4 k_a^2g^2 -\frac{32 E^m_a}{3}-\frac{128}{27}=0.  \label{cubic_form_v2f}
	\end{equation}
	Numerical analysis verifies that the discriminant of this algebraic cubic curve is zero because of an ordinary double point at $z^m_a$:
	\begin{eqnarray}
	\Delta &=&  64(\tfrac{16}{3} + 8 E^m_a)^3 -16(108 k_a^2g^2 +288 E^m_a+128 )\nonumber\\
	&=& 0,
	\end{eqnarray}
	from which one can also solve for the energy of complex bion $[\mathcal I \bar{\mathcal I}]$ correlated events.
	\begin{figure}
		\centering
		\includegraphics[width=0.8\linewidth]{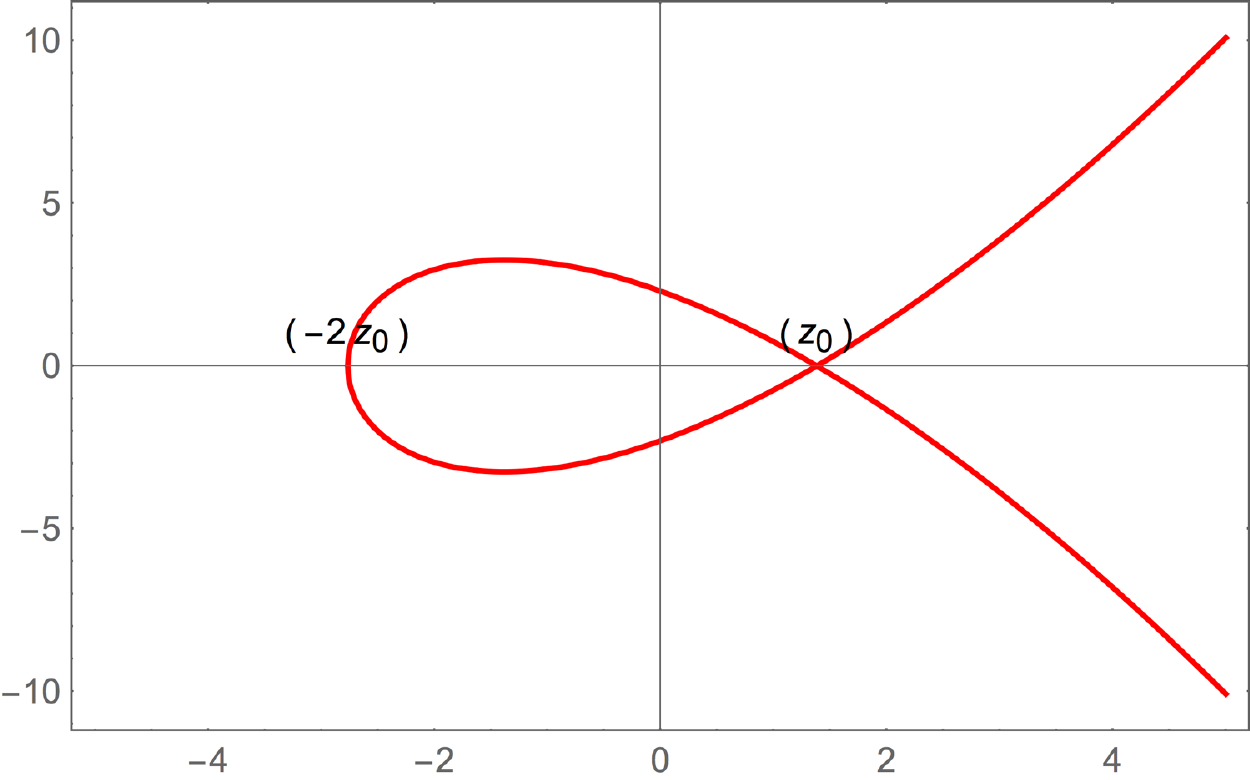}
		\caption{ Graph of the reduced spectral curves $\mathfrak{E}_a(\mathbb C)$ over the $\text{char}=0$ field of 
			complex numbers. The closed orbit here shows an exact periodic non-BPS solution of equations of motion
			in the tilted double-well potential known as \textit{complex bion} $[\mathcal I \bar{\mathcal I}]$, that is a complex configuration on a thimble attached to its critical point at infinity \cite{Behtash:2018}. There is an ordinary double point, i.e., node, at $z_0$ originated from the singularity of the curves (\ref{spec_dw}). Unlike smooth elliptic Weierstrass curves, $\mathfrak{E}_a(\mathbb C)$ have genus 0 in view of Riemann-Hurwitz formula. }
		\label{nodalcurve}
	\end{figure}
	Same analysis indicates that there is a simple turning point at $z= -2z^a_0$ which is a branch point 
	of the spectral curve (\ref{cubic_form_v2f}), and a double turning point at $z = z^a_0$, being in turn the ordinary double point of $\mathfrak{E}_a (\mathbb C)$, see
	Fig.~\ref{nodalcurve}. Therefore, the final form of the algebraic curve encoding complex bion can be written as the nodal cubic curve
	\begin{equation}
	E_0:\, y^2= (z-z^a_0)^2(z+2z^a_0). \label{cb_curve}
	\end{equation}
	which defines a moduli space $M_{k_a,g}$ for all possible non-perturbative configurations contributing to the ground-state
	of the tilted double-well system. The genus of this curve is 0 and over $\mathbb C$ it indeed is the Riemann sphere or complex projective space $\mathbb{CP}^1$ with the two poles $z=0$,$\infty$ identified. Technically, a torus with a complex structure constant $\mathbb{t}$, $T^2=\mathbb C /(\mathbb Z + \mathbb{t} \mathbb Z)$ becomes homeomorphic to a nodal cubic curve at the limit $\mathbb{t}\rightarrow i\infty$ as one cycle goes away. There is a holomorphic map $\mathbb{CP}^1\rightarrow \mathbb{CP}^2$ defined by $h_a:w\mapsto [z_a(w),y_a(w), 1]$ where
	\begin{equation}
	z_a(w) =z^a_0+\frac{12wz^a_0}{(w - 1)^2},\,\text{and} \,\,y_a(w) =\frac{4w(w + 1)(3z^a_0)^{\tfrac{3}{2}}}{(w - 1)^3},
	\end{equation}
	that maps $\mathbb{CP}^1$ onto nodal/singular curve $E_0$ which is injective away from the points $0,\infty \in  \mathbb{CP}^1$ at which the double point singularity of $E_0$, namely, $[z^a_0,0,0]$ is reached via $h_a$. 
	 
	The orbit seen in Fig.~\ref{nodalcurve} is
	a homology cycle (a line generated as a result of flow by steepest descent) attached to the north pole (critical point) of $S^2$, starting there at time $\tau\rightarrow-\infty$, and finally reaching the south pole at time $\tau=\tau^\star$. In spherical coordinates, this flow line is actually a semi great circle $\phi= \phi_0$ with the conserved quantity being
	\begin{equation}
    \text{circumference} =\pi k_a 
	\end{equation}
	where $k_a$ is the scale of the Riemann sphere for $\mathfrak{E}_a$. This is exactly equal to the HTA calculated using the integral formula given in (\ref{period_gamma_dw}). Upon taking $k_a$ to be an integer, we get a Bohr-Sommerfeld type quantization for the HTA and no resurgence will take place in the sense of \cite{Dunne:2015eaa}. Thus, the orbit in Fig.~\ref{nodalcurve} clearly represents the circle of $E_0\cong S^2\vee S^1$ and $k_a$ in this sense is the scale of the $S^1$. It is hence expected that any quantum theory with $E_0$ as its quantized moduli space of classical vacua accepts a possible non-BPS exact solution contributing non-perturbatively to the vacuum energy and breaking supersymmetry. 
	
	More insight in this regard is drawn from the thimble analysis of near-supersymmetric $\mathbb{CP}^{k-1}$ quantum mechanics in the quasi moduli space of kinks and anti-kinks \cite{Fujimori:2017oab}. There it is demonstrated using explicit calculations for $k=2$ case that the theory does entail complex bion solutions but they do not have a $\text{mod}\,\pi$ HTA. We elucidate this result in the context of current section by mentioning that the length of every great circle of $\mathbb{CP}^{1}\cong S^2$ corresponding to a non-perturbative solution in the supersymmetric limit (i.e. $\epsilon=1$ in \cite{Fujimori:2017oab}) is naturally equal to $2\pi$ in units of $\epsilon$ with an integer winding number for higher order (real) bions. Also, it is obvious that the spectral curve $y=\sqrt{V(|z|)+2E^m}$
	does not develop a cusp singularity at the energy level $-E^m$ associated with global minimum of the potential $V(|z|)$ because $y$ only depends on  $|z|\ge 0$ ($|z|$ being the modulus of the inhomogeneous coordinate $z$) contrary to $\mathfrak{E}_a (\mathbb C)$ described by  (\ref{spec_dw}).
    	
	If we insist to know the Picard-Lefschetz theory of a complex bion, it is apparent that the singularity in Fig.~\ref{nodalcurve} is a setback. But it is a cusp catastrophe that could be well examined using
	catastrophe theory, which is a potentially powerful replacement for Picard-Lefschetz theory \cite{Arnold1}.
	
	A recent development concerning the spectral curves in the $A_1$ and $A_{N-1}$ theories of class $\mathcal{S}$ in $4d$ $\mathcal{N} = 2$ theories is the interesting concept of ``BPS graph'' whose nodes correspond to the roots of the spectral curve of the Hitchin system related to the theory \cite{Gabella:2017hpz}. Therefore it is believed on general grounds that the collision of roots in the BPS graph associated with the potential curve $y^2=(z-z^a_0)^2(z+2z^a_0)$ is related to complex bions. It would be nice to understand what exactly 
	this connection is, but we leave that to future work.
	\subsection{The sign of HTA} \label{sub:signofAHTA}
	In general the question that one has to address regarding the HTA is that its sign seems 
	to be arbitrary. One way of resolving this issue is to consider if anomalies would arise because of a change in the Lefschetz thimble integrals. We recall that the topological invariant here is merely a relative homology or more suitably a relative cohomology   
	$H^1(\Gamma_{\mathbb{C}},G)$ where $\Gamma_{\mathbb{C}}$ is the path (field) space of the quantum mechanics or
	quantum field theory and $G\subset \Gamma_{\mathbb{C}}$ is the union of all (disjoint) good regions. 
	In Picard-Lefschetz theory, upon changing the relative cycles $\mathcal{J}_a$ to $\mathcal{J}_a^*$ in the expression 
	\begin{equation}
	Z=\int_{\Gamma_\mathbb{C}}\mathcal{D}\tilde{x}\, e^{-S(\tilde{x})} = \sum_{a} n_a\int_{\mathcal{J}_a}d\tilde{x}\,e^{-S(\tilde{x})}=\sum_{a} n_aZ_a,
	\end{equation}
    $Z^*$ would make sense as long as $\theta_a={\theta}^*_a\mod 2\pi\mathbb{Z}$ since
	$Z^*_a=\exp{i(\theta_a-\theta^*_a)}Z_a$. We then say that $\mathcal{J}_a,{\mathcal{J}}^*_a$ subject to the same boundary conditions are
	in the same relative (co)homology class and therefore are path homotopic $\mathcal{J}_a\cong{\mathcal{J}}^*_a$. 
	
	Otherwise, this would imply that there is an anomaly ${\rm Ano}_a=(\exp{i(\theta_a-\theta^*_a)}-1)Z_a$ that must
	be removed by introducing a boundary term in $S(\tilde{x})$ that is anomalous in retrospect. This generically requires the integration manifold
	$\mathcal{C}=\sum_a\mathcal{J}_a$ to have a boundary but if $\mathcal{C}$ is closed (thus defining an ordinary homology as will be the case in the upcoming sections), it is not clear how this would be fixed (One cure might come from the insertion of punctures on the boundary and possibly inside $\mathcal{C}$).
	
	For integer values of $k_a$ there are no anomalies arising from the sign of HTA. For all other values of $k_a$, the anomaly of $Z^*_a$'s can be
	canceled by insertion of appropriate terms at the boundary as discussed above. The consequence of this breaking of supersymmetry with the help of gauged linear sigma model formulation of $2d$ $(2,2)$ $\mathbb{CP}^{N-1}$ theory on $S^2$ was explored in a recent work \cite{Dorigoni:2017smz} using ideas from resurgence theory. But to our knowledge, there is no understanding of path integral of the theory in the presence of supersymmetry-breaking terms. 
    
	\subsection{Holomorphic 1-forms and higher genus}
	Higher genus algebraic curves that develop complex bions $[\mathcal I \bar{\mathcal I}]^n$ ($n\ge 1$) are also obtained by generalizing
	$\mathfrak E_a(\mathbb C)$ to higher degree ($d>2$) -hyperelliptic- spectral curves with singular points. Notice that for curves with simply $E_a^m$ replaced by a generic $E_a$ in \eqref{spec_gen}, the genus formula is given by
	\begin{equation}
	\mathbf{g}=d-1,
	\end{equation}
	which is equal to the dimension of space of holomorphic 1-forms i.e., $\frac{z^{d-n}dz}{y}$ for $n=1,\dots \mathbf{g}$ or the number of handles on the corresponding Riemann surface $\mathcal{S}_{E_a}(\mathbb{C})$. However, $E_a(\mathbb C)$ locally looks like $y^2\sim z^d+\dots$ -physically corresponding to the famous ``Argyres-Douglas'' point in Seiberg-Witten type theories, for which genus formula is then replaced by
		\begin{equation}
	\mathbf{g}_s=d/2-1, {\rm for~even}~d,\quad \mathbf{g}_s=(d-1)/2, {\rm for~odd}~d.
	\end{equation}
	In such situations, $(d-1)/2$ for odd $d$ (and $(d/2)$ for even $d$) of the sections of the fibers $V_{E_a}= H^1(\mathcal{S}_{E_a}(\mathbb{C}),\mathbb C)$ (with $E_a$ being the energy level) of the locally trivialized flat bundle $V$ over the Riemann surface of genus $\mathbf{g}$ vanish, giving rise to $\mathbf{g}\rightarrow \mathbf{g}_s$. Because complex bions occur exactly at the degeneration point $E_a=E_a^m$, 
	one has to come up with a suitable basis -a linear combination of holomorphic 1-forms on $\mathcal{S}_{E_a}(\mathbb{C})$- to integrate over
	along the branch cuts, which can be given as
	\begin{equation}
	\omega_{c}=2 \frac{P(z)dz}{y}=2 \sum_{n=1}^{\mathbf{g}}\frac{c_nz^{d-n}dz}{y}\label{eq:corr}
	\end{equation}
     where $c_n$ are chosen such that $P(z-z_m^a)=0$, and the factor $2$ accounts for the fact that the motion is periodic so integrating along the branch cut connecting two turning points is worth half the actual amount of cohomological correction. 

	 Therefore for singular curves one has to keep in mind that the cohomology classes of the (classical) ``WKB'' 1-form $\omega\sim g^{-1}\sqrt{y^2}dz$ in theories with $\mathbf{g}_s>0$ should receive extra ($g$-independent) corrections (modulo an overall sign factor), namely
	\begin{equation}
	\widetilde{\omega}=\omega+ \omega_c.\label{wkb}
	\end{equation}
	Upon specifying the 1-cycle, the integral of $\widetilde{\omega}$ should produce the correct HTA.  
	For $[\mathcal{I}\bar{\mathcal{I}}]$, as discussed above, this cycle is taken to be the branch cut connecting a complex conjugate pair of turning points. For a bound state
	of bions, namely $[\mathcal{I}\bar{\mathcal{I}}]^{n}$ this cycle is split into the branch cuts connecting 
	each of $n$ pairs of neighboring complex conjugate turning points.
	
	Eq. (\ref{wkb}) suggests that the sections of {\it sheaf}
	of holomorphic 1-forms is indeed what generates the complex phases of bions. 
	So the proper semi-classical treatment of singular spectral curves should replace the cotangent bundle by a richer object, {\it cotangent sheaf}. In fact, WKB approximation fails to capture the full topology of algebraic curves of higher genus in the presence of singular points though it might explain some things via the information hidden in the fixed points (turning points) of the WKB 1-form. 
	This is however not the case for an smooth affine curve, where topological recursion has been proven to indeed reconstruct the WKB expansion of the quantum spectral curve \cite{Bouchard:2016obz}.   
	\subsection{$\mathbf{g}_s=1$ example: triple-well}
    With the superpotential $W'(z) = 4z^3 - 3 z$ in eq.~\eqref{spec_gen} and focused only on $k_a>0$, we make the case of a $\mathbf{g}_s=1$ curve that admits a complex bion  given by 
    \begin{equation}
    z^a_{[\mathcal{I}\bar{\mathcal{I}}]}(\tau)=\frac{\omega_a}{2 \sqrt{3- i\sqrt{\omega_a^2-9} \cosh(2\omega_a\tau)}},
    \end{equation}
    where $\omega_a=\sqrt{9 + 12 g k_a}$ is the natural frequency at the global minimum of $V(z)$, i.e., $z=0$. Computing the action of this solution, one finds
    \begin{equation}
     S^a_{[\mathcal{I}\bar{\mathcal{I}}]}=\frac{3\omega_a}{8 g}+\frac{3k_a}{2}\log \left(\frac{\sqrt{\omega_a+3}}{\sqrt{\omega_a-3}}\right)+\frac{3i k_a \pi}{4}.
    \end{equation}
	The imaginary piece of this action should be equal to $H_a$ via the formula \eqref{period_gamma_dw},
	\begin{equation}
	H_a=\frac{1}{ g}\int^{z^T_a}_{\bar{z}^T_a} dz \sqrt{\omega_a^2z^2-24 z^4+16 z^6}, \label{period_gamma_dw_new}
	\end{equation}
	and a quick computation proves this for a complex conjugate pair of turning points, say $z_a^T=\tfrac{1}{2} \sqrt{3+2 i \sqrt{3g k_a}}$. Now, we form the $\omega_c$, that is just
	\begin{equation}
	\omega_c = 2\frac{z^2dz}{\sqrt{\omega_a^2z^2-24 z^4+16 z^6}}.
	\end{equation}  
	We then compute that
	\begin{equation}
	\int_{\bar{z}_a^{T}}^{z^T_a} \omega_c = 2\times \frac{i\pi}{8} = \frac{i\pi}{4}.
	\end{equation}
	Adding this correction to \eqref{period_gamma_dw_new} one obtains the HTA as
	\begin{equation}
	\theta_a = \pi(3k_a+1)/4,
	\end{equation}
	that for a supersymmetric theory, $k_a=1$, yields $\theta=\pi$ as expected.
	
	For $k_a<0$, the potential has two global minima which we pick either one and
	take the complex turning points in the immediate neighborhood and compute $H_a$ from \eqref{period_gamma_dw}, that gives
	$-3i\pi k_a/2$ that has to be corrected. The only non-trivial cohomological correction comes from
	\begin{equation}
	\omega_c = 2\frac{\left(z^2-\tfrac{1}{4}(2+\sqrt{1+4 g k_a})\right)dz}{y},
	\end{equation} 
	where once again it has been established that $\mu_M=2$ and $y^2=(4z^3 - 3 z)^2+3gk_a(4z^2-1)+2E^m_a$. Then the HTA becomes
	\begin{equation}
	\theta_a = -\pi (3k_a+1)/2,
	\end{equation}
    which for $k_a=-1$ reproduces $\theta=\pi$ as well. 
	\section{Index Formula From Picard-Lefschetz Theory: $\mathcal{N}=4$ Quantum Mechanics} \label{sec:Neq4QM}
	$\mathbf{Reminder}$: In the rest of this paper, we will adopt the notation $\kappa=2\pi i$ for simplicity. Our aim in this section is to write down a formula like (\ref{PLFormula}) that correctly computes the refined (categorified) Witten
	index
	\begin{equation}
	\mathcal{I}(y, \zeta ) := \text{tr}(-1)^F e^{-\beta H}y^{J} \label{indexformulaNeq4}
	\end{equation}
	for the supersymmetric quantum mechanics with four real supercharges, and a $U(1)$ $R$-symmetry
	$J$ that appears as a left-moving $U(1)$ $R$-symmetry in the original theory that will be taken to be $2d$ $\mathcal{N}=(2,2)$ gauge theory. Excited states will not be playing a role in the quasi-topological index we are seeking in what follows 
	so there will not be any dependence on $\beta$ anywhere. The index depends on both the $R$-charges of chiral fields and the Fayet-Iliopoulos (FI) parameter $\zeta$. The latter enters the index formula in view of a $Q$-exact deformation term in the $\mathcal{N}=(2,2)$ Lagrangian. Hence, Witten index is categorified in terms of the $R$-symmetry gauge fugacity $y$ defined by
	\begin{equation}
	y  =\exp \tfrac{1}{2}\kappa\left(\oint_{S^1}A_R\right)\equiv e^{\tfrac{1}{2}\kappa z},
	\end{equation}
	where $A_R:= \frac{z}{2\pi}$ is the $U(1)_R$ gauge field. 
	
	This formalism will totally avoid the JK residue operation used in \cite{Cordova:2014oxa} since we are considering
	relative homology cycles $\mathcal{J}_i$ as constituent components of the closed integration cycle $\gamma$ and focus on the thimble
	integrals within the fundamental domain of the moduli space $X$ -interchangeably called $u$-space from now on- parametrized by 
	\begin{equation}
	u :=-v_\tau+i\sigma, \label{u-var}
	\end{equation}
	where we have set the circumference of the Euclidean circle to $2\pi$. $v_\tau,\sigma$ are the constant gauge and real scalar fields (of the vector multiplet)  that parametrize the moduli space of the supersymmetric (or BPS) configurations for the $\mathcal{N}=(2,2)$ gauge theory. We choose the topological sector $v_\tau=0$ and rather follow a common exercise in Picard-Lefschetz theory that requires complexification of $\sigma$ such that \footnote{We thank Danielle Dorigoni for pointing this out.}
	\begin{equation}
	u\rightarrow \tilde{u}\equiv i\sigma_1 -\sigma_2 \in \mathbb{C} \label{com_u_var}
	\end{equation}
	where the complexified field is $\tilde{\sigma} = \sigma_1 + i\sigma_2$ with $\sigma_{1,2}\in \mathbb{R}$. In what follows we
	drop the tilde and assume that $\sigma_2$ direction is compact and $\sigma_2\sim \sigma_2+2\pi$ so that $\tilde{\sigma}$ still takes values in $\mathbb{C}/\mathbb{Z}$. We can avoid this assumption at the expense of changing the topology of the moduli space, but we are going to enforce such a change by introducing a regularization anyways to have controllability over the modified Lefschetz thimbles. 
	 We also remind the reader that the anti-holomorphic variables are washed away from the full 1-loop determinant after writing it as a total derivative term in $\bar{u}$ (see eq. (3.20) of Ref. \cite{Benini:2013nda}) \footnote{This requires putting $D=0$ in the chiral 1-loop determinant of $(2,2)$ theory. In an upcoming work,
	we investigate holomorphizing $\sigma$ in $\mathcal{N}=2$ supersymmetric quantum mechanics that avoids this condition 
	and gives rise to a much refined analysis of Lefschetz thimbles.}. 
	
	The quantum mechanics in consideration has the total gauge group $G=U(1)^\alpha $ unless otherwise stated \footnote{A generic group in the literature of gauged quiver quantum mechanics happens to mostly be of the form $G=\prod_{i} U(n_i)$ for $n_i\in \mathbb{Z}^+$ where $n_i$ are the coefficients for the contribution of each elementary BPS state charge $c^Q_i$ present in the $\mathcal{N}=4$ quiver quantum mechanics to the BPS charge of a one-particle $4d$ $\mathcal{N}=2$ system \cite{Alim:2011ae}:
		\begin{equation}
		c^{4d} = \sum_i m_i c^Q_i.\label{BPSindex}
		\end{equation} The numbers $m_i$ determining the basis for the space of BPS particles $\it should$ coincide with the magnitude of relative intersection coefficients that is built out of Lefschetz thimbles satisfying equations (\ref{FlowEqs}) with the effective 1-loop action of the $\mathcal{N}=4$ quiver system. In this work, we will proceed to verify that the flow diagram in the $u$-space accurately captures the BPS configuration contributing to the index with $m_i$ as given in (\ref{BPSindex}).}. Since the chiral matter is chosen to be in the bifundamental representation of $G$, allowing an overall $U(1)$ to be decoupled in the process, arrows will indicate the bifundamentals in a quiver graph and nodes will represent a $U(1)$ quiver gauge group. 
	
	We now generalize the discussion of non-compact moduli space of $U(1)$ gauge theory and its constituent Lefschetz thimbles
	 in Sec.~\ref{sec:intro} to some higher-rank gauge groups. The general premise then is that only the saddle points of the effective 1-loop determinant at imaginary-infinity will contribute to the $\chi_y$ genus and thus to the corresponding refined Witten index of the $1d$ gauged quiver system. From Picard-Lefschetz theory, $\chi_y$ genus can be written down as\footnote{In case of having $\#$ decoupling $U(1)$'s, one has to set $\alpha\rightarrow\alpha-\#$.} 
	\begin{eqnarray}
	\mathcal{I}(z,\zeta) &=& \frac{1}{\kappa^{\alpha}}\sum^{2^\alpha}_{j=1} n_j(\zeta)\int_{\mathcal{J}_j}\mathcal{Z}^{1\text{L}} d\vec{u} \nonumber\\
	&=& \frac{1}{\kappa^{\alpha}}\sum^{2^\alpha}_{j=1} n_j(\zeta) \lim_{\vec{u}\to \vec{u}^\star_j} \mathcal{Z}^{1\text{L}}(y,\vec{u})\nonumber\\&\equiv& \frac{1}{\kappa^{\alpha}}\sum^{2^\alpha}_{j=1} n_j(\zeta)  Z_j(y).\label{index_main}
	\end{eqnarray}
	where $d\vec{u} = \prod_{a}du_a$, $\alpha$ is the total number of $U(1)$ factors in $G$ of the quiver system, i.e., rank of $G$, $2^\alpha$ corresponds to the total number of critical points distinguished by the integration of $\mathcal{Z}^{\text{1L}}$ over $j$th $\alpha$-dim Lefschetz thimble, $\mathcal{J}_j$, attached to the $j$th critical point at $\vec{u}^\star_j$. The number of Lefschetz thimbles is determined by the number of different limits at imaginary infinity directions. For example, for $G=U(1)^4$, one finds $\vec{u}^\star_j= i\vec{v}_j\infty$ where $\vec{v}_j$ is a 4-vector with components $+1$ and/or $-1$, and correspondingly there are $2^4=16$ $4$-dim $\mathcal{J}$-cycles that construct the integration cycle. The limits in (\ref{index_main}) give the thimble integrals exactly that themselves measure a relative homology as discussed in Sec.~\ref{sec:Review}. 
	
	On supersymmetric grounds, as in this paper, if the moduli space $\mathcal{M}$ is compact, one can write a refined index formula of the form
	\begin{equation}
	\mathcal{I}(y,\zeta)=\sum_{p,q=0}^{D} h^{\mathcal{M}}_{(p,q)}(\zeta)(-1)^{p-q} y^{2p-D}.\label{index_ModuliSpace}
	\end{equation}
	Here, we are talking about a compact moduli space $\mathcal{M}$ of complex dimension $D$ endowed with a K\"ahler structure that entails all we need about the ground state spectrum of the underlying theory. The numbers $h^{\mathcal{M}}_{(p,q)}$ are the Hodge numbers of the bigraded algebra corresponding the Dolbeault cohomology groups. A closer look to our index formula (\ref{index_main}) suggests that it actually is similar in nature to (\ref{index_ModuliSpace}). In Picard-Lefschetz theory, the $m$-dim homology cycles $\mathcal{J}$ are labeled ``relative'' that lift to $H_m(X,X_{\tau^\star})$ in non-compact manifolds $X$ which may contain singularities\footnote{The quality of being relative for a good cycle means that for any cycle $\mathcal{J}\in X$ for some (punctured) non-compact manifold $X$, we need to find its ends at large flow time limit $\tau\rightarrow \tau^\star$ in good regions $G_i,G_j \in X_{\tau^\star}$ for $i\ne j$ where $X_{\tau^\star}$ is defined to contain all good regions relative to the whole of $X$ accessible at large flow time.}, whereas the cohomology of $\mathcal{M}$ in (\ref{index_ModuliSpace}) is global. So a direct comparison might not be possible between $h^{\mathcal{M}}_{(p,q)}$ and intersection numbers. However, we will see that a slight modification of this last formula would make it feasible to find a relation between the two. 
	
	\section{Linear Abelian Quiver Quantum Mechanics (dyon chains)} \label{sec:LAQM}
	
	Physically, quiver quantum mechanics appears as the low energy theory of a set of wrapped $D$-branes which encodes information on the dynamics of single and multicentered BPS black hole geometries in $4d$ $\mathcal{N} = 2$ supergravity. It also arises in $2d$ $\mathcal{N}=(2,2)$ gauge theories. The gauged quantum mechanics governs the low energy sector of the full quantum field theory, which is then useful for index calculations and understanding the dynamics of BPS bound states. A famous example that has been studied heavily in \cite{Denef:2002ru,Lee:2012sc} is dyon chains that will be our focal point in the rest of this paper. 
	
	\begin{equation*}
	\textbf{2-node quiver: $\mathcal{N}=4\,\, \mathbb{CP}^{k-1}$ model} \label{sec:2nodequiver}
	\end{equation*}
	Let us study the 2-node linear supersymmetric quiver quantum mechanics also known as $\mathbb{CP}^{k-1}$ derived from $4d$ $\mathcal{N}=2$ $SU(2)$ Yang-Mills theory. It is a massive rank-one Abelian theory that does not flow to a fixed point at IR and hence is not superconformal. The left $R$-symmetry is anomalous unless $z\in \mathbb{Z}/k$ for which the 1-loop determinant becomes single-valued. We choose to keep $z$ generic 
	since $R$-anomaly removal as we shall see, causes Stokes phenomenon to kick in that leads to technical complications. Setting $z=0$ is also dangerous as discussed in \cite{Benini:2013xpa} where the authors had to introduce flavor holonomies or an extra chiral multiplet to make sense out of an ill-defined 1-loop determinant. 
	
	There are $k$ bifundamental chiral fields between the nodes, represented by $k$-overlapping arrows in Fig.~\ref{fig:2node-quiver}. 
	\vspace{.2cm}
	\begin{figure}[H]
		\centering
		\includegraphics[width=0.3\linewidth]{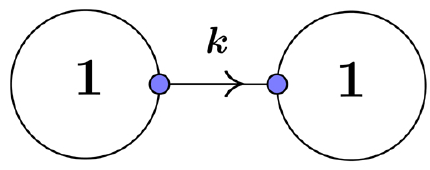}
		\caption{ A $\mathbb{CP}^{k-1}$ model. These are 2-node linear quivers with rank $1$ Abelian gauge groups at each node. $k$, the number of arrows, also shows the number of bifundamentals.}
		\label{fig:2node_quiver}
	\end{figure}
	\subsection{General strategy} \label{strategy}
	Let $G_i$ denote the $i$th zero $\epsilon$-balls of radii $\epsilon\ll1$ centered at the zeros of the 1-loop determinant. The flow lines 
	reaching an $\epsilon$-ball are guaranteed to be `stuck' at a zero of the 1-loop determinant. We denote by $\braket{G_i|G_j}$ with $i<j$ any possible good cycle in the slightly extended domain $\mathfrak{E}:=\{u\in X|-1\le\Re (u-z)< 1\}$ compared to the fundamental one, and aim to count the number of $\mathcal{J}$-cycles contributing to the integration cycle $\gamma$ so that the $\chi_y$ genus  captures the correct spectrum of BPS states. Physically, the $G_i$ is the location of the $i$th fermion zeromode in the moduli space at which the 
	Lefschetz thimbles end. 
	
\begin{figure}
		\centering
		\includegraphics[width=0.6\linewidth]{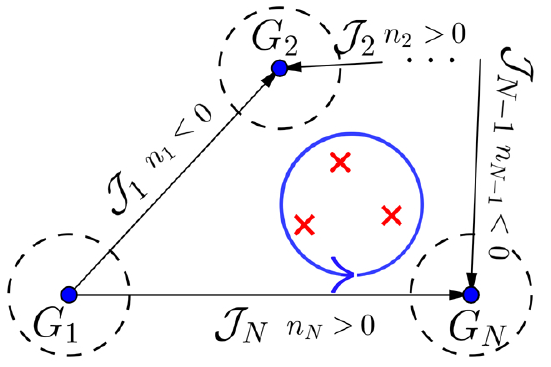}
		\caption{To fix the orientation for each thimble we start from one zero $\epsilon$-ball and continue counter-clockwise toward 
			the other zero $\epsilon$-ball until we come back to the initial zero $\epsilon$-ball. In other words, $\gamma 
			=n_1\braket{G_1|G_2}+n_2\braket{G_2|G_3}+\dots+n_{N-1}\braket{G_{N-1}|G_N}+n_{N}\braket{G_{N}|G_1}$ where
			the sign of $n_i$ makes the orientation of $\braket{G_{i}|G_{i+1}}\equiv\mathcal{J}_i$-cycle align that of $\gamma$. 
			This is basically similar to Kirchhoff's loop rule for a given electric network. }
		\label{gamma_contour}
	\end{figure}

	In dealing with situations in which integration cycle $\gamma$ is closed, on the contrary to open integration cycles, one needs to assemble a means of identifying orientations for Lefschetz thimbles because they do not have natural orientations. The result of this integral leads to an algebraic invariant -here a polynomial in the fugacity of $U(1)_R$ gauge field, $y$- that measures the relative homology $H_1(X,X_{\tau^\star})$ on the space $X=\mathbb{C}^\times\cong\mathbb{C}\setminus\{0\}\cong
	\mathbb{C}/\mathbb{Z}$ where $\cong$ means homeomorphic \footnote{For a quantum mechanical quiver system of rank-$\alpha$ gauge group $G$, the $u$-space is simply $X=({\mathbb{C}^\times})^\alpha$ \cite{Cordova:2014oxa}.}. The point $u=0$ coincides with a \emph{fundamental} pole of $\mathcal{Z}^{\text{1L}}$ and the $\mathbb Z$ is taken to be the subgroup of translations
	by integer multiples of the fundamental pole that will constitute the set of poles of $\mathcal{Z}^{\text{1L}}$ on the real $u$-direction.
	Computing this integral is usually a burdensome task for even a simple space like $\mathbb{R}^2$, let alone more complicated manifolds of higher dimensions. But the key tool that allows us to do computations in the current work is the exactness of saddle point approximation thanks to localization and quiver decoupling for higher-rank overall gauge group. Above all else lies the fact that there are might be finite saddles that do not contribute to the index, letting us only focus on those cycles that pass through the rims at infinite imaginary directions of $u$-space. Considering the fact that we know a \textit{priori} that thimble integrals would yield expressions that have the same overall sign in terms of powers of $y$, we prescribe the rule pictured in Fig.~\ref{gamma_contour}.
	
	The starting point is to plot the solutions to flow equation (\ref{FlowEqs}) for the following effective action: 
	\begin{equation}
	S_{\rm eff}(u,y):=-\ln(\mathcal{Z}^{\text{1L}}). \label{2node_quiver_action}
	\end{equation}
	
\begin{figure}
		\centering
		\includegraphics[width=.9\linewidth]{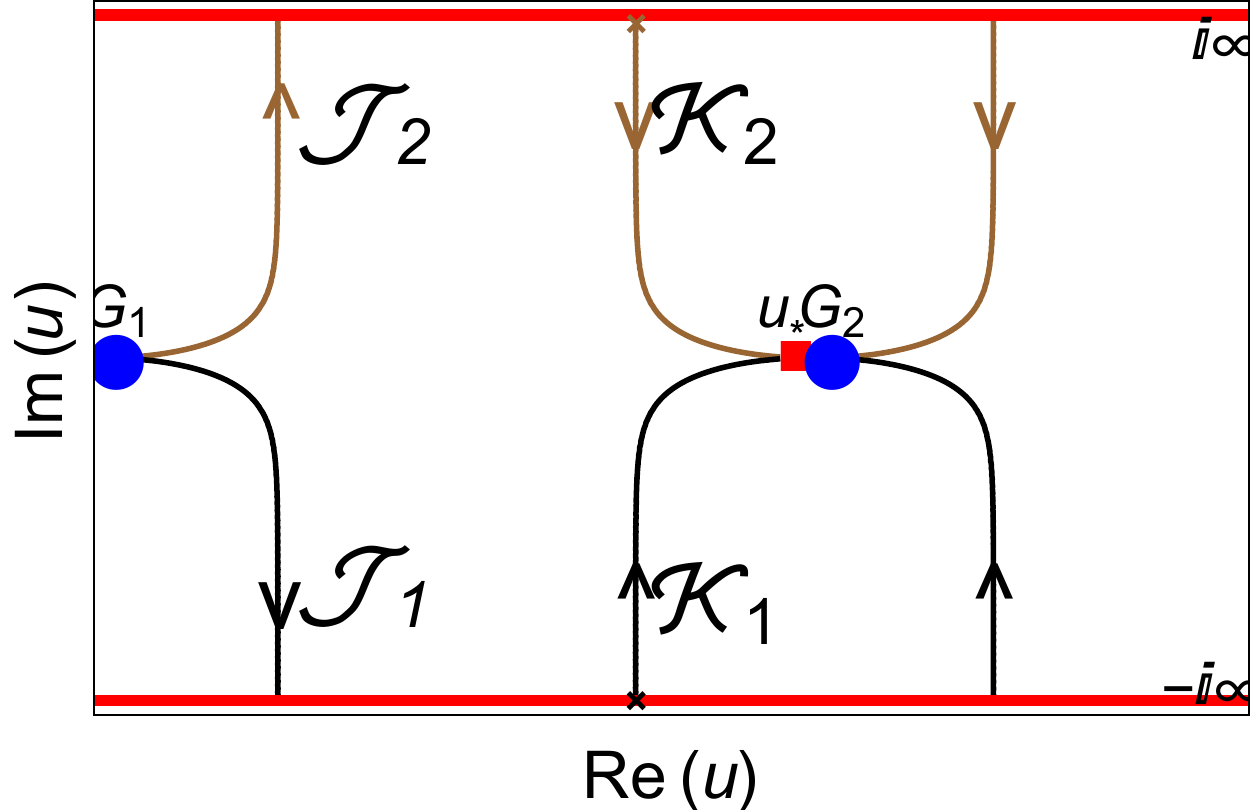}\label{2node-quiver}
		\caption{The plot of Lefschetz thimbles for $\mathbb{CP}^1$ model. Here, $z=0.05$. The red lines show the saddle rims at imaginary-infinity. The tails of Lefschetz thimbles end at the zero $\epsilon$-balls $G_1,G_2$ where the 1-loop determinant is guaranteed to vanish.
		The cross signs label the intersection points, giving $n_1=-n_2=1.$ The integration cycle is closed, which is in turn enforced by the orientation of the $\mathcal{J}_1,\mathcal{J}_2$ in $\gamma$, and thus by the intersection numbers. }
		\label{fig:2node-quiver}
	\end{figure}

\begin{figure}
		\centering
		\includegraphics[width=0.8\linewidth]{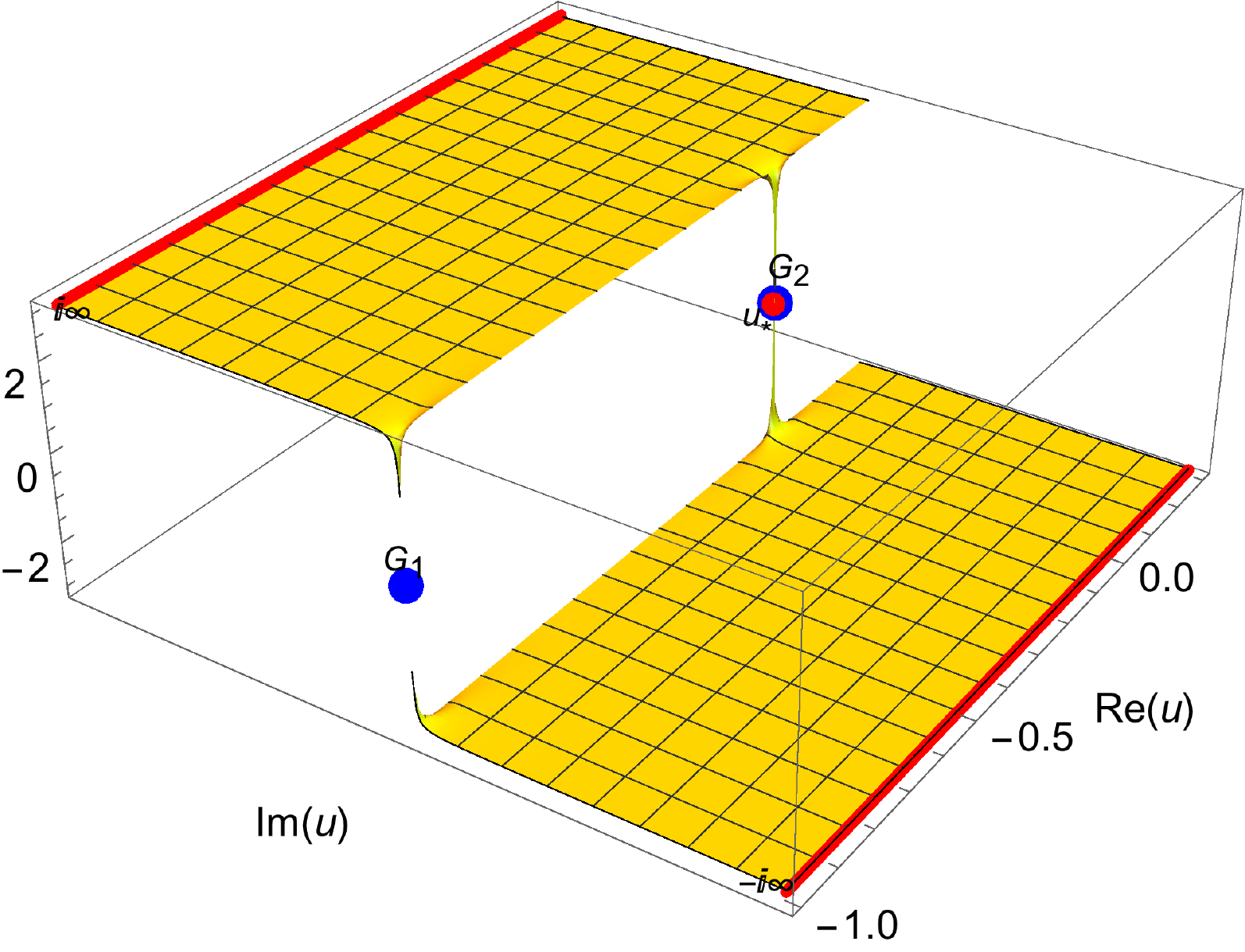}
		\caption{The imaginary part of $S_{\rm eff}(u)$ for the linear Abelian 2-node quiver with $z=0.005$. The branch 
			cut is $[z-1,z]\setminus [0,z]$. Saddle rims on both $- i\infty,i\infty$ ends are shown with red lines. The 
		imaginary parts of $S_{\rm eff}(u)$ over Lefschetz thimbles are $\pm 3.1$ which tend to $\pm \pi$ as $z\rightarrow 0$.}
		\label{fig:2node_quiver1}
	\end{figure}

	To do this, we first solve the flow line equations (\ref{FlowEqs}) and count the possible number of good cycles in $\mathfrak{E}$. It turns out that this number is 2 as seen in Fig.~\ref{fig:2node-quiver}. We recall that the Lefschetz thimbles ${\mathcal J}_{1,2}$
	seem to be the only ones contributing to the index \eqref{index_main}, but indeed there are infinitely many copies of 
	the same thimbles each starting from a point at imaginary-infinity with the initial values
	\begin{equation}
		\lim_{\tau\rightarrow-\infty} u(\tau)= r\pm i\infty\in {\rm saddle~rims},
	\end{equation}
	 for real $r$ which by path homotopy, we aim to refine such a degeneracy\footnote{A note is due to clear a possible misleading point here. The relative homology cycles can happen to actually yield the same invariant after thimble integration over them. Thus, two independent $\mathcal{J}$-cycles might in fact have an identical invariant, and yet be considered in the index formula. In general, the number of independent $\mathcal{J}$-cycles is $2^\alpha$ where $\alpha$ is the rank of total p $G$ or the number of nodes because $G=U(1)^\alpha$.  }. So in the complex $u$-plane, the saddle points $r_{ij}\pm i\infty$, attached to cycles $\braket{G_i|G_j}$ for $|i-j|<2$, are only distinguishable for infinite $r_{ij}$, which for a bounded $\mathfrak{E}$ is not feasible\footnote{We note that the boundedness of $\mathfrak{E}$ along real direction is necessary for the applicability of Picard-Lefschetz theory in this situation. If one lifted the boundedness requirement, the Lefschetz integration would be undefined for $\Re (u)\rightarrow\pm\infty$.}. Therefore, $\braket{G_1|G_2}$ and $\braket{G_2|G_3}$, where $G_3\subset\mathfrak{E}\setminus\mathfrak{F}$ are indeed copies of the same cycles unless there is a pole or branch-cut in between. The minimal set of all the independent cycles that can build the contour of integration in the moduli space is dubbed from now on the \textit{irreducible $\mathbf{J}$-set}, whose cardinality is $2^\alpha$ for a gauged quantum mechanics with $\alpha$ being the rank of the gauge group. In other words, the Lefschetz thimble integrals over $\mathbf{J}$ form a vector space of dimension $2^\alpha$ that consists of a necessary and sufficient number of Lefschetz thimbles in the fundamental domain $\mathfrak{F}:=\{u\in \mathbb{C}^\times|0\le\Re (u-z)<1\}$ where $\mathfrak{F}\hookrightarrow \mathfrak{E}$, i.e., $\mathfrak{F}$ is only a retract of $\mathfrak{E}$. This is best understood by examining the saddle points of the 2-node quiver effective action (\ref{2node_quiver_action}), which, too, are those of the 1-loop determinant. 
	We show this in Fig.~\ref{fig:2node_quiver1}.
	
	Compared to JK residue operation, a big advantage comes into play when the location and number of poles become immaterial. This, in fact, comes up in our analysis naturally by just focusing on the hunt for an irreducible $\mathbf{J}$-set, which in the current case is $\mathbf{J}=\{\mathcal{J}_1,\mathcal{J}_2\}$. The poles are replaced by branch points of the effective action and
	the branch cut is everywhere along $[z-1,z]\setminus [0,z]$. This is shown in Fig.~\ref{fig:2node-quiver}. 
	
	It is possible to apply this
	technology to higher rank gauge groups but one should not integrate out $D$-field and rather complexify $\sigma$ in the gauge multiplet which also removes the degeneracy of saddle points \footnote{Work in progress.}.
	
	For an Abelian 2-node quiver model with $k$ bifundamentals, we have the 1-loop meromorphic form 
	\begin{equation}
	\mathcal{Z}^\text{\text{1L}}du=-\pi \sin(\pi z)^{k-1} (\cot(\pi z)-\cot(\pi u))^k du.\label{2nodequiver_cr} 
	\end{equation}
	Since  $u^\star_1=-i\infty$ and $u^\star_2=i\infty$, the thimble integrals over $\mathcal{J}_1,\mathcal{J}_2$ may be given by the 
	saddle point approximation as (modulo a zero Hessian)
	\begin{eqnarray}
	{Z}_1(y)&=&\lim_{u\to -i \infty } \frac{-\pi(\cot(\pi z)-\cot(\pi u))^{k}} { \sin(\pi z)^{1-k}}=\kappa\frac{ y^{-k}}{y^{-1}-y},\nonumber\\
	\quad {Z}_2(y)&=&\lim_{u\to i \infty }  \frac{-\pi(\cot(\pi z)-\cot(\pi u))^{k}} {\sin(\pi z)^{1-k}}=\kappa\frac{y^k}{y^{-1}-y}\nonumber.
	\end{eqnarray}
	Therefore, the formula (\ref{index_main}) produces the following index for the 2-node quiver quantum mechanics:
	\begin{subequations}
		\begin{align}
		\mathcal{I}(y,\zeta)&= n_1(\zeta)\frac{y^{-k}}{y^{-1}-y}+n_2(\zeta) \frac{y^k}{y^{-1}-y}\label{index_CPN_intersection}\\
		&=
		\Theta(\zeta) \frac{y^{-k}}{y^{-1}-y}- \Theta(\zeta) \frac{y^k}{y^{-1}-y}\nonumber\\
		&= \begin{cases}
		\frac{y^{-k}}{y^{-1}-y}-\frac{y^k}{y^{-1}-y}\equiv y^{1-k}\sum_{i=0}^{k-1}y^{2i},& \text{if } \zeta>0\\
		0.             &\text{if } \zeta<0 
		\end{cases}
		\label{index_2node}
		\end{align}
	\end{subequations}
	(Here, use was made of the formula (\ref{FIn}). For more information, the reader is encouraged to look at its derivation in Sec.~\ref{subsec:FIdep}.) The piecewise FI parameter dependence of $n_i(\zeta)$ in general is just a direct consequence of the fact that they do not show up inside the trace function of (\ref{indexformulaNeq4}) explicitly. It should be stated, however, that this dependency is rather shown to be in a less clear way on the contour of integration $\gamma$ in the JK residue operation. The Picard-Lefschetz theory does seemingly suggest that this contour is of the form
	\begin{equation}
	\gamma(\zeta) = n_1(\zeta)\mathcal{J}_1+ n_2(\zeta)\mathcal{J}_2. \label{gamma_PLtheory}
	\end{equation}
	Vividly, any jump in the index is related immediately to the jumps in $n_i(\zeta)$ caused by either Stokes phenomena or moving through FI chambers or both. This latter one will be apparent in the cases where the index in the other FI chambers is non-zero such as generalized \textit{XYZ} model. In other words, it is possible that applying Lefschetz decomposition may end up with an \textit{on-the-wall} index and any probable wall crossing phenomenon is lurking here (see Sec.~\ref{sec:GenXYZ}). 
	
	This index is obviously invariant under $z\rightarrow -z$ or, equivalently, $y\rightarrow y^{-1}$ exchange symmetry that can be understood in terms of Lefschetz thimbles by noticing that all it does is for the zero $\epsilon$-balls to be flipped around imaginary axis in Fig.~\ref{fig:2node-quiver} without 
	thimble integration affected. Because the same pole is again encircled by the thimbles at work here, the JK residue operation will yield the same result as well. 
	
	Next, we should match this result with the formula (\ref{index_ModuliSpace}) for $\mathcal{M}=\mathbb{CP}^{k-1}$ with $D=k-1$. Let us first simplify things further by noting the Hodge numbers in (\ref{index_ModuliSpace}) can be extracted from the Betti numbers in the following way:
	\begin{equation}
	\begin{split}
	\mathcal{I}(y,\zeta)&=\sum_{p=0}^{k-1} b_{p}(\zeta) \frac{y^{2p-k+1}-y^{2p-k+3}}{1-y^2}\\&=\sum_{p=0}^{k-1}\sum_{q=0}^{k-1} (-1)^{p-q}h^{\mathbb{CP}^{k-1}}_{(p,q)} \frac{y^{2p-k+1}-y^{2p-k+3}}{1-y^2},\label{betti_CPK}
	\end{split}
	\end{equation}
	where the Betti numbers are 
	\begin{equation}
	b_{p}(\zeta)=\sum_{q=0}^{k-1} (-1)^{p-q}h^{\mathbb{CP}^{k-1}}_{(p,q)}(\zeta),\label{bettidim}
	\end{equation}
	are the Betti numbers of the cohomology of $\mathbb{CP}^{k-1}$. Notice that for our later purposes, a slight manipulation in the original form of (\ref{index_ModuliSpace}) was made to cast it in a more comparable form to the one derived using the Picard-Lefschetz theory in (\ref{index_CPN_intersection}). 
	Equating (\ref{betti_CPK}) with the index formula obtained in (\ref{index_CPN_intersection}) imposes the condition
	\begin{equation}
	b^{\mathbb{CP}^{k-1}}(\zeta) = n_1(\zeta) = -n_2(\zeta).\label{conditionbetti}
	\end{equation}
	where $b_0=b_1=\dots=b_{k-1}\equiv b^{\mathbb{CP}^{k-1}}.$
	
	Eq.~\eqref{conditionbetti} is a very elegant correspondence between Betti numbers of the Dolbeault cohomology groups over the compact moduli space of the supersymmetric vacua in the 2-node quiver and the intersection coefficients of the $\mathcal{J}$-cycles in the relative homology drawn from the function $S_{\rm eff}$ over a non-compact 2-manifold $X=\mathbb{C}^\times$ which is the $u$-space. This actually is a \textit{finite} dimensional moduli space of solutions to $\mathcal{N}=2$ supersymmetric locus (BPS equations) drawn from the original infinite dimensional path space of the $2d$ $\mathcal{N}=(2,2)$ supersymmetric gauge theory by localization. The power of this correspondence is that even if one does not know the intersection numbers, which happen to be so complicated to obtain in more general cases, the condition (\ref{conditionbetti}) determines the relation between $n_1,n_2$. Mathematically, $\gamma $ can be lifted to an element of $H_1(\mathfrak{F})$ where $H_1(\mathfrak{F})$ is an ``ordinary'' homology of $\mathfrak{F}\subset \mathbb{C}^\times$ while the constituent cycles still fall into relative homology groups. $\gamma$, a closed loop encircling one singularity at $u_*=0$, gives the betti number $b_1=1$ for $H_1(\mathfrak{F})$. This is the statement that the Poincar\'e function for the relative homology $H_1(\mathfrak{F},\mathfrak{F}_{\tau^\star})$ is in fact a Poincar\'e series because the vector multiplet contribution to 1-loop determinant is $\propto \frac{1}{1-y^2}$. Finally, this correspondence is $H_1(\mathfrak{F})\cong H^1(\mathbb{CP}^{k-1})$ regardless of the value of $k$ \footnote{Note that, $k$ appears in the exponent of the chiral multiplet contribution in the 1-loop determinant. Hence, modulo the $U(1)$ vector multiplet components that are not related to the saddle points in ${u}$-space e.g., eq.~(\ref{thimble_1_mod}), Lefschetz thimbles are independent of $k$.}. 
	
	The isomorphism $H_1(\mathfrak{F})\cong H^1(\mathbb{CP}^{k-1})$ needs a little bit of explanation. In the path integral of the supersymmetric gauge theory, the space to be integrated over is initially infinite dimensional. Localization brings it down to a finite dimensional integral over a specific contour, $\gamma$, in the space of saddle points of the low-energy theory. These saddles are simply the solutions to  supersymmetric BPS conditions that form a moduli space. $\gamma$ is fixed by Picard-Lefschetz theory to be (\ref{gamma_PLtheory}), which is a middle-dimensional cycle- e.g. a smooth 1-manifold in $\mathfrak{F}$ for a quiver system of $\alpha=2$ reduced by decoupling a $U(1)$ factor. Because the constituent pieces of this contour are the thimbles extended all the way to infinity along imaginary directions, $\gamma$  could be shrunk into a compact submanifold $\mathfrak{W}\subset \mathfrak{F}$ with $ \{G_1,G_2\}\cap \mathfrak{W} \ne {\emptyset}$ by way of introducing a regulator. It should be added that it is not possible to include any of zero $\epsilon$-balls in their entirety in $\mathfrak{W}$ as they vary with respect to $y$ and may grow out of $\mathfrak{W}$. This simply induces a \textit{deformation} retract $\mathfrak{W}\hookrightarrow \mathfrak{F}$ and thus $H_1(\mathfrak{F})\cong H_1(\mathfrak{W})$. The induction by the regulator is not unique in the sense that the localized path integral, that is, the refined Witten index $\mathcal{I}(y,\zeta)$, does not depend on the choice of regulator as long as it does not change the pole structure.
	
	To fully compute the Betti numbers $b_p$, one always needs a generating function that is also difficult to obtain due to the topological complications of the moduli spaces. In simple cases such as $\mathcal{M}=\mathbb{CP}^{k-1}$ or Grassmannians, where closed loops are absent in the quiver, this problem has been settled by Reineke in \cite{Reineke:2002}. There, counting supersymmetric ground
	states of quantum mechanics on the moduli space of a 2-node quiver with $k$ arrows, and dimension vector $(1, n)$ associated with nodes of the quiver is shown to be related to the cohomology of a Grassmannian $\mathcal{M}=\mathbf{Gr}(n,k)$ in an explicit way:
	\begin{equation}
	P(y)\equiv\sum^{n(k-n)}_{p=0} b_p y^{2p}=\frac{\prod_{j=1}^{k-1} (1-y^{2j})}{\prod_{j=1}^{n} (1-y^{2j})\prod_{j=1}^{k-n} (1-y^{2j})}.\index{generating_function}
	\end{equation}
	Now multiplying all sides by $y^{-k+1}$ and putting $n=1$ yields the index for $\mathbb{CP}^{k-1}$ model. From this, we find $b^{\mathbb{CP}^{k-1}}=1$ which was expected from the analysis of Lefschetz thimbles earlier. The $\zeta>0$ is a geometric phase for $\mathcal{N}=4$ and thus the $\mathbb{CP}^{k-1}$ model has a non-zero index in this phase, whose space of supersymmetric ground states is of dimension
	\begin{equation}
	h^{\mathbb{CP}^{k-1}}_{(p,q)}(\zeta>0)=
	\begin{cases}
	1& \text{if } p=q=0,\dots,k-1\\
	0             &\text{if otherwise,}
	\end{cases}
	\label{N=4index}
	\end{equation}
	\cite{Hori:2014tda}. Introducing this in (\ref{bettidim}) implies correctly the result $b^{\mathbb{CP}^{k-1}}=1$ and affirms the Picard-Lefschetz analysis.
	
	If we were to follow things starting from the Picard-Lefschetz theory side, we would have hit a snag to get the cohomology of the moduli space. The reason for this is straightforward to understand. Consider the expansion of the integration cycle (\ref{gamma_PLtheory})
	is given. Using saddle point approximation, the values of 1-loop determinant at saddles $u_1^\star,u_2^\star$, i.e., $Z_1,Z_2$ give thimble integrals over $\mathcal{J}_1,\mathcal{J}_2$ (see Subsec.~\ref{subsec:thimbleCal} for an explicit calculation). $Z_1,Z_2$ do not, however, have any finite expansion in $y$ because of the divergence at $y=\pm 1$ caused by the singularities of the vector multiplet contribution but apparently the combination does have a Laurent expansion once $n_1$ and $n_2$ are set to be $+1$ and $-1$, respectively. Since the dependence on $\zeta$ is piecewise, the intersection coefficients in the relative homology of $\mathcal{J}$-cycles must encode the cohomological data of the moduli space of supersymmetric quiver vacua and their locations. 
	
	Still, we need to look for means of generating these coefficients and clarifying what their link to FI parameters is. In the following, we try to put the idea of wall crossing phenomenon and Stokes jumps into perspective together to address these questions. For a detailed mathematical account of 
	these subjects, we guide the reader to consult the beautiful text by Kontsevich and Soibelman \cite{Kontsevich:2008fj}.
	
	\subsection{Regularization of infinite saddles using FI parameters: Morsification}\label{regularization} 
	
	Even though matching the relative homology of the $\mathcal{J}$-cycles derived from the Picard-Lefschetz theory with the cohomology of the moduli space of stable quiver representations derived using algebraic topological tools is very instructive as it stands, one may be concerned about the saddle rims at infinity and whether or not studying the Picard-Lefschetz theory is legitimate in such cases due to the degeneracy of saddle points.
	
\begin{figure}
	\centering
	\includegraphics[width=0.8\linewidth]{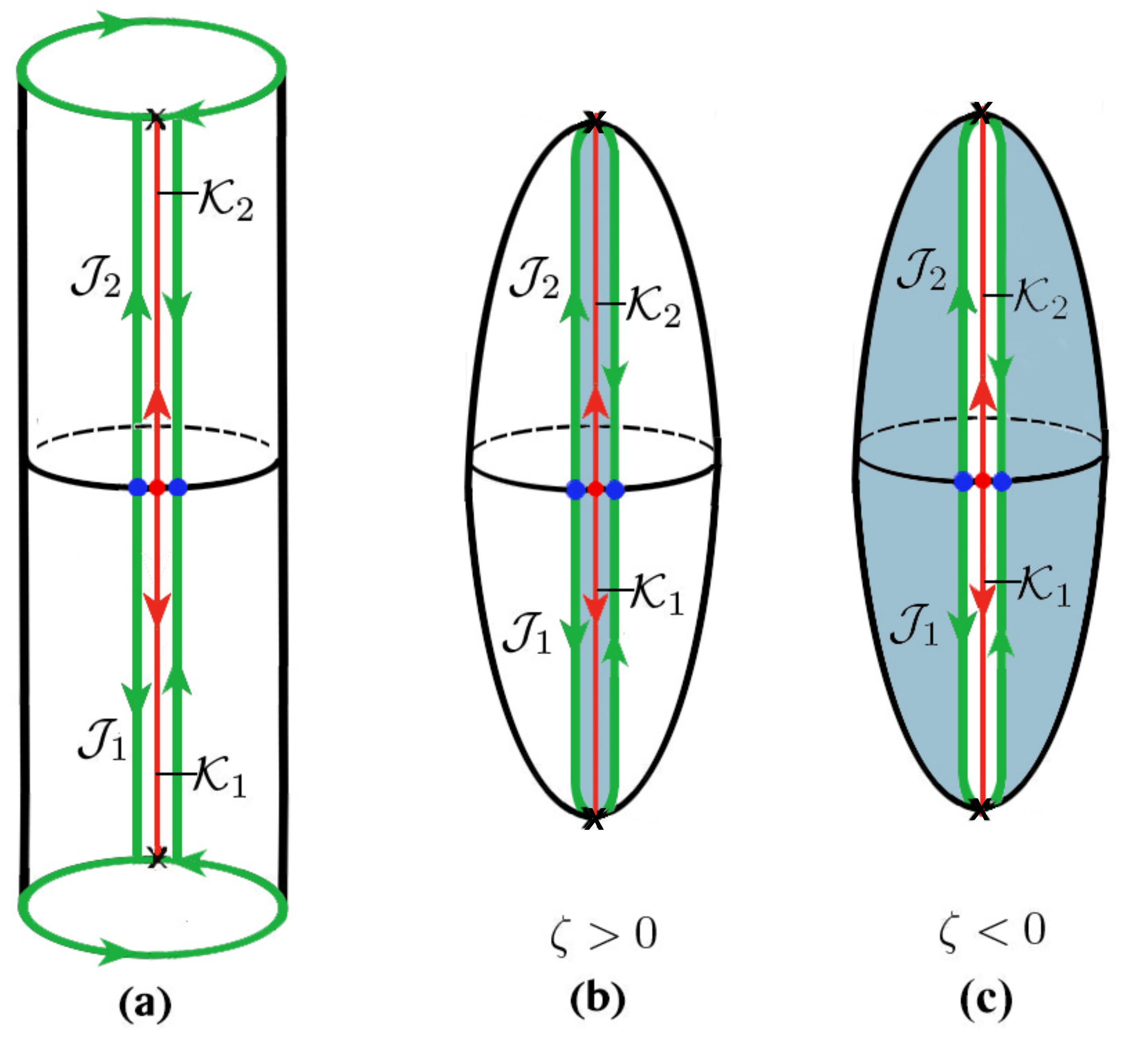}
	\caption{ (a) Moduli space of the $U(1)$ gauge theory (b) The Lefschetz thimbles of the compactified theory by adding a gauge-invariant 
		regularization term in the phase $\zeta>0$. (c) The Lefschetz thimbles seen from the back of the sphere
		do not encircle the pole so the index is zero.
		The saddle rims have been 
		reduced to fixed points of the action of $G=U(1)$ on the sphere \protect\cite{Witten:1992xu}. Those are the non-degenerate isolated critical points of the regularized effective
		action. A basis for homology cycles $\mathcal {J}_i$ in $(a)$ is determined modulo path homotopy. The first ordinary homology is seen to be of rank $2$ upon regularization. }
	\label{Pathspace}
\end{figure}

	As a result we should seek out an off-the-wall interpretation of the index which encodes data on the BPS configuration inside of the FI chambers (Note: this is different from the quiver invariants intrinsic to the wall of marginal stability that will be referred to as on-the-wall index here. See Sec. \ref{sec:GenXYZ}). Taking this as a fact, given that $\gamma$ is essentially sitting on a Stokes ray, going away from the ray gives rise to Stokes jumps responsible for any possible change in the relative homology of the $\mathcal{J}$-cycles:
	\begin{equation}
	\mathcal{J}_i \rightarrow \mathcal{J}_i + \sum_{j\ne i}\sigma_{ij} \mathcal{J}_j, \quad
	n_i \rightarrow n_i + \sum_{j\ne i}\eta_{ij}n_j,\label{Stokes1}
	\end{equation}
	where $\sigma_{ij}$ and $\eta_{ij}$ are integers. To guarantee a smooth variation across the Stokes rays, these constants have to satisfy 
	the relation
	\begin{equation}
	\sigma_{ij}+\eta_{ji}+\sum_l \eta_{li}\sigma_{lj}=0.
	\end{equation}
	for every ${i\ne j}$.
	Fix a parameter $\mu$ that triggers (\ref{Stokes1}) in a two-chamber theory. Although we now have a homology cycle, say $\mathcal{C}_\mu$ off the Stokes line to the right of it $(\mu\rightarrow 0^+)$ or else $(\mu\rightarrow 0^-)$, that is locally constant as prescribed by
	 $\sigma_{ij}$ and $\eta_{ij}$, globally the presence of poles would make a difference in the homology of cycles. Stokes jumps are immediately accompanied by variation of FI parameters responsible for building different chambers in the moduli space of solutions to the BPS equations- $u$-space. Thus, the index experiences contributions from a perhaps different BPS configuration depending on how $\mathcal{J}_i$ change, so
	 \vspace{-.3cm}
	\begin{equation}
	\gamma=n_+(\zeta)\,\mathcal{C}_+ + n_-(\zeta)\,\mathcal{C}_- \label{wall-cross-gamma}
	\end{equation}   
	where $n_\pm(\zeta)=\Theta(\pm \zeta)$ are the intersection coefficients of the regularized theory, which gives us two phases
	shown in Fig. \ref{Pathspace}. This means that as soon as $\mu =0^+$ or $\zeta>0$, 
	$\gamma = \mathcal{C}_+$ which is a non-vanishing cycle as opposed to the other phase in which $\zeta<0$ when
	$\gamma = \mathcal{C}_-$ is a vanishing cycle. 

The jumps through the Stokes ray are hence the reason for variations of contours $\mathcal{C}_-\leftrightarrow \mathcal{C}_+$ in a locally homologically invariant way but globally distinct. The homology cycles $\mathcal{C}_\pm$
	are associated with two chambers defined by the conditions that coefficients $n_\pm(\zeta)$ hold onto. Nonetheless, one still needs
	to clarify  what $\mu$ is and how it is related to FI parameters. This is precisely
	analogous to the Stokes jumps across the singularity rays in the Borel plane and picking up
	residues as Stokes rays are crossed \cite{Cherman:2014ofa}.

	In case $D$-field is first integrated out, the wall crossing effects become highly nontrivial as critical points in terms of $u$ degenerate into
	saddle rims. This comes with the caveat that the condition $D=0$ is enforced for holomorphicity of the function $h(\tau, z, u, D) g(\tau, z, u, D)$ of
	the $(2,2)$ theory \cite{Benini:2013nda}.  There is however an escape route that allows the thimble analysis to be done in a finite region more systematically and could explain the $\mu$-dependence of $n_i$ explicitly with the help of $\mu$. 
	Assuming that $\gamma$ is on a Stokes ray, one can add an arbitrary smooth gauge-invariant regulator of the form $\mu(\vec{\zeta})R(\vec{u})$ to (\ref{2node_quiver_action}), where $R(\vec{u})$ vanishes at the singularities of $\mathcal{Z}^{\rm 1L}$ in $X$ and $\mu(\vec{\zeta})=0$ characterizes the wall in $\zeta$-space, forcing simultaneously the condition for degeneracy to come back and thus for $\gamma$ to lie on a Stokes ray in the $u$-space. Giving up gauge-invariance in $X \setminus \mathfrak{F}$, we can define the monomial 
	\begin{equation}
	R(\vec{u})= \prod_{i}(H_i(\vec{u}))^2\label{eq:reg}
	\end{equation} 
	where $H_i(\vec{u})=0$ define the poles or, in general, singularity hyperplanes (higher-rank) in $\mathfrak{F}$. Thus, $\mu(\vec{\zeta})$ is a regularization parameter for the $u$ variables that allows the effective action to have its non-isolated degenerate saddle points re-incarcerated as isolated nondegenerate ones within a finite region. On Morse theory grounds \cite{Witten:2010cx}, it means that the effective action can be Morsified, meaning that the \emph{regularized} function $h:=-\Re(\ln\mathcal{Z}_{\text{reg}}^{\text{1L}})$ has finitely many isolated nondegenerate critical points of Morse index 1, so it is a perfect Morse function. Then, the rank of $H_1(\mathfrak{F},\mathfrak{F}_{\tau^\star})$ 
	is exactly the number of critical points of $h$ in $\mathfrak{F}$, equal to the number of $\mathcal J$-cycles in $\mathbf{J}$-set and all these cycles are middle dimensional. 
		\begin{figure}
		\centering
		\includegraphics[width=1\linewidth]{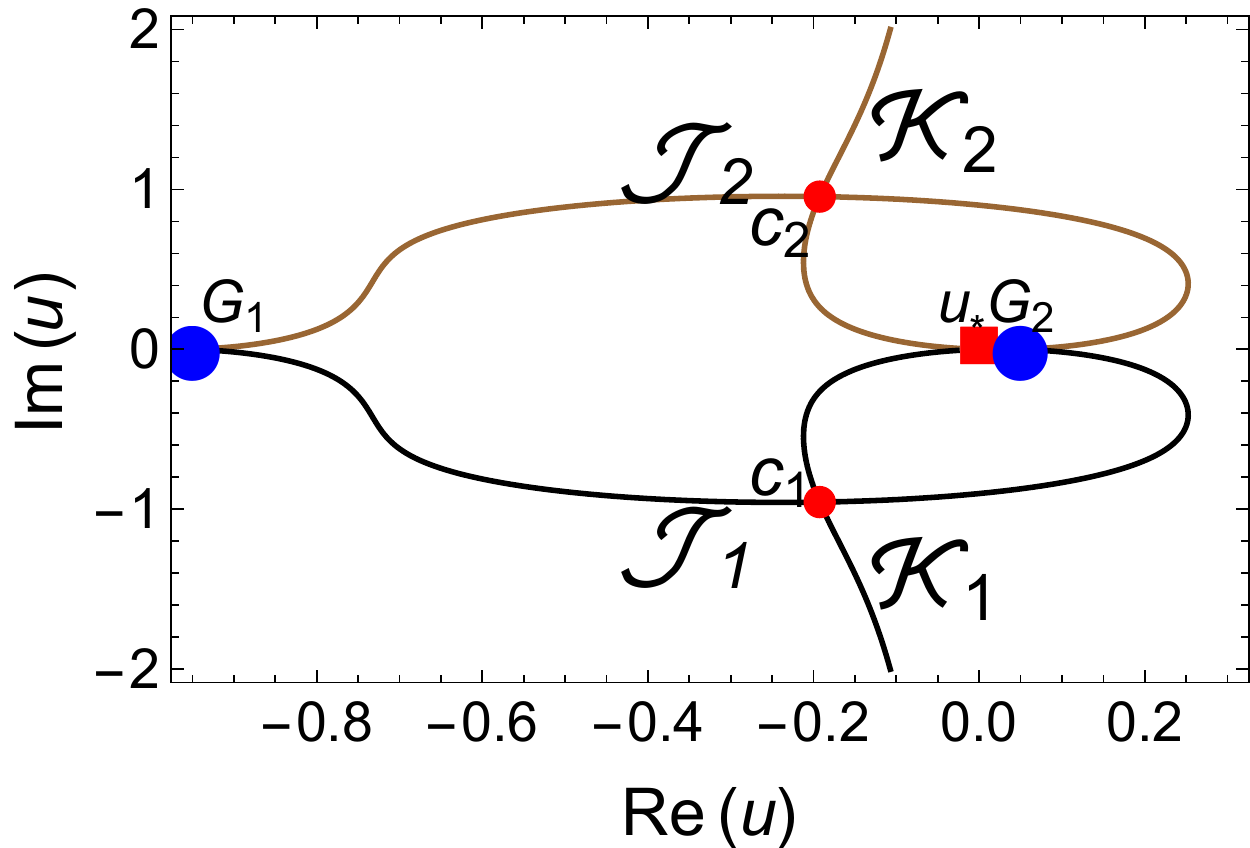}
		\caption{The moduli space of the regularized $\mathbb{CP}^1$ model following regularization of its effective action. Here, the 
			regulator is $\frac{\mu}{2} u^2$ with $\mu=0.01$ and $z=0.05$ so $R$-symmetry is anomalous. $c_1$ and $c_2$ correspond to the critical points. The total gauge group is $G=U(1)^2$ and the FI term is $\zeta(D)=\zeta_1D_1+\zeta_2D_2$. The condition $\zeta_1+\zeta_2 =0$ decouples one $U(1)$ and amputates one node. We take $\mu=|\zeta_2|=|-\zeta_1|$. The regularization term is not invariant under $u\rightarrow u+1$ so the regularized theory
			is not replicated along the compact direction of $u$-space. Because of this regularization, the model has a true Morse
			interpretation and the superposition of $\mathcal{J}$-cycles forms an ordinary homology -inherited from a Morse homology-, whose associated  invariant quantity is
			a non-vanishing index in the chamber $\zeta\equiv\zeta_2>0$ and $0$ otherwise.
			 Rank of $H_1(\mathfrak{F},\mathfrak{F}_{\tau^\star})$ is, therefore, two.}
		\label{fig:2-node_reg}
	\end{figure}

	For a $0+1$-dim $U(1)$ supersymmetric quantum gauge theory such as $\mathbb{CP}^{1}$, the Morsification results in 
	the flow lines depicted in Fig. \ref{fig:2-node_reg} where the refined index follows as
	\begin{eqnarray}
	\mathcal{I}(\zeta,y)
	&=&\kappa^{-1}\sum_{\pm}n_\pm(\zeta)\int_{\mathcal{C}_\pm} e^{-\mu R(u) - S_\text{eff}(u)} du\nonumber\\
	&\equiv&-\kappa^{-1}\sum_i  \oint_{u=u^P_i} e^{- S^{\rm reg }_\text{eff}(u)} d{u}\nonumber
	\label{WittenInd}
	\end{eqnarray} 
	where $\mathcal{C}_{\mu(\zeta)}$ is a closed $\mu$-deformed integration cycle that may or may not encircle the poles $u^P_i$
	depending on $\zeta$. We note that adding a regulator of the form \eqref{eq:reg} caps the boundaries of the moduli space into a compactified space that is topologically a sphere.
	The only effect of this regularization from the relative homology point of view is to remove the degeneracy so the Lefschetz thimbles in $\mathbf J$ do not change as expected. However, the homology of $\gamma$ separates into two classes, defining the phase $\zeta>0$ where a pole is included in $\gamma$, and the phase
	$\zeta<0$ where $\gamma$ is homologous to a vanishing cycle.
	
	What if we have more than two chambers? Then $\mu$ should be analytically continued to complex plane i.e. $\mu(\zeta)e^{i\theta}$ and by changing  $\theta$ the cycle $C_{\mu(\zeta)e^{i\theta}}$ will be rotated such that the integration cycle on the Stokes ray, $\gamma$, is approached by $\mathcal{C}_{0e^{i\theta}}$. The $0$ in the subscript is the bifurcation or junction of all the segments of the wall and phase factor $e^{i\theta}$ 
	allows a variation of the integration cycle around this point that fall into different chambers in the corresponding $\zeta$-space. Therefore, we can take $\theta$ as describing a single point in the $\zeta$-space of FI parameters.
		
	\subsection{Intersection numbers in quiver quantum mechanics} \label{subsec:intersection}
\subsubsection{Witten index check}
 This method is based on the fact that a correct integration cycle $\gamma$ has to reproduce the Witten index (which is the Euler characteristic of the target manifold in certain cases such as $\mathbb{CP}^{k-1}$ and $\mathbf{Gr}(n,k)$ theories), as a limiting case of the refined index once all the fugacities of the gauge fields are turned off \cite{Grassi:2007va}. Also, the index formula we are advocating in this work still does not seem to give a solid relation between the FI parameters and intersection numbers and little is known about the relative homology techniques that one may in practice employ to compute these numbers, which in principle determine the degeneracies of BPS states (see footnote \ref{foot1}). In this regard, the theory of spectral networks have been successful in determining these degeneracies and the geometry governing them in $4d$ $\mathcal{N}=2$ theories \cite{Gaiotto:2012rg}. 
	
	The Picard-Lefschetz theory prescribes for general thimble configurations, the following abstract formula:
	\begin{equation}
	n_i=\braket{\gamma|\mathcal{K}_i},\label{ni}
	\end{equation}
	which, including the sign, actually counts the number of $\mathcal{K}$-cycles going through a specific pole encircled in $\gamma$ and crossing $\gamma$ at $i$th $\mathcal{J}$-cycle. Since the flow equations are highly non-linear in the examples we study, this formula is of little practical use for $\alpha>1$ as plotting Lefschetz thimbles of dimension $\geq2$ turns out to be beyond hope. So in general we do not have a practical tool for understanding the Morse flows and so on. 
	
	The situation, however, is not as bad as it sounds. For example, we might guess the coefficients $n_i$ by looking at the Witten index. For this, we go back to the formula (\ref{index_CPN_intersection}) and check if the formula is finite once $y\rightarrow1$ or, equivalently, $z\rightarrow0$. This gives 
	\begin{equation}
	\mathcal{I}(\zeta)\underset{y\rightarrow 1}{\longrightarrow}
	\tfrac{1}{0}(n_1(\zeta)+n_2(\zeta)).
	\end{equation} 
	So upon choosing $n_1(\zeta)=-n_2(\zeta)$, we get a speculative Witten index, finite yet undetermined. Similarly, for a  generic Witten index,
	the constraint 
	\begin{equation}
	\sum_{i=1}^{2^\alpha}n_i(\zeta)=0,\label{n_constraint}
	\end{equation}
	has to be fulfilled for finiteness. Moreover, for $\mathcal{M}$ with a K\"ahler geometry of complex dimension $D$, the Witten index derived from (\ref{index_ModuliSpace}) is defined in the geometric phase as \cite{Witten:1982im}
	\begin{equation}
	\mathcal{I}=\sum^{D} _{p=0} b_p = \chi(\mathcal{M}).
	\end{equation}
	This already puts a strong constraint on what the coefficients are. Since $n_i=\pm 1$ for a convergent thimble integral, our formula in the geometric phase gives
	\begin{eqnarray}
	\chi{(\mathcal{M})}&=&\lim_{ y\rightarrow 1}\frac{1}{\kappa^{\alpha}}\sum^{2^\alpha}_{j=1} n_j Z_j(y)\label{eulerchar}
	\end{eqnarray}
	The right-hand side must have a finite limit so the the coefficients $n_j$ only take certain values for satisfying this. Given that we do not have 
	prior knowledge of the Witten index for some $\mathcal{M}$, $n_j$ have to be unique up to an overall sign for the index $\chi(\mathcal{M})$. 
	We propose that the overall sign should be consistent with that of each thimble integral once written as a series in $y$. For example, the constraint (\ref{n_constraint}) forces
	$n_2=-n_1$ for the $\mathcal{N}=4$ decoupling Abelian 2-node quiver with $k$-bifundamentals and $\zeta>0$, so
	\begin{equation}
	\chi(\mathbb{CP}^{k-1}) = n_1 k.
	\end{equation}
	Taking into account the overall sign of each Lefschetz thimble integrals ($=y^{-k+1}\sum_{i=0}^\infty y^{2i}$ and $=y^{k+1}\sum_{i=0}^\infty y^{2i}$), namely, $+1$, we conclude that $n_1=1$. 
	
	The downside of this check is that the divergence in the vector multiplet contribution to 1-loop determinant simply does not disappear from the index. Alternatively, the vector multiplet determinant is derived from a twisted chiral multiplet of axial $R$-charge $2$ and vanishing vector $R$-charge. Chiral multiplet of this $(2,2)$ theory, i.e., $\Sigma(\sigma,\lambda,F_{12}+iD)$ in the adjoint representation, entails an auxiliary scalar field $D$ that comes with the field strength in a linear form. Taking the $\chi_y$ genus limit, $\text{tr} \Sigma^2$ is proportional to \cite{Benini:2013nda}
	\begin{equation}
	\frac{y}{1+y^2}
	\end{equation}
	which is divergent at $y^2=-1.$ For $\mathcal N = (2, 2)$ $SU(2)$ gauge theory with $k$ fundamental chiral multiplets, this contribution to 
	the refined Witten index will cause the index to be singular given an even $k$, leading to the failure of the Witten index check.

	\subsubsection{An alternative check} If we go back to (\ref{ni}), we notice that $\mathcal K$-cycles have their boundaries lie at the singularities of the $u$-space $X$, which combined with the fact that $\braket{\mathcal{K}_i, \mathcal{J}_j}=\delta _{ij}$ provides the best clue on how 
	to obtain the intersection coefficients. So tracking the $\mathcal{K}$-cycles ending at the only singularity of $\mathfrak F \subset X$, made it possible in the simplest case of $\mathbb{CP}^{k-1}$ model to get $n_1=-n_2=+1$: 
	$$\braket{\gamma,\mathcal{K}_1}=+1,\quad \braket{\gamma,\mathcal{K}_2}=-1.$$
	As $x=e^{\kappa u} $, this singularity at $u=0$ (mod $\mathbb{Z}$ due to $u\sim u+1$) becomes the point $x=1$ in the $x$-parametrization of $u$-space, that will be used in what follows. The saddle rims at $\Im(u)=\pm\infty$, or similarly $|x|=0,\infty$, together with the only singularity point divide $\mathfrak{F}$ into two regions $D_1$ and $D_2$ such that $\mathcal K_{1,2}\subset D_{1,2}$ and $x=1\in D_1\cap D_2$. The main conclusion to draw was that $\text{sign}(n_1)=-
	\text{sign}(n_2)$. Same strategy will be generalized to closed quivers with rank-one quiver gauge groups.
	
	Let $G$ be the total gauge group of the gauged quantum mechanics with rank $\alpha$. We will also assume that $G$ is a product of unitary
	groups with bifundamental representations and for simplicity take $U(1)$ quiver gauge groups at every node in an $N$-node quiver. Then, every
	node is labeled by $1$, which means that the total gauge group is $G= U(1)^\alpha$ where $N=\alpha$. In $4d$ $SU(M)$ SYM, these $N$ nodes 
	all together with $k_{N-1}$ bifundamental fields determine the bound states of a collection of $N\le M$ distinct dyons. Here, we will only consider the quivers with rank-one gauge groups at all nodes. The following approach is a thimble-based geometrization of the quiver mutation/reduction  \cite{Alim:2011ae} that were used in \cite{Cordova:2014oxa} to compute the refined Witten index. 
	
	We first decouple a $U(1)$ factor of the gauge group, sitting at the $l$th node where there are a total of $\alpha$ nodes.
	Now we have a linear chain of nodes exactly like $\to\hspace{-.5em}\bigcirc \hspace{-.5em}\to\cdots$ with $\alpha-1$ nodes. Pick out, next, another node, say the ${l+\tilde{l}}$th, corresponding to the gauge holonomy, say ${\bf x}$, such that the total number of remaining distinct (bi)fundamental chiral
	multiplets (arrows) on the reduced quiver is at least $\alpha-1$. For instance, in closed quivers, the ${l+\tilde{l}}$th node might be the one in the immediate neighborhood
	of the ${l}$th. On the contrary, in linear nodes, this might be the last node appearing on the reduced quiver. Then we calculate the {\it asymptotic 1-loop functions}
	\begin{eqnarray}
	{Z}_{\infty}(y,x_i)&:=&\lim_{{\bf x}\to  \infty } \,\mathcal{Z}^{1\text{L}},\nonumber\\
	\quad {Z}_{0}(y,x_i)&:=&\lim_{{\bf x}\to 0 } \,  \mathcal{Z}^{1\text{L}}.\label{eq:asymp}\nonumber
	\end{eqnarray}
	where $i$ runs over $1,\dots,\widehat{l},\dots,\widehat{{l+\tilde{l}}},\dots, \alpha$\footnote{The case $\alpha = 2$ is straightforwardly checked by plotting the Lefschetz thimbles as we did for the $\mathbb{CP}^{k-1}$ model.} with $\widehat{l}$ meaning that $l$ is omitted. In a, now, $\alpha-2$ dimensional path space $X$ of gauge holonomies parametrized by $x_i=e^{\kappa u_i}$, these functions determine a couple of 
	$\alpha-2$ complex dimensional unbounded manifolds of co-dimension $2$  defined by
	$$F_{{\infty,0}}(x_i,y)={Z}_{{\infty,0}}(x_i,y)\hookrightarrow \left(\mathbb{C}^\times\right)^{\alpha-2}.$$ Let us call $\widetilde{F}(y,x_i)$ the 
	restriction of $F$ to the submanifolds parametrized by only the real part of $x_i$. We now consider the maps
	\begin{equation}
	\i_j :\mathcal J_j \to \widetilde{F}_{{\infty}}, \quad  \i_k:\mathcal J_k \to \widetilde{F}_{{0}},
	\end{equation}
	with $j\in J$ and $k\in K$, where $\i_j,\i_k$ are inclusion maps by restriction to $\Re(x_i)\in {\left(X^{\times}\right)}^{\alpha-2}$ with singularity insertions at each bosonic zeromode of $\mathcal{Z}^{1\text{L}}$. The result of thimble integrals are now just a real homologically invariant function of $y$, $\widetilde{Z}_{j,k}(y)=\int_{\mathcal J_{j,k}} \mathcal{Z}^{\text{1L}}(x_i,y) \prod_{a \ne l,l+\tilde{l}}^{{\alpha}} dx_a$. We also bear in mind that the spectrum of Ramond sector is invariant under charge conjugation and therefore $\mathcal I(\zeta,y^{-1})=\mathcal I(\zeta,y)$. In our current treatment, this simply means that 
	\begin{equation}
	y\longleftrightarrow y^{-1} \Longleftrightarrow \widetilde{F}_0\longleftrightarrow\widetilde{F}_\infty.\label{surface_switch}
	\end{equation}
	
	Now we assume that $\text{sign}(n_j)=-\text{sign}(n_k)$ which is a choice we make that could be reversed as well. The important thing is that the 
	 sign change is based on a simple observation that the difference in $\Im(S_{\rm eff})$ along $\mathcal{J}$-cycles flowing to either ${\bf x}\rightarrow 0$ or ${\bf x}\rightarrow \infty$ subspace of the moduli space is mod $2\pi$.
	
	 Then, in the context of quiver systems with the aforementioned features, the following conjecture holds good:
	
	\noindent {\bf Conjecture}: {\it  $\mathcal J$-cycles along the Stokes rays either on $\widetilde{F}_{\infty}$ or $\widetilde{F}_{0}$ will have opposite intersection numbers 
		as all the singularity curves, surfaces or hypersurfaces are crossed. Using (\ref{surface_switch}), we have that if \text{rank}$(G)=\alpha$ is odd, then $\mathcal J$-cycles along the Stokes rays on $\widetilde{F}_{\infty}$ and $\widetilde{F}_{0}$ will have opposite intersection numbers against one another, and the same for even $\alpha$ regardless of the singularity content of moduli space.}
	
	The remaining $2^{\alpha-1}-|J|-|K|$ of intersection coefficients are immediately zero. Here, $|J|$ is the cardinality of the index set $J$. 
	
	We have all pieces needed to determine the index modulo an overall sign. The refined Witten index in the presence of $R$-symmetry fugacities is given by
	\begin{equation}
	\boxed{\kappa^{\alpha-1}\mathcal I (y,\zeta) = \sum_{j=1}^{|J|} Z_j(y,\zeta)-\sum_{k=1}^{|K|} Z_k(y,\zeta),}\label{indexformula}
	\end{equation}
	where $\zeta$ dependence has been integrated with the thimble integrals. For overall sign, we will stick with the same proposal prescribed 
	below (\ref{eulerchar}). As a quick consistency check, we can see that for a $2$-node quiver, this formula correctly
	produces  $\mathcal I (y,\zeta) =\kappa^{-1}(Z_1- Z_2)=y^{-k+1} \sum_{i=0}^{k-1} y^{2i}$ for $\zeta>0$. 
	Also, for $\zeta<0$ it produces a vanishing result since there is no singularity contained in the shaded region of Fig. \ref{Pathspace}-(c)
	and the integral over $\mathcal J_1$ cancels that of $\mathcal J_2$, meaning that $\gamma$ is a vanishing cycle.
	
	We conjecture that if $|K|=|J|$, we can write the index formula (\ref{indexformula}) for a quiver with total gauge group $G\cong U(1)^\alpha$ and an overall $U(1)$ factor decoupled as
	\begin{equation}
	\mathcal I (y,\zeta) = 	\kappa^{1-\alpha}\sum_{i=1}^{2^{\alpha-1}} (-1)^{t_{i-1}}Z_i(y,\zeta)\label{indexformula_v1}
	\end{equation}
	where $t_i$ is the Thue-Morse sequence $t_{i\ge 0} = 0 1 1 0 1 0 0 1 1 0 0 1 0\dots$, and taken to be as the 
	generating function for the intersection coefficients $n_i$ in the theory. Here, the order of $y$ in the numerator of $Z_i$
	is nondecreasing with increasing $i$. Note that the magnitude of intersection coefficients $(-1)^{t_{i-1}}$ is one, being in accord with the coefficients of expansion of a stable BPS bound state of the quiver theory
	in terms of elementary BPS states \cite{Alim:2011ae}.
	
	One may continue such reductions by repeating the procedure explained before \eqref{eq:asymp} to produce {\it hyper} asymptotic 1-loop functions,
    \begin{eqnarray}
    {Z}_{\infty}(y,x_l)&:=&\prod_{\tilde{l}\ne l}\lim_{{{ x}_{\tilde{l}}}\to  \infty }\,\mathcal{Z}^{1\text{L}},\nonumber\\
      {Z}_{0}(y,x_l)&:=&\prod_{\tilde{l}\ne l}\lim_{{{ x}_{\tilde{l}}}\to  0}\,\mathcal{Z}^{1\text{L}}.\label{eq:asymp1}\nonumber
    \end{eqnarray}
    The role of these functions is simply to determine where the thimbles contribute and where they possibly do not have any contribution and
    as expected, the critical points being at infinity and exactness of saddle point approximation nearly allows one to fully decompose 
    the moduli space $\mathcal M$ as
    \begin{equation}
   \mathcal M=\prod_{i=1}^{\alpha-1} \mathbb{CP}^{k_{i}-1},
    \end{equation}
   that is the classical Higgs branch of supersymmetric vacua in the quiver \cite{Denef:2002ru}. This means that in the relative homology, the homological decomposition is carried out like 
    \begin{equation}
   	\int_{\mathcal{\gamma}}\prod_{i=2}^{\alpha} dx_i=\int_{\gamma_2}dx_2\int_{\gamma_3} dx_3\cdots \int_{\gamma_{\alpha}} dx_{\alpha}
    \end{equation}
    once for instance $l=1$ and $\tilde{l}=2,\dots,\alpha$ where
    \begin{equation}
    \gamma_1 = \mathcal{J}_1-\mathcal{J}_{2},\quad \gamma_2 = \mathcal{J}_3-\mathcal{J}_{4},\quad\dots \label{eq:reducedJ}
    \end{equation}
    and the total number of $\mathcal{J}$-cycles are $2^{\alpha-1}$. It is significant to point out that  this decomposition is only allowed if breaking the middle-dimensional $(=\alpha-1)$ relative homology group $H_i(X,X_\tau^{\star})$
   does not receive nontrivial contributions from rank $\ne1$ homologies. Therefore, if a quiver reduction is possible then one may wish to form the original integration cycle in terms of lower dimensional cycles appearing in \eqref{eq:reducedJ} which
   are $1d$ but notice that $\gamma$ is $\alpha-1$ dimensional. Also, the number of $+1$ and $-1$ signs is equal as suggested in \eqref{indexformula_v1}.

\begin{figure}[H]
		\centering
		\includegraphics[width=.8\linewidth]{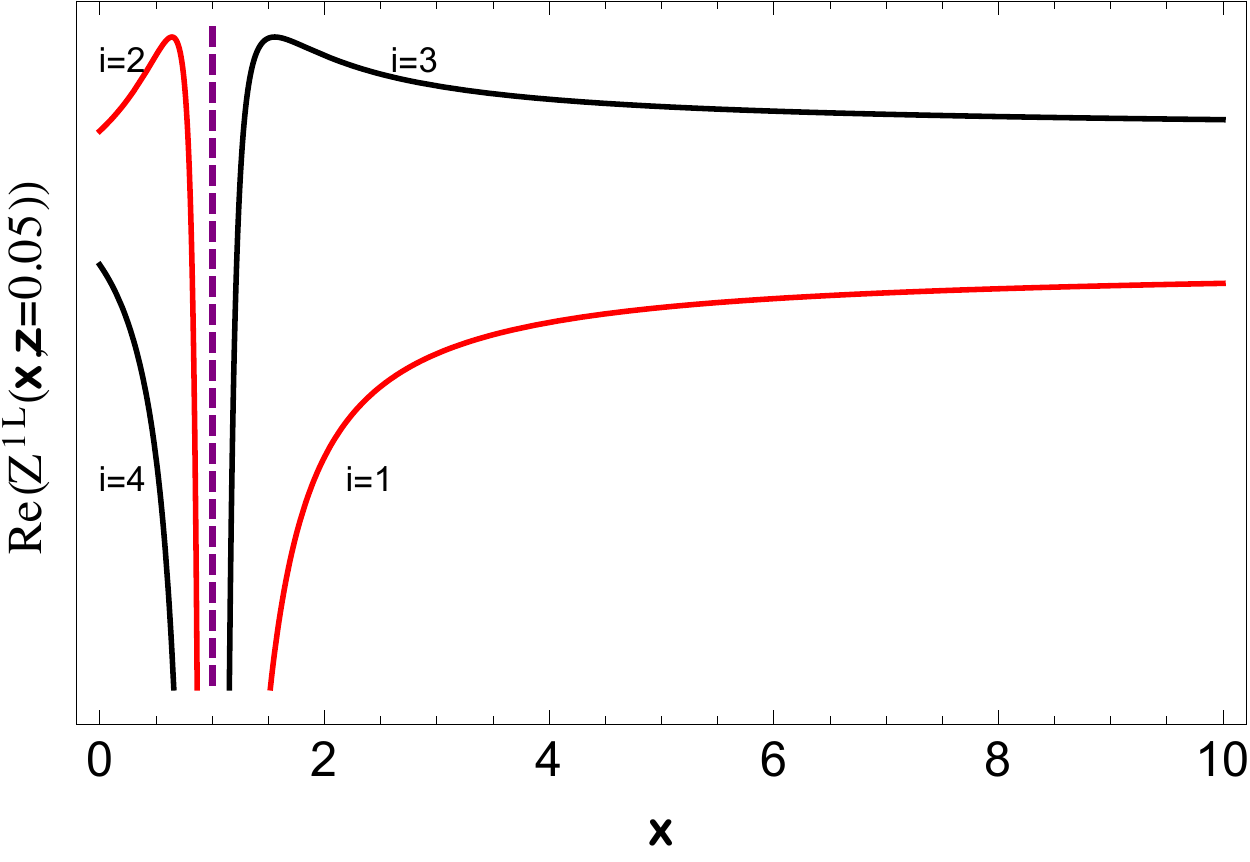}
		\caption{The real part of asymptotic 1-loop functions $Z_{0},Z_{\infty}$ for the 3-node quiver model is exactly the real part of
			the 1-loop determinant of $\mathbb{CP}^{k-1}$ for either $k=k_1$ or $k=k_2$ depending on which node is to be mutated. Here we have set
			 $k_2=2k_1=4$ and $x=x_2=e^{\kappa u_2}$. All possible contributions
			  to the refined index of the 3-node model with $z=0.05$ come from integrals over $\mathcal J$-cycles whose intersection
			  coefficients, $n_i$, are given by ${{n_1}}=-{{n_2}}=+1, {n_3}=-{{n_4}}=-1.$ The jump discontinuity takes place at $x_2=1$. }
		\label{fig:threenode}
	\end{figure}

\vspace{.4cm}
    \noindent {\bf An example: 3-node quiver.} 
	Let us consider the 3-node linear quiver with $k_i$ bifundamental chiral multiplets between the $i$th and the $i+1$th nodes for $i=1,2$. We decouple
	a $U(1)$ from the quiver in a way that the 3-node $U(1)$ quiver model, the 1-loop determinant is given by 
	\begin{eqnarray}
	\mathcal{Z}^\text{1L}&=&\frac{\kappa^2 y^{-k_1-k_2+2} \left(\frac{y^2-x_2}{1-x_2}\right)^{k_1}\left(\frac{x_2 y^2-x_3}{x_2-x_3}\right)^{k_2}}{\left(1-y^2\right)^2}.
	\end{eqnarray}
	The thimble integrals are given by
	\begin{eqnarray}
	{Z}_1(y)&=&\lim_{\substack{x_2\to  \infty \\x_3\to \infty }} \,\mathcal{Z}^{1\text{L}}=\kappa ^2y^{-k_1-k_2+2}\left(1-y^2\right)^{-2}  ,\nonumber\\
	\quad {Z}_2(y)&=&\lim_{\substack{x_2\to \infty \\x_3\to 0} } \,  \mathcal{Z}^{1\text{L}}=\kappa ^2y^{-k_1+k_2+2}\left(1-y^2\right)^{-2}  ,\nonumber\\
	\quad {Z}_3(y)&=&\lim_{\substack{x_2\to 0 \\x_3\to \infty} } \,  \mathcal{Z}^{1\text{L}}=\kappa ^2y^{k_1-k_2+2}\left(1-y^2\right)^{-2}  ,\nonumber\\
	\quad {Z}_4(y)&=&\lim_{\substack{x_2\to 0 \\x_3\to 0}} \, 
	\mathcal{Z}^{1\text{L}}=\kappa ^2y^{k_1+k_2+2}\left(1-y^2\right)^{-2} .\nonumber
	\end{eqnarray}
	Now using the formulae
	\begin{subequations}
		\begin{eqnarray}
		\gamma(\zeta)&=& \sum_{i=1}^{4}n_i(\zeta)\mathcal J_i(y),\\
		\mathcal{I}(y,\zeta)&=& \kappa^{-2}\sum_{i=1}^{4}n_i(\zeta)Z_i(y),
		\end{eqnarray}
	\end{subequations}
\begin{figure}[H]
	\centering
	\includegraphics[width=0.5\linewidth]{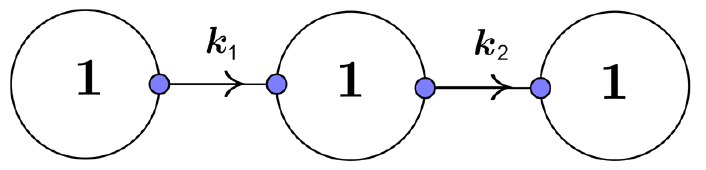}
	\caption{ A generalization of $\mathbb{CP}^{k-1}$ model. These are 3 nodes that all together form a linear quiver with rank $\alpha=1$ Abelian gauge 
		groups at each node. $k_i$, the number of arrows, also shows the number of bifundamentals.}
	\label{fig:3node_quiver}
\end{figure}
	the Witten index constraint forces
	\begin{eqnarray}
	\mathcal{I}(y,\zeta)&=& \frac{-(n_2+n_3+n_4)y^{-k_1-k_2+2}+n_2y^{-k_1+k_2+2}}{\left(1-y^2\right)^2}\nonumber \\&+& \frac{n_3y^{k_1-k_2+2}+n_4y^{k_1+k_2+2}}{\left(1-y^2\right)^2}\nonumber \\
	&=& \frac{y^{-k_1-k_2+2}}{\left(1-y^2\right)^2}\big(-n_2-n_3-n_4+n_2y^{2k_2}\nonumber\\
	&+& n_3y^{2k_1}+n_4y^{2k_1+2k_2}\big),\label{index_3node}
	\end{eqnarray}
	where we have dropped the $\zeta$ from the coefficients for simplicity. For a finite Witten index, the numerator must divide the denominator. The expression in the parenthesis can be factorized as $(a+n_3y^{2k_1})(b+n_2y^{2k_2})$ where 
	\begin{equation}
	-a b=n_2+n_3+n_4, \quad n_3n_2=n_4, \quad a n_2= n_2, \quad b n_3= n_3, 
	\end{equation}
	which in the non-vanishing chamber, immediately give $$-1=n_2+n_3+n_3n_2.$$ Finally, under $k_1\leftrightarrow k_2$ exchange, the index (\ref{index_3node}) remains unchanged so $n_2 =n_3$.
	Therefore, $n_2=n_3=-1$ and $n_1=n_4=1$ as shown in Fig. \ref{fig:threenode}, giving the correct nonzero index with the overall plus sign 
	\begin{eqnarray}
	\mathcal{I}(y)&=& \left(y^{-k_1+1}\sum_{i=0}^{k_1-1} y^{2i}\right)\left(y^{-k_2+1}\sum_{i=0}^{k_2-1}y^{2i}\right).\label{ind_3node}
	\end{eqnarray}
	
	One may check straightforwardly that the formula (\ref{indexformula_v1}) exactly produces this result that is instantly generalizable to any longer Abelian linear $N$-node quiver $(N>3)$.
	\section{Oriented closed quivers}  \label{sec:ocq}
	\subsection{$XYZ$ model}\label{subsec:XYZ}
\begin{figure}[H]
		\centering
		\includegraphics[width=0.3\linewidth]{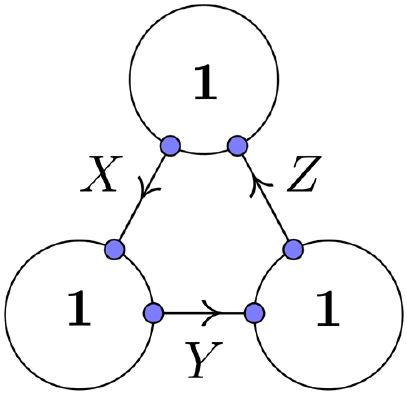}
		\caption{Triangular quiver has three rank-one quiver gauge groups at each vertex and the bifundamentals are $X,Y,Z$. }
		\label{fig:triangularquiver}
	\end{figure}

	The triangular quiver with the superpotential $W=(XYZ)^k$ is depicted in Fig.~\ref{fig:triangularquiver} for a positive integer number $k$. The
	$R$- charges satisfy \begin{equation}
	R_X+R_Y+R_Z = 2/k. \label{charges}
	\end{equation}
	Decoupling the first $U(1)$, the 1-loop determinant is given as
	\begin{equation}
	\mathcal{Z}^{\text{1L}}=\frac{-\kappa^2 y^{-1} \left(y^2-x_1 y^{R_X}\right)  \left(x_1 y^2-x_2 y^{ R_Y}\right)\left(y^{R_Z }-x_2 y^{2}\right)}{\left(1-y^{2 }\right)^2 \left(1-x_1 y^{R_X}\right) \left(x_1-x_2 y^{R_Y}\right) \left(x_2-y^{R_Z}\right) }
	\end{equation}
	Calculating the thimble integrals, one finds
	\begin{eqnarray}
	{Z}_1(y)&=&\lim_{\substack{x_2\to 0 \\x_3\to \infty} } \,  \mathcal{Z}^{1\text{L}}=\kappa ^2y\left(1-y^2\right)^{-2} ,\nonumber\\
	{Z}_2(y)&=&\lim_{\substack{x_2\to 0 \\x_3\to 0}} \, 
	\mathcal{Z}^{1\text{L}}=\kappa ^2y\left(1-y^2\right)^{-2} ,\nonumber\\
	{Z}_3(y)&=&\lim_{\substack{x_2\to  \infty \\x_3\to \infty }} \,\mathcal{Z}^{1\text{L}}=\kappa ^2y^3\left(1-y^2\right)^{-2}   ,\nonumber\\
	{Z}_4(y)&=&\lim_{\substack{x_2\to \infty \\x_3\to 0} } \,  \mathcal{Z}^{1\text{L}}=\kappa ^2y^3\left(1-y^2\right)^{-2}  .\nonumber
	\end{eqnarray}
	
	The Witten index check in determining the intersection coefficients in this case is doomed as the degree of polynomials in the denominators of $Z_i$'s is higher than that of the polynomials in the numerators. Instead, for $XYZ$ model and in general cases, we employ the alternative method and verify that the formula (\ref{indexformula}) yields the correct index only by imposing $k=1$ in (\ref{charges}). 
	
\begin{figure}
		\centering
		\includegraphics[width=0.8\linewidth]{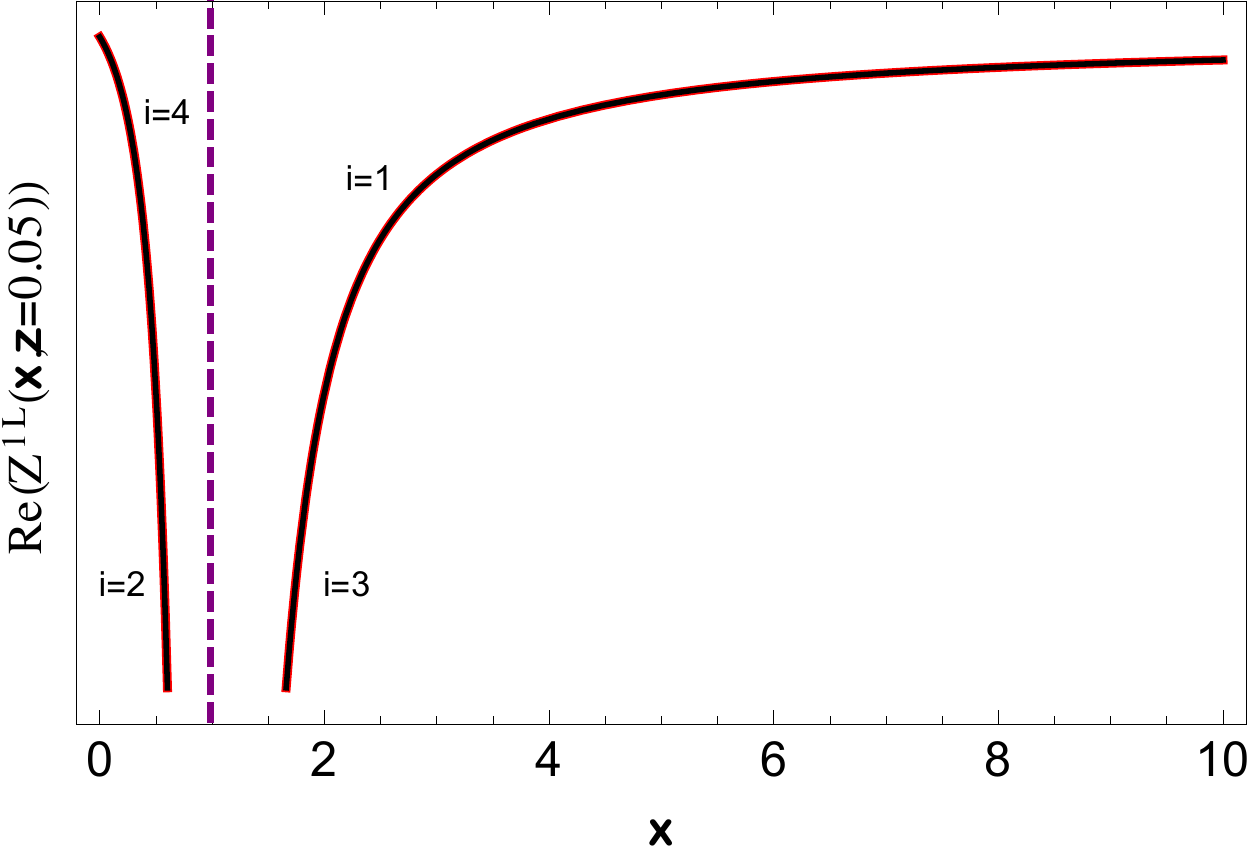}
		\caption{All four possible contributions of thimble integrals in the refined index of $XYZ$ model in the 
				$R$-charge configuration $(0,0,1)$ or $(0,1,0)$ where $k=1$. The asymptotic 1-loop functions
				are overlapping and the relative sign makes the index vanish. Here, the conjecture gives ${{n_1}}=-{{n_2}}=+1, {{n_3}}=-{{n_4}}=-1.$ The jump discontinuity occurs
			at $x=x_1=y^{-1}$.}
		\label{fig:XYX}
	\end{figure}

	Because in this model a $U(1)$ factor of the gauge group is decoupled, so as shown in Fig.~\ref{fig:XYX} there are a total of four $\mathcal J$-cycles, $\mathcal J_1,\mathcal J_4$ coming with 
	$n=+1$ and $\mathcal J_2,\mathcal J_3$  with $n=-1$. We already know $\mathcal{J}_j$ and $\mathcal J_k$ are non-homologous because $\mathcal J_k\in F_{\infty}$ and $\mathcal J_j\in F_{0}$. The singularity ray described by $x_3^{2} = y^{R_Z}$ will not allow
	$\mathcal J_2,\mathcal J_3$ to fall in the same homology class and so will be the case with $\mathcal J_1,\mathcal J_4$. Putting a $U(1)$ aside, 
	we have $2^2=4$ thimbles in the irreducible $\mathbf J$-set.
	
	The integration cycle is given by
	\begin{equation}
	\gamma = \mathcal J_1-\mathcal J_2-\mathcal J_3+\mathcal J_4.
	\end{equation} 
	This result is consistent with eq. (4.26) of \cite{Cordova:2014oxa} for $k=1$: 
	\begin{eqnarray}
	\mathcal I(y) &=& \lim_{k\to 1}[y(1-y^{2/k})^{-1} - y^{2/k-1}(1-y^{2/k})^{-1}]\nonumber\\
	&=&\lim_{k\to 1}[(y-y^{2/k+1}) - (y-y^{{2/k}+1})](1-y^{2/k})^{-2}\nonumber\\	&=&\frac{y}{(1-y^{2})^2}-\frac{y}{(1-y^{2})^2}-\frac{y^3}{(1-y^{2})^2}+ \frac{y^3}{(1-y^{2})^2}\nonumber\\
	&=& \kappa^{-2}(Z_1-Z_2-Z_3+Z_4)=0.
	\end{eqnarray}	
	This verifies the authors' hypothesis about the genericity and quasi-homogeneity of superpotential dictating the relationship between $U(1)_R$ 
	charges of chiral fields in models where holomorphic gauge invariant monomials are turned on in chiral fields. 
	\subsection{$4d$ $\mathcal N=2$ $SU(3)$ Yang-Mills} \label{ymquiver}
	Finally, we want to examine a non-trivial example concerning the BPS states of $4d$ $\mathcal N = 2$ $SU(3)$ Yang-Mills theory with rank-one 
	quiver gauge groups. The BPS particle is a $W$-boson just like Seiberg-Witten which is stable and ground states of the quiver quantum mechanics exist at weak coupling regime of the theory. Lefschetz thimble decomposition of $\gamma$ is found to be
	\begin{equation}
	\gamma = \sum_{i=1}^{8} (-1)^{t_{i-1}}\mathcal J_i.
	\end{equation}

	The Refined Witten index of the theory in the non-vanishing chamber is then
	\begin{eqnarray}
	\mathcal I(y)&=&\kappa^{-3} \sum_{i=1}^{8} (-1)^{t_{i-1}}Z_i\nonumber\\
	&=& \kappa^{-3}(Z_1-Z_2-Z_3+Z_4-Z_5+Z_6+Z_7-Z_8)\nonumber\\
	&=&(y^{-1}-2y+y^3-y^3+2y^5-y^7)\left(1-y^2\right)^{-3}\nonumber\\
	&=& y+y^{-1}. \label{index4d}
	\end{eqnarray}
	There are four chambers in this theory in two of which the index is given by (\ref{index4d}) and zero in the other two. In the regularized scheme,
	there is a 4-way junction point in $u$-space around which $\theta$ changes so we sweep all possible chambers in the $\zeta$-space:
	\begin{equation}
	\gamma=\tilde{n}_1(\zeta)\,\mathcal{C}_{0e^{i\theta_1}} + \tilde{n}_2(\zeta)\,\mathcal{C}_{0e^{i\theta_2}}+\tilde{n}_3(\zeta)\,\mathcal{C}_{0e^{i\theta_3}}+\tilde{n}_4(\zeta)\,\mathcal{C}_{0e^{i\theta_4}}\label{wall-cross-gamma2}
	\end{equation} 
	where $\theta_4>\theta_3>\theta_3>\theta_1$. Every $\tilde{n}_i(\zeta)$ depends on the equation of the corresponding chamber in terms of $\Theta(\vec{\zeta})$ and all of them are normalized to 1.
	\begin{figure}[H]
	\centering
	\includegraphics[width=0.6\linewidth]{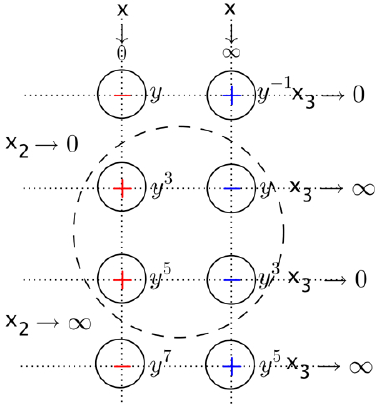}
	\caption{Intersection coefficients of the quiver quantum mechanics mimicking the BPS states
		of $4d$ $\mathcal N=2$ $SU(3)$ SYM for all Lefschetz thimbles in the $\mathbf J$-set. The duality $y\rightarrow y^{-1}$ amounts to a rotation of 
		coefficients around $x_1\equiv\bf{x}$-axis by $180$ degrees. Self-dual saddle contributions $\propto\pm y^3$
		will cancel out because rank of the reduced quiver is $3.$ As expected, the coefficients in each row have opposite signs (along a Stoke curve
		parallel to $x_3$-axis).}
	\label{fig:SU(3)SYM}
\end{figure}

	\section{More on $\mathcal N=4 \, \mathbb{CP}^{k-1}$ model}\label{sec:2nodetheory}
	In this section, we first elucidate how monodromies of $U(1)_R$ gauge
	arise as conserved quantities along Lefschetz thimbles. It is then shown that 
	the thimble integrals an be calculated exactly by slightly going off the Stokes lines that
	appear as $R$-anomaly removal condition is imposed. Finally, an implication in terms of 
	wall crossing is given.
	\subsection{Monodromies, $R$-anomaly removal and Stokes lines} \label{subsec:thimbleCal}
	When the rank of gauge group is one, instead of solving flow equations which are highly non-linear, we focus on the alternative definition of Lefschetz thimbles using the conservation of Hamiltonian flow governed by the effective action $S_{\text{eff}}(u,y)$ along the thimble $\mathcal{J}_i$, namely
	\begin{equation}
	\Im(S_{\text{eff}}(u,y))=\Im(S_{\text{eff}}(u^\star_i,y))=\text{const.} \label{PLcond}
	\end{equation}
	where $u_i^\star\in \mathcal{J}_i$ is a critical point. Note that the auxiliary field $D$ has already been integrated out so that the effective action does not contain the FI parameter, being now carried over into the intersection coefficients $n_i(\zeta)$ (see Subsec.~\ref{subsec:FIdep}). So wall crossing phenomena will be independent of the thimbles themselves but will arise rather cleverly in the way thimbles form vanishing closed homological cycles or else.

	For our example of an Abelian linear 2-node quiver with $k$ bifundamentals and rank-one gauge fields at each node, one has 
	\begin{equation}
	\begin{split}
	S_\text{eff}(u,y)&:=-\ln(\mathcal{Z}^{\text{1L}}(u,y))\\&=
	-\ln\left(-\pi \sin(\pi z)^{k-1} (\cot(\pi z)-\cot(\pi u))^k\right),
	\end{split}
	\end{equation}
    We want to calculate the integrals 
	\begin{equation}
	Z_{1,2}(y)=\int_{\mathcal{J}_{1,2}}\mathcal{Z}^{\text{1L}}(x,y)\frac{dx}{x}=\kappa\int_{\mathcal{J}_{1,2}}\mathcal{Z}^{\text{1L}}(u,y){du}. \label{Thimble_Int_2node}
	\end{equation}
	Starting first with $\mathcal{J}_1$, we find that this Lefschetz thimble is described by 
	\begin{equation}
	\begin{split}
	&\{u\in \mathfrak{F}\subset\mathbb{C}^\times|\Im(\ln\left(-\pi \sin(\pi z)^{k-1} (\cot(\pi z)-\cot(\pi u))^k\right)\\&=-\arg\left(-e^{-i k\pi z} \csc(\pi z)\right)\}.\label{thimble_1}
	\end{split}\raisetag{1.3em}
	\end{equation}
	Here, we assume that $\mathcal{J}_1$ is attached to the $-i\infty$ saddle rim in the $u$-plane. Now 
	for generic positive, real, but small $z$ we have
	\begin{equation}
     S_{\rm eff}\big|_{\mathcal{J}_1}\approx\log z+i \pi  k z-i \pi,\label{thimble_1_mod}
	\end{equation}
	that is divergent at $z=0$. But the imaginary part of this expansion to all orders is just $i \pi  k z-i \pi$, that defines the monodromy
	of the $U(1)_R$ around the compact direction $\Re(u)=-\sigma_2$,
	\begin{equation}
	e^{-i\Im (S_{\rm eff})|_{\mathcal{J}_1}} = - y^{-k}.
	\end{equation}
	Therefore, the condition \eqref{thimble_1} turns out for any generic real $z$ to be
	\begin{equation}
     -\arg \left(-\csc (\pi  z) \frac{\sin ^k(\pi  (i\sigma_1-\sigma_2-z))}{\sin ^k(\pi  (i\sigma_1-\sigma_2))}\right) =-\pi  (1-k z),\label{thimble_1_mod1}
	\end{equation}	
	 where $u=i\sigma_1-\sigma_2$ as in \eqref{com_u_var}
	 It then immediately follows that
	 	\begin{equation}
	 S_{\rm eff}\big|_{\mathcal{J}_2}\approx\log z-i \pi  k z+i \pi,\label{thimble_2_mod_f}
	 \end{equation}
	which yields the equation for $\mathcal{J}_2$:
	\begin{equation}
	-\arg \left(-\csc (\pi  z) \frac{\sin ^k(\pi  (i\sigma_1-\sigma_2-z))}{\sin ^k(\pi  (i\sigma_1-\sigma_2))}\right)=\pi  (1-k z).\label{thimble_2_mod_f1}
	\end{equation}
	This suggests that the relation
	\begin{equation}
    e^{-i\Im (S_{\rm eff})|_{\mathcal{J}_2}} = - y^{k}.
    \end{equation}
    holds which is again a conserved quantity along $\mathcal J_2$ that describes the monodromy
    of the $U(1)_R$ gauge field around the compact direction $\Re(u)=-\sigma_2$.
    
     Let us consider the elliptic genus for a $U(1)$ theory with $k$ chiral multiplets. One knows that on $T^2$, there are monodromies in the now $D$-independent function $g(\mathbb{t},z,u)=\mathcal{Z}_V(\mathbb{t},z)\prod_j\mathcal{Z}_{\Psi,Q_j}(\mathbb{t},z,u)$ because the left-moving $R$-symmetry is anomalous. This is readily translated in the $g$-function as
     \begin{eqnarray}
     g(\mathbb{t},z,u+l_1 + l_2 \mathbb{t}) &=& e^{2\pi izl_2k}g(\mathbb{t},z,u)\nonumber\\&\equiv& y^{2l_2k}g(\mathbb{t},z,u), \label{eq:anom}
     \end{eqnarray}
     $\text{for}\, l_1,l_2\in \mathbb{Z}$. So one has to set $z\in \mathbb{Z}/k$ to have a single-valued $g(\mathbb{t},z,u)$. This requirement basically means that $y^{2k}= 1$.
     
    The analysis followed in the last subsection is valid for any $z$ except $z=0$ where the 1-loop determinant is ill-defined. However, there is a gauge invariant way to cure this which requires turning on flavor holonomies \cite{Benini:2013xpa}. Doing this does not change the index on flat manifolds
    but renders $R$-symmetry unbroken, thus allowing $z$ to be arbitrary. Once this is achieved,
    we can analytically continue to $z=0$ and proceed.
  
    It is obvious from eqs.~\eqref{thimble_1_mod},\eqref{thimble_2_mod_f} the value $z=1/k$ on the principal branch of logarithm in the effective action
    gives 
    \begin{equation}
    \Im(S_\text{eff}(-i\infty,y))=\Im(S_\text{eff}(i\infty,y))=0, \label{stokes}
    \end{equation}
    and by going to other branches one obtains all values of $z$ allowed by $R$-anomaly removal constraint. Eq. \eqref{stokes} is 
    the condition for an Stokes line (ray) to appear along which integration cycle becomes ambiguous. 
    
     Flow lines connected to saddle points along imaginary-infinity rims are initiated to flow out of these regions at $\tau\rightarrow-\infty$, and 
     depending on which side of the pole they are located,
     they lie on either $ST_1:\Re(u)=z-1$ or $ST_2:\Re(u)=z$ that happen to emerge in 
     $\mathbb{CP}^{k-1}$ model as soon as $R$-anomaly removal condition is imposed. Along these rays \eqref{stokes}
     fixes $y^{2k}=1$, the exact same condition described under eq.~\eqref{eq:anom} obtained from
     single-valuedness of the 1-loop determinant.
     
    \subsection{Lefschetz thimbles and relation to Hori-Kim-Yi wall crossing formula}	 \label{subsec:HKY}
    In a seminal paper by Hori-Kim-Yi on $1d$ Witten index \cite{Hori:2014tda}, the authors gave a prescription for wall crossing
     under the name ``simple wall crossing'' that refers to the change of index
     across the phase boundary supporting a mixed branch of rank one. The $U(1)$ case, e.g., $\mathcal{N}=4$ $\mathbb{CP}^{k-1}$ theory 
     has the simple form
     \begin{equation}
     \Delta \mathcal{I}
      =\frac{1}{\kappa}
      \left( \oint_{x\rightarrow0} -\oint_{x\rightarrow\infty} \right) g(x,y) \frac{dx}{x}, \label{eq:HYK}
     \end{equation}
     where $g(x,y)$ is the product of all one-loop determinants in $\mathcal{N}=2$ supersymmetric quantum mechanics.
	
	In our setup, first we add a small real number $\delta\ll 1$ to $1/k$ to avoid hitting the Stokes lines. A solution of (\ref{thimble_1}) is written as the union (more technically, path composition in $2d$) of two flow lines $\mathcal{J}_1=J^{-\infty}_1\cup J^{-\infty}_2$ where
	\begin{equation}
	\begin{cases}
	\tau\rightarrow -\infty: &  \partial J^{-\infty}_1=\partial J^{-\infty}_2=-i\infty\\
	\tau>-\infty:  & \Re(J^{-\infty}_1)= z,\, \Re(J^{-\infty}_2)= z-1.
	\label{sols}
	\end{cases}
	\end{equation}
	Since the real part of $u$ is left untouched at $\tau\rightarrow-\infty$, there is an action of $U(1)$ on the infinity-boundary  of
	$\mathcal{J}_1$ described by $x=\infty e^{\kappa\, \Re(u)}$ where $z-1\le\Re(u)\le z$ and this does
	 precisely correspond to the ``big-circle'' boundary of the $u$-space in $x$-parametrization illustrated as a contributing
	element of the simple wall crossing formula in \eqref{eq:HYK}.
	
	Similarly, $\mathcal{J}_2$ is described as a union of two flow lines $\mathcal{J}_2=J^{\infty}_1\cup J^{\infty}_2$ where
	\begin{equation}
     \begin{cases}
     \tau\rightarrow -\infty: &  \partial J^{\infty}_1=\partial J^{\infty}_2=i\infty\\
     \tau>-\infty:  & \Re(J^{\infty}_1)= z,\, \Re(J^{\infty}_2)= z-1.
     \label{sols1}
     \end{cases}
     \end{equation}
	 Again, there is an action of $U(1)$ on the infinity-boundary  of
	$\mathcal{J}_2$ described by $x=0 e^{\kappa\, \Re(u)}$ where $z-1\le\Re(u)\le z$ and this does
	precisely correspond to the ``small-circle'' boundary of the $u$-space in $x$-parametrization illustrated as the first
	element of the simple wall crossing formula in \eqref{eq:HYK}.
	
	The associated $\mathcal{K}$-cycles are given as
	\begin{subequations}
		\begin{alignat}{4}
		\mathcal{K}_1 &=\{u:\Im(u(\tau\rightarrow-\infty))=\infty, \, \Re(u)=0 \}\nonumber\\ 
		\mathcal{K}_2 &=\{u:\Im(u(\tau\rightarrow-\infty))=-\infty,\, \Re(u)=0 \}\nonumber \label{thimble_K2mod}
		\end{alignat}
	\end{subequations}
	where $\tau$ changes from $-\infty$ to $\tau^\star$.
	
	Both $\mathcal J_1$ and $\mathcal J_2$ are codimension 1 submanifolds of $\mathbb C^{\times}$ which are
	union of all the possible solutions to the flow equations with non-Cauchy initial values that at
	$\tau\rightarrow-\infty$ they start at the saddle rims and flow to zero $\epsilon$-balls at $\tau^\star$.
	
	On $\mathcal{J}_1$, (\ref{Thimble_Int_2node}) can be re-written as
	\begin{eqnarray}
	\tilde{Z}_{1}(y) &=& \int_{\mathcal{J}_{1}} \frac{(x'-y^2+1)^k}{(x'+1)x'^k} dx'\nonumber\\
	&=& \int_{\mathcal{J}_{1}}\frac{(1+\frac{1-y^2}{x'})^k}{x'+1} dx'\nonumber\\
	&=&\kappa\sum_{n=0}^{\infty} \left(\begin{array}{c}
	k\\
	n
	\end{array}\right)(1-y^2)^n
	\int_{\mathcal{J}_{1}}  \frac{du}{(e^{\kappa u}-1)^n}, \label{Thimble_J1}
	\end{eqnarray}
	where $\tilde{Z}_{1}(y) =y^{k-1}(1-y^2)Z_{1}(y)$ and use was made of the binomial expansion since obviously $|x'|=|x-1|>|1-y^2|$.  
	Because only the $n=0$ term survives in (\ref{Thimble_J1}), one is left with
	\begin{equation}
	Z_1(y) = \kappa\frac{y^{1-k}}{1-y^2}.
	\label{Thimble_J1_vf}
	\end{equation}
	
	Similarly, on $\mathcal{J}_2$ (\ref{Thimble_Int_2node}) is easily calculated to be  
	\begin{eqnarray}
	\tilde{Z}_{2}(y) &=& \kappa\int_{\mathcal{J}_{2}} \frac{(e^{\kappa u}-y^2)^k}{(e^{\kappa u}-1)^k} du\nonumber\\
	&=&  \kappa y^{2k} \int_{z}^{z+1} \frac{(1-\epsilon e^{\kappa r}y^{-2})^k}{(1-\epsilon e^{\kappa r})^k}dr\nonumber\\
	&=&\kappa y^{2k}\frac{\sum_{n=0}^{\infty} {\tiny\left(\begin{array}{c}
			k\\
			n
			\end{array}\right)}(\epsilon y^{-2k})^n}{\sum_{n=0}^{\infty} {\tiny \left(\begin{array}{c}
			k\\
			n
			\end{array}\right)}\epsilon^n}\overset{\epsilon\rightarrow0}{=} \kappa y^{2k}
	, \label{Thimble_J2}
	\end{eqnarray}
	and finally 
	\begin{equation}
	Z_2(y) = \kappa \frac{y^{1+k}}{1-y^2}.
	\label{Thimble_J2_vf}
	\end{equation}
	Therefore, apart from having to determine a couple of coefficients, the index takes the preliminary form of
	\begin{eqnarray}
	\mathcal{I}(y,\zeta)=n_1(\zeta)\frac{y^{1-k}}{1-y^2}+ n_2(\zeta)\frac{y^{1+k}}{1-y^2}.
	\label{index_CPK}
	\end{eqnarray}
	The simplification that comes with a deviation from the anomaly removal condition with a small $\delta\ll 1$ is that one can explicitly form 
	the flow lines with non-Cauchy initial values and calculate the thimble integrals, as we did, which receive only 
	contributions from the boundaries $x=0 e^{\kappa\, \Re(u)},\infty e^{\kappa\, \Re(u)}$, and the integrals along imaginary directions
	 cancel out between $J^{-\infty}_{1,2}$ and $J^\infty_{1,2}$. Hence, in the index calculations using Picard-Lefschetz theory, there seems to be a natural way of avoiding the complications of thimble integrals by implementing smooth local deformation of the Lefschetz thimbles off the Stokes lines whose well-studied counterpart in the context of JK residue operation (at least for the class of theories considered in this paper) lies, for example, where some flavor gauge fields are turned on. 
	\subsection{Piecewise FI parameter dependence}\label{subsec:FIdep}
	But how do we know FI parameter dependence e.g., $\zeta$-dependency, of intersection coefficients? We will only focus on rank-one gauge group $G$ that applies to the present case. 
	
	Localization helps the infinite-dimensional path integral of the field theory to reduce to an integral of the 1-loop determinant over the finite-dimensional moduli space of supersymmetric configurations, $\mathcal{M}$ which inherits the same topological structure as that of the spacetime torus $T^2$ for elliptic genus, which is a complex torus $\mathbb{C}/(\mathbb{Z}+\mathbb{t} \mathbb{Z})$ (modulo group automorphisms). It is of real dimension $2$ or equivalently, a $1d$ complex torus. Taking the limit $\mathbb{t}\rightarrow i\infty$ reduces $\mathcal{M}$ to the $u$-space $X=\mathbb{C}/\mathbb{Z}$ with a non-compact imaginary direction. The 1-loop determinant is basically a meromorphic top-form on $\mathcal{M}$, and therefore considering the poles of the $\mathcal N=(2,2)$ chiral and vector multiplet contribution to the full 1-loop determinant, which are present in localization locus, elliptic genus is given by
	\begin{equation}
	\mathcal{I}_{T^2}=\lim_{e\rightarrow 0}\int_\Gamma \frac{dD}{\kappa D} e^{\frac{-D^2}{2e^2}-\zeta D} \oint_\mathcal{-\gamma} g(\mathbb{t},z,u,D)  du ,\label{ellipticg}
	\end{equation}
	with $g(\mathbb{t},z,u,D)$ being a holomorphic function in $u$ expressed as the product of vector and chiral 1-loop determinants,
	\begin{equation}
	g(\tau,z,u,D)=\mathcal{Z}_V(q,y)\prod_j\mathcal{Z}_{\Psi,Q_j}(\mathbb{t},z,u,D,\zeta), \label{eq:g-func}
	\end{equation}
	and again $u$ takes values in $\mathcal{M}$. The poles of the function $g$ form a subset $\mathcal{M}_{\rm sing}\subset \mathcal{M}$. Residues at the poles $u_j^P\in\mathcal{M}_{\rm sing}$ as instructed by JK residue operation will enter the index formula through a correct choice of integration cycle $\gamma$. However, there are also poles in the $D$-plane that entirely lie on the imaginary axis. It is shown in \cite{Benini:2013xpa,Benini:2013nda,Hori:2014tda} by a relatively straightforward computation that in the $D$-plane, there is only a pole $D_j^P$ for each chiral multiplet of charge $Q_j$ that approaches the real axis as $u\rightarrow u_j^P$ at a rate of $\epsilon^2$ where $\epsilon$ shows the radius of an infinitesimally small closed loop going around every pole in $u$-plane. The rest of these poles stay far away from the real axis in $D$-plane. Thus, picking a deformed contour $\Gamma$ for the $D$-integral that avoids singularities of $\mathcal{M}_{\rm sing}$ and $D_j^P$ at the same time away from the point $D=0$ is feasible but subject to taking the localization limit $e\rightarrow0$ in (\ref{ellipticg}) after the limit $\epsilon\rightarrow0$ is performed. This, therefore, will necessitate a choice of some Cartan subalgebra-valued ``displacement vector'' $\delta \in \mathfrak{h}_\mathbb{C}$ where $\mathfrak{h}_\mathbb{C}$ is the complexified $D$-plane, such that 
	\begin{equation}
	\Gamma:\,\,\Re(D)=\mathbb{R},\,\, \Im(D) =\delta. \label{imD}
	\end{equation}
	The final result of JK residue operation will not depend on this vector anyways. 
	
	Now at this stage, before integrating out $D$ and taking the localization limit, we modify the FI term using an analytically continued version of ``Higgs scaling'' \cite{Hori:2014tda}:
	\begin{equation}
	\begin{split}
	\mathcal{I}_{T^2}&=\lim_{e\rightarrow 0}\int_\Gamma \frac{dD'}{\kappa D'} e^{\frac{-D'^2}{2e^2}-\frac{1}{e^2}\zeta' D'} \oint_\mathcal{\gamma} g(\mathbb{t},z,u,D')  du \nonumber\\
	&=\lim_{e\rightarrow 0}e^{-\frac{1}{2e^2}|\zeta'|^2}\int_{\Gamma'} \frac{dD'}{\kappa(D'+\zeta')} e^{\frac{-D'^2}{2e^2}} \nonumber\\ &\times\oint_\mathcal{\gamma} g(\mathbb{t},z,u,D'+\zeta') du \label{ellipticg_mod}
	\end{split}
	\end{equation}	
	where $\zeta'= \zeta e^2$ is held fixed and $D\rightarrow D'$. Here, $\zeta'$ is a pure imaginary number and thus it is assumed that $\zeta'\in \mathfrak{t}$ with $\mathfrak{t}$ being the Lie algebra of the maximal torus of $G$. For brevity we
	will drop the primes in the following. 
	
	Now everything said in the paragraph before (\ref{imD}) remains equally valid except the contour deformation in $ \mathfrak{h}_\mathbb{C}$ is done away from the point $D=-\zeta$, so
	\begin{equation}
	\Gamma':\,\, \Re(D)=\mathbb{R},\,\,\Im(D)=\delta - \zeta. \label{imD_mod}
	\end{equation}
	Taking the limit $\epsilon \rightarrow 0$ will move the very close poles $D_j^P$  to the point $D=-\zeta$, and therefore the normalized elliptic genus (\ref{ellipticg_mod}) may be equally computed by 
	\begin{equation}
	\mathcal{I}_{T^2}=\lim_{e\rightarrow 0}e^{-\frac{1}{2e^2}|\zeta|^2}\int_{\Gamma'} \frac{dD}{\kappa (D+\zeta)} e^{\frac{-D^2}{2e^2}} \oint_\mathcal{\gamma} g(\mathbb{t},z,u) du \label{ellipticg_mod_v1}.
	\end{equation}
	For a theory with a total $U(1)$ gauge group, we can simply take $\delta$ to be some small positive number so the $D$-integral in (\ref{ellipticg_mod}) will produce a factor
	\begin{equation}
	i \pi-i \pi \text{erf}\left( \zeta/\sqrt{2e^2}\right), 
	\end{equation}
	which upon taking localization limit $e\rightarrow0$, yields
	\begin{equation}
	\kappa\Theta(\zeta),\label{coeff_jump}
	\end{equation}
	where $\Theta(\zeta)$ is the Heaviside step function, equal to 1 if $\zeta \ge 0$ along imaginary direction and zero otherwise. The appearance of $\Theta(\zeta)$ is in line with the expectation that the index should depend on FI parameter in a discontinuous fashion. So taking $\zeta$ to be pure imaginary or real will not change the result of index. This suggests that analytic continuation of $\zeta$ to pure imaginary numbers, does indeed keep the index intact and therefore FI chambers will stay the same. Elliptic genus is then given as
	\begin{equation}
	\mathcal{I}_{T^2}= \Theta(\zeta)\oint_\mathcal{\gamma} g(\mathbb{t},z,u) du \label{ellipticg_mod_v2},
	\end{equation}
	which is a special case of formula (3.62) in \cite{Benini:2013xpa}. Translating (\ref{ellipticg_mod_v2}) into relative homology language, one immediately finds 
	\begin{equation}
	\mathcal{I}_{T^2}= \sum_in_i(\zeta)\int_{\mathcal{J}_i}g(\mathbb{t},z,u) du \label{ellipticg_PL}.
	\end{equation}
	This equation is  gist of the correspondence between Lefschetz thimbles in the present work and the JK residue operation in \cite{Benini:2013nda,Benini:2013xpa}. We observe that as expected, the only dependence of the index on $\zeta$ is via the intersection coefficients $n_i(\zeta)$ not the thimble integrals. 

	 On the contrary to $1d$, in $2d$ theories with a compact target space one can turn on a theta term which, for instance, may lead the elliptic genus at $\theta=0$ to differ from that for $\theta \neq 0$ at
	 $\zeta\ne0$ where in the latter there would be no jump in the index \cite{Witten:1993yc}. Making a non-compact direction in moduli space $\mathcal M$ is essentially done by considering the dimensional reduction of the elliptic genus on a circle known as ``$\chi_y$ genus'', which in our formalism results in
	\begin{equation}
	\mathcal{I}=\sum_in_i(\zeta)\int_{\mathcal{J}_i}\lim_{\mathbb{t}\rightarrow i\infty}g(\mathbb{t},z,u) du \label{chiy_PL}.
	\end{equation}
	where $\mathcal{J}_i$ are 1-manifolds in $X=\mathbb{C}^\times$. We note that first the limit is applied and then thimble integrations are carried out. $\chi_y$ genus enjoys phase transitions as $n_i$ may jump by varying $\zeta$. 
	
	In the case of a 2-node quiver with total quiver gauge group $U(1)^2$, there are two domains or ``phases'' separated by a wall at $\zeta_2=-\zeta_1=0$ in $\mathfrak{t}=\mathfrak{h}$ (Abelian) which allows us to set $\delta = \zeta_2\equiv \zeta$  located on the imaginary line $i\mathbb{R}\subset \mathfrak{t}$. So the intersection coefficients jump at $\zeta = 0$, motivating us to write
	\begin{equation}
	n_i(\zeta):=n_i\Theta(\zeta),\quad n_i\in \mathbb{Z}.\label{FIn}
	\end{equation}
	This is the simplest example of wall crossing in $1d$ rank-one supersymmetric quantum theories.
	
	\section{Jumping On and Off The Wall: Generalized $\mathbf{XYZ}$ Model} \label{sec:GenXYZ}
	The main lesson we will be learning in this section is that the integration cycle formula 
	\begin{equation}
	\gamma = \sum_{i=1}^{J} n_i(\zeta) \mathcal{J}_i 
	\end{equation}
	where $J$ shows the cardinality of $\mathbf{J}$-set, can also define a moduli space analogue of the wall of marginal stability. It means that when $\gamma$ lies on a Stokes ray, then the refined Witten index computation gets in some sense 
	carried out  \textit{on the wall}. If the index takes only an identical non-vanishing 
	function of $y$ in some chamber(s) and zero in the remaining chambers, as in all examples covered in Sec.~\ref{sec:Neq4QM},
	 an {\it off-the-wall} computation will tell about the index in those non-vanishing chambers, and there is nothing special about this. For example, in the 2-node quiver system expounded
	at length in the present paper with the total quiver group $G=U(1)^2$, the wall given by
	$\zeta_2=-\zeta_1=0$, is seen as if it distributes a non-vanishing contribution to the  $\zeta_2>0$ chamber (off-the-wall index) and zero to the  $\zeta_2<0$ chamber (check the explanation under Fig. \ref{fig:2-node_reg}). 
	
    The Stokes lines connect two vacua to each other and therefore integration over them 
   after an appropriate Morsification would result in a Witten index that is merely the difference between the indices in the two chambers in which 
   the vacua live. Indeed all supersymmetric wave functions have support on the Stokes lines before Morsification. In \cite{Hori:2014tda} the careful examination of general $1d$ $\mathcal{N}=2$ suspersymmetric quantum systems with $U(1)$ gauge group in the Coulomb branch has shown that
   for large values of $\Im(u)=\sigma$, there is a normalizable supersymmetric ground state supported around either $|\sigma|\sim {1}/\zeta\gg0$ for $\zeta>0$. Since we are so close to the wall, $\zeta=0^+$, the ground states run away to infinity i.e. the boundaries of the wall.
   The $\sigma>0$ and $\sigma<0$ correspond to the Coulomb branches, the union of which ($\sigma$ line) is a Stokes line connecting the 
   vacua at imaginary-infinity rims. We note that in studying the Coulomb branch, the large $\sigma$-approximation 
   of the supersymmetric theory leads to a non-holomorphic $\sigma$ dependence that yields two effective theories. This fact, which is exactly 
   the main assertion of \cite{Hori:2014tda} in regards with the wall crossing formula, completely agrees with the result obtained in Subsec. \ref{subsec:HKY}.
   
    On the other hand, jumping back on the wall is special. There are some key remarks due:
    
    \noindent{\bf (i)} Treating 
	the index on the walls of marginal stability was basically done in \cite{Alexandrov:2014wca,Pioline:2015wza} in four-dimensional field theories on $\mathbb{R}^{3,1}$ with
	$\mathcal{N} = 2$ supersymmetry. The authors' proposed index is smooth across the walls of marginal stability and hence also applies on the wall as well. This index along with the concept of quiver invariants \cite{Lee:2012sc,Lee:2012naa} or similarly single-centered indices \cite{Manschot:2013sya} are different from the index on the Stokes wall here unless in the Coulomb branch where 
	the wall is basically the Coulomb branch itself \cite{Hori:2014tda,Ohta:2015fpe}. In that case, the intersection of all
	 Stokes lines happen to coincide at imaginary-infinity critical points, that within the context of current approach, are degenerate and located on the boundaries of the moduli space of BPS configurations. The rigorous thimble construction out of these saddle rims requires regularization of the effective action as explained in Subsec. \ref{regularization}.
	  It should be stressed out that by putting $D=0$ in the $g$-function \eqref{ellipticg}
     and considering the effective theory in terms of $u$ (e.g. all the resulting quantum mechanical systems upon dimensional
     reduction), one proves that the entire set of saddle points run away to the imaginary-infinity
     rims.
      
    \noindent{\bf (ii)} When all the FI parameters vanish simultaneously, there might exist some square-normalizable
     wave functions at the intersection of all marginal stability walls that make up the quiver invariant. Equivalently, the 
     intersection of Stokes lines always contains a critical point in the Lefschetz thimble construction but whether or not it contributes 
     to the index and/or quiver invariant is not clear. Since such critical point is at infinity, and is obviously degenerate, 
      the on-the-wall index in higher-rank quivers will be very complicated to compute. Nonetheless, at the intersection point, going off the wall
      leaves out possibly a set of critical points with their Lefschetz thimbles that do not undergo Stokes phenomena. This set of thimbles are
       specific to the wall and integrating over them defines the analogue of 
        quiver invariant. Notice that after Morsification, they might contribute to the
        Witten indices. However, it is expected that exactly on the Stokes line/surface such a contribution exists but has a different
        value.
	
We want to briefly weigh in on the most curious example of a closed quiver that is a generalization of $XYZ$ model in Subsec.~\ref{subsec:XYZ} with a cubic superpotential \cite{Cordova:2014oxa} that will put this idea into perspective. 
	Since the index is non-zero in at least two FI chambers, studying this case can shed more light on the interpretation of $\gamma$ and what it means to make sense out of an ill-defined homology cycle. 
	The quiver is made of a total of $2p+2$ chiral multiplets with field content of $[\mathbf{X},\mathbf{Y},\mathbf{Z}]$ in the bifundamental
	representation of the gauge group for $p\ge1$:
	\begin{equation}
	\mathbf{X}= (X_1,X_2)\,\,\,\mathbf{Y}=(Y_1,\dots,Y_p)\,\,\,\mathbf{Z}=(Z_1,\dots,Z_p).
	\end{equation} 
	A cubic superpotential requires $R_{\mathbf{X}}+R_{\mathbf{Y}}+R_{\mathbf{Z}}=2$ where $R_{\mathbf{i}}$ stands for the $R$-charge of 
	each field and all chiral fields in the same multiplet carry the same charge assigned to them. 
	The 1-loop determinant is given as
	\begin{equation}
	\mathcal{Z}^{\text{1L}}=\frac{\kappa^2 y^{4+2p} \left(1-x_1 y^{R_{\bf X }-2}\right)^2 \left(\frac{x_2-y^{R_{\bf Z}-2}}{x_2-y^{R_{\bf Z}}}\frac{ x_1-x_2 y^{R_{\bf Y}-2}}{x_1-x_2 y^{R_{\bf Y}}}\right)^p}{\left(1-y^2\right)^2 \left(1-x_1 y^{R_{\bf Z}}\right)^2}
	\end{equation}
	The integration cycle is 
	\begin{equation}
	\gamma = \sum_{i=1}^4 n_i(\zeta)\mathcal{J}_i \equiv \mathcal{J}_1-\mathcal{J}_2-\mathcal{J}_3+\mathcal{J}_4. \label{XYZwall}
	\end{equation}
	Therefore, one finds by calculating the thimble integrals
	\begin{eqnarray}
	\quad {Z}_1(y)&=&\lim_{\substack{x_1\to 0 \\x_2\to 0}} 
	\mathcal{Z}^{1\text{L}}=\kappa ^2y^{4-2p}\left(1-y^2\right)^{-2} ,\nonumber\\
	{Z}_2(y)&=&\lim_{\substack{x_1\to  \infty \\x_2\to \infty }} \,\mathcal{Z}^{1\text{L}}=\kappa ^2\left(1-y^2\right)^{-2},\nonumber\\
	\quad {Z}_3(y)&=&\lim_{\substack{x_1\to 0 \\x_2\to \infty} } \,  \mathcal{Z}^{1\text{L}}=\kappa ^2y^{4}\left(1-y^2\right)^{-2}  ,\nonumber\\
	\quad {Z}_4(y)&=&\lim_{\substack{x_1\to \infty \\x_2\to 0} } \,  \mathcal{Z}^{1\text{L}}=\kappa ^2y^{2p}\left(1-y^2\right)^{-2}.\nonumber
	\end{eqnarray}
	that off-the-wall index is
	\begin{eqnarray}
	\mathcal{I}&=&y^{-2 p} \left(1-y^2\right)^{-2}\left(y^{4}-y^{2 p}-y^{4+2p}+y^{4p}\right)\nonumber\\
	&=&\sum _{j=1}^p (j-1) \left(y^{-2 (p-j)}+y^{2 (p-j)}\right)-p. \label{eq:GXYZ}
			\end{eqnarray}
	Comparing this result to that of \cite{Cordova:2014oxa}, we find that, as expected, it is made out of the superposition of BPS configurations of the quiver theory in two chambers sharing the same 
	wall $\gamma$. This is because the boundaries of the $\mathcal J$ cycles
	lie in different chambers so technically by integrating over these thimbles we are interpolating between two vacua.
	For one thing, it is insightful to cast \eqref{eq:GXYZ} into the form
	\begin{eqnarray}
	\mathcal{I}&=&\frac{y^{-2 p} \left(y^4-y^{2 p}-y^{2 p+4}+y^{4 p}\right)}{\left(1-y^2\right)^2} + p - p\nonumber\\
	&=& {\left(1-y^2\right)^{-2}\big[y^{4-2p}+(p-1)+(p-1)y^{4}+y^{2 p} - 2p y^2}\nonumber \\
	&-& (p+py^{4} - 2p y^2)\big]. \label{eq:self-dual}
	\end{eqnarray}
	
	The appearance of a self-dual term cannot be an artifact of 
	adding a regularization term in the effective action because of gauge-invariance at least in the fundamental domain. However, it perhaps is best understood once ${\mathbb{C}^\times}\times{\mathbb{C}^\times}$ is regarded as a topological complex group (or a product space), and the integration cycle $\widetilde{\mathcal{C}}$ is then constructed, off the Stokes rays, in the form of 
	\begin{equation}
	\widetilde{\mathcal{C}} =  \sum_{i,j} n_{ij}\widetilde{\mathcal{J}_i}\times\widetilde{ \mathcal{J}_j}.
	\end{equation}
	This is similar to the isomorphism between the complex group $H$ and the product $H\times H$ in the context of analytic continuation of gauge theories with compact gauge groups or simply defining an integration cycle over a product space. 
	We just add that the action of the closed subgroup $G$ readily factorizes over ${\mathbb{C}^\times}\times{\mathbb{C}^\times}$.  Here, we have defined the $\widetilde{\mathcal{J}}$-cycles in such a way that the thimble integrals are rescaled 
	according to $\widetilde{Z_i}^2=(y-y^{-1})^2{Z_i}$ which defines the index to be \cite{Witten:2010cx}
	\begin{equation}
	\mathcal{I}(\zeta) = \kappa^{-2}\left(y-y^{-1}\right)^{-2}\sum_{i,j} n_{ij}(\zeta)\widetilde{{Z}_i}\times\widetilde{{Z}_j}.
	\end{equation}
	We note that $[\widetilde{{Z}_i},\widetilde{{Z}_j}]=[\widetilde{\mathcal{J}_i},\widetilde{\mathcal{J}_j}]=0$ for all $i,j$, and the overall factor is just the vector multiplet contribution which is $U(1)\times U(1)$-invariant as well. 
	
	We know that a decomposition of middle homology into all possible
	homologies may be done via the relative homology version of the K\"unneth formula as
	\begin{flalign}
	 & H_2(\widetilde{X}_i\times\widetilde{X}_j,\widetilde{X}_i^{\tau^{\star}}\times \widetilde{X}_j\cup \widetilde{X}_i\times \widetilde{X}_j^{\tau^{\star}})\nonumber \\&\hspace{1cm}\cong H_2(\widetilde{X}_i,\widetilde{X}^{\tau^{\star}}_i)\otimes H_0(\widetilde{X}_j,\widetilde{X}^{\tau^{\star}}_j)\nonumber \\
	 &\hspace{1cm} \oplus  H_1(\widetilde{X}_i,\widetilde{X}^{\tau^{\star}}_i)\otimes H_1(\widetilde{X}_j,\widetilde{X}^{\tau^{\star}}_j)\nonumber\\
	 &\hspace{1cm} \oplus  H_0(\widetilde{X}_i,\widetilde{X}^{\tau^{\star}}_i)\otimes H_2(\widetilde{X}_j,\widetilde{X}^{\tau^{\star}}_j)+\dots\,.
	\end{flalign} 
	It is very difficult to compute these homologies and one needs handle decomposition techniques in general. Nonetheless, {\it assuming} a
	regularized effective action, one can immediately get a perfect Morse function with only critical points of index one, following the 
	regularization \eqref{eq:reg}, that trivializes this problem in the sense that the only nontrivial group that contributes to the middle relative homology is actually $H_1(\widetilde{X}_i,\widetilde{X}^{\tau^{\star}}_i)\otimes H_1(\widetilde{X}_j,\widetilde{X}^{\tau^{\star}}_j)$.
	Therefore, it is in this latter sense that the integration cycles over the product space take the form
	\begin{eqnarray}
	\widetilde{\mathcal{C}}_{0e^{i\theta_2}} &=& \widetilde{\mathcal{J}_1}^2+(p-1)~\widetilde{\mathcal{J}_2}^2 \nonumber\\
	&+& (p-1) \widetilde{\mathcal{J}_3}^2 +\widetilde{\mathcal{J}_4}^2- 2p~ \widetilde{\mathcal{J}_2}\times\widetilde{\mathcal{J}_3},\nonumber\\
	\widetilde{\mathcal{C}}_{0e^{i\theta_3}} &=& p~\widetilde{\mathcal{J}_2}^2 +p~\widetilde{\mathcal{J}_3}^2- 2p~ \widetilde{\mathcal{J}_2}\times\widetilde{\mathcal{J}_3}\equiv\widetilde{\mathcal{C}}_{0e^{i\theta_1}}, \nonumber
	\end{eqnarray}
	where in light of the prescription (\ref{wall-cross-gamma}) we have defined
	\begin{equation}
	\gamma= \tilde{n}_1(\zeta)\widetilde{\mathcal{C}}_{0e^{i\theta_1}}  + \tilde{n}_2(\zeta)\widetilde{\mathcal{C}}_{0e^{i\theta_2}} +  \tilde{n}_3(\zeta)\,\widetilde{\mathcal{C}}_{0e^{i\theta_3}},
	\end{equation}
	with $\tilde{n}_i$ being in general a function of $\Theta(\zeta)$ in the $i$th chamber of $\zeta$-space described by
	\begin{eqnarray}
	\tilde{n}_1 &=&\Theta(\zeta_1+\zeta_2)\Theta(\zeta_2),\quad\tilde{n}_2 =\Theta(-\zeta_1-\zeta_2)\Theta(-\zeta_1) ,\nonumber\\
	\tilde{n}_3 &=&\Theta(-\zeta_2)\Theta(\zeta_1)\nonumber.
	\end{eqnarray}
	 There are a total of
	three chambers. The jumping 
	of cycles in transitioning from
	chamber 2 to 3 is then given by the relations
	\begin{eqnarray}
	\widetilde{\mathcal{J}_1}^2&\rightarrow &\widetilde{\mathcal{J}_1}^2-\widetilde{\mathcal{J}_4}^2+\widetilde{\mathcal{J}_2}^2,\quad\,	\widetilde{\mathcal{J}_4}^2\rightarrow\widetilde{\mathcal{J}_4}^2-\widetilde{\mathcal{J}_1}^2+\widetilde{\mathcal{J}_3}^2\nonumber\\
	\widetilde{\mathcal{J}_2}^2&\rightarrow&\widetilde{\mathcal{J}_2}^2,\quad \widetilde{\mathcal{J}_3}^2\rightarrow\widetilde{\mathcal{J}_3}^2. \label{eq:XYZjumps1}
	\end{eqnarray} 
	Similarly, the only 
	jumps happening when going from the chamber 2 to 1, are  
	\begin{eqnarray}
	\widetilde{\mathcal{J}_1}^2&\rightarrow &\widetilde{\mathcal{J}_1}^2-\widetilde{\mathcal{J}_4}^2+\widetilde{\mathcal{J}_3}^2,\quad\,	\widetilde{\mathcal{J}_4}^2\rightarrow\widetilde{\mathcal{J}_4}^2-\widetilde{\mathcal{J}_1}^2+\widetilde{\mathcal{J}_2}^2\nonumber\\
	\widetilde{\mathcal{J}_2}^2&\rightarrow&\widetilde{\mathcal{J}_2}^2,\quad \widetilde{\mathcal{J}_3}^2\rightarrow\widetilde{\mathcal{J}_3}^2.
	\label{eq:XYZjumps2}
	\end{eqnarray} 
	There are no jumps when one enters the chamber 1 coming from 3 or vice versa. 
	
	The fact that the two cycles $\widetilde{\mathcal{C}}_{0e^{i\theta_1}}$,$\widetilde{\mathcal{C}}_{0e^{i\theta_3}}$ are 
	identical tells us that setting
	\begin{equation}
		\widetilde{\mathcal{C}}_{0e^{i\theta_1}}\sim (\widetilde{\mathcal{J}}_2-\widetilde{\mathcal{J}}_3)^2\equiv
		\widetilde{\mathcal{C}}_{0e^{i\theta_3}} \label{eq:C12}
	\end{equation}
	causes a smooth transition between the associated chambers, thus contributing to the off-the-wall $1d$ Witten index identically,
	which is already implied by the form of \eqref{eq:GXYZ}.  But what about when
	$\theta=0$ (exactly $\zeta=0$)? As is clear from eqs. \eqref{eq:XYZjumps1}-\eqref{eq:XYZjumps2}, the Lefschetz thimbles $\widetilde{\mathcal{J}_{2}},\widetilde{\mathcal{J}_{3}}$ do not undergo 
	a jump when crossing the Stokes lines and, consequently, nor does the combination \eqref{eq:C12} on the product space. This suggests that on the Stokes wall one has a general cycle of the form
	\begin{equation}
	W(\theta=0):=\sum_{l=1}^{|W|}\,m_l \widetilde{\mathcal{C}}_l\,,\quad \widetilde{\mathcal{C}}_l =\left(\sum_{i=1}^{{2^\alpha}}n^l_i(0)\widetilde{\mathcal{J}}_i\right)^\alpha ,
	 \label{eq:formulaOmegaS}
	\end{equation}
	on a product space $({\mathbb{C}^\times})^\alpha$ where $|W|$ is the total number of different repeated Witten indices in FI chambers and $m_l$ are
	 intersection numbers. Then 
	the $\Omega_S$ intimately measures the on-wall-index aka quiver invariant as explained in the beginning of this section via the integration 
	over the cycle \eqref{eq:formulaOmegaS},
	\begin{equation}
	\Omega_S=\sum_{l=1}^{|W|} m_l\int_{\widetilde{\mathcal{C}}_l}\prod_{i=1}^\alpha du_i\, \mathcal{Z}^{1L}(\vec{u}).
	\end{equation} 
	
	This observation is consistent with the computation of the quiver invariants for $(N,1,1)$ quivers in \cite{Lee:2012naa} 
	where a nonzero invariant is always accompanied by having at least two identical Witten indices in
	some chambers. In general, if both identical indices are an integer, then $|W|=1$ and $\Omega_S=m_1$
	 where integration over $\widetilde{\mathcal{C}}_1$ equals one. In the present case, $\widetilde{\mathcal{C}}_1=(\widetilde{\mathcal{J}}_2-\widetilde{\mathcal{J}}_3)^2$, and $\Omega_S=p-1$ \cite{Lee:2012naa}. Unlike the coefficients $n_i^l(0)$ which can be computed using the smoothness of the cycles $\widetilde{\mathcal C}_l$ under the Stokes phenomena, we do not know how to obtain the intersection numbers $m_l$, and in general whether or not they should be considered as just input data as in the Kontsevich-Soibelman
	 wall-crossing formulas \cite{Kontsevich:2008fj}, is not yet known. 
     	
     	In higher dimensional gauge theories, it would also be interesting to study the on-the-wall index from the viewpoint of domain walls and their intersections \cite{Carroll:1999wr,Gibbons:1999np}.

	\section{Knots and Lefschetz thimbles} \label{sec:knots}
	A possible interesting feature of supersymmetric theories is their relation to knot homologies and more recently Khovanov homology \cite{Gaiotto:2011nm,Witten:2016qzs,Witten:2011zz,WItten:2011pz,Fuji:2012pi,Gukov:2007ck,Galakhov:2017pod}.
	In math literature, it is known from a classic work by Ozsvath and Szabo \cite{Szabo:2004} that any null-homologous knot $K$ in $S^3$ (or in general any closed, 
	oriented three-manifold) can be given a Floer-homology description where, roughly speaking, homology cycles are flow lines of the Morse function over the three-manifold.  Knot invariants in the form of a categorification of Alexander polynomials are then obtained by calculating the knot Floer homology groups of the
	three-manifold by applying different surgeries along the knot such as crossing resolutions and connected sums. Floer homology is an infinite-dimensional analog of finite-dimensional Morse homology, which is best suitable for studying invariants of $3d$ quantum field theories \cite{Gukov:2016gkn}. 
	
	A knot is a circle embedded in a three dimensional space. Higher dimensions give a lot of freedom to deform knots. On the other hand, a codimension 1 knot turns out to be very restricting. However, taking knots in a $3d$ space and projecting them onto a $2d$ space seems to provide a crucial link between Lefschetz thimbles and the flow dynamics of quiver quantum mechanics in the following sense.
	
	 Let $G$ be a gauge group that is either a unitary group or product of them. The dimensional reduction of elliptic genus on $S^1$ produces a moduli space for all supersymmetric configurations to which the path integral localizes. Call this moduli space $\Sigma$ and observe that it is of complex dimension $\alpha=\text{rank} (G)$. To be specific, suppose for instance, $\Sigma$ is the $u$-space of an $\mathcal{N}=4$ quiver theory with total gauge group $G= U(1)^\alpha$ and bifundamental matter where an overall $U(1)$ factor is decoupled so that $\tilde{\alpha}=\alpha-1$. Then $\Sigma$ has certain singularities that periodically happen so we will instead take $\Sigma^F\subset\Sigma$, the fundamental domain in $\Sigma$. The zero $\epsilon$-balls are centered at the fermion zeromodes. Take an alternating mirror symmetric knot $K$ in $\mathbb{R}^3$ such that $\pi_1(\mathbb{R}^3 \setminus K)\cong \mathbb Z$. Suppose the $\tilde{\alpha}$th relative homology $H_{\tilde{\alpha}}(\Sigma^F, \Sigma^F_{\tau^\star})$ is defined
	and $\mathcal{J}_i$ are $\tilde{\alpha}$-cycles with homologies $ H_{\tilde{\alpha}}(\Sigma^F, \Sigma^F_{\tau^\star})$. Again, $\Sigma^F_{\tau^\star}$ is a zero $\epsilon$-ball	to be reached at $\tau>\tau^\star$ by Lefschetz thimbles $\mathcal {J}_i\subset \Sigma^F$. We define the effective action $S_{\text{eff}}=-\ln(\mathcal Z^{1\text{L}})$ in which $\mathcal Z^{1{\text L}}(x_1,\dots,\hat{l},
	\dots x_\alpha)$ corresponds to the 1-loop determinant of the gauged quiver quantum theory with $l$th node removed. The complex Morse function is $h_\mathbb{C}=S^{\rm reg}_{\text{eff}}$ with the usual regularization in effect.
	
\begin{figure}[H]
		\centering
		\includegraphics[width=0.4\linewidth]{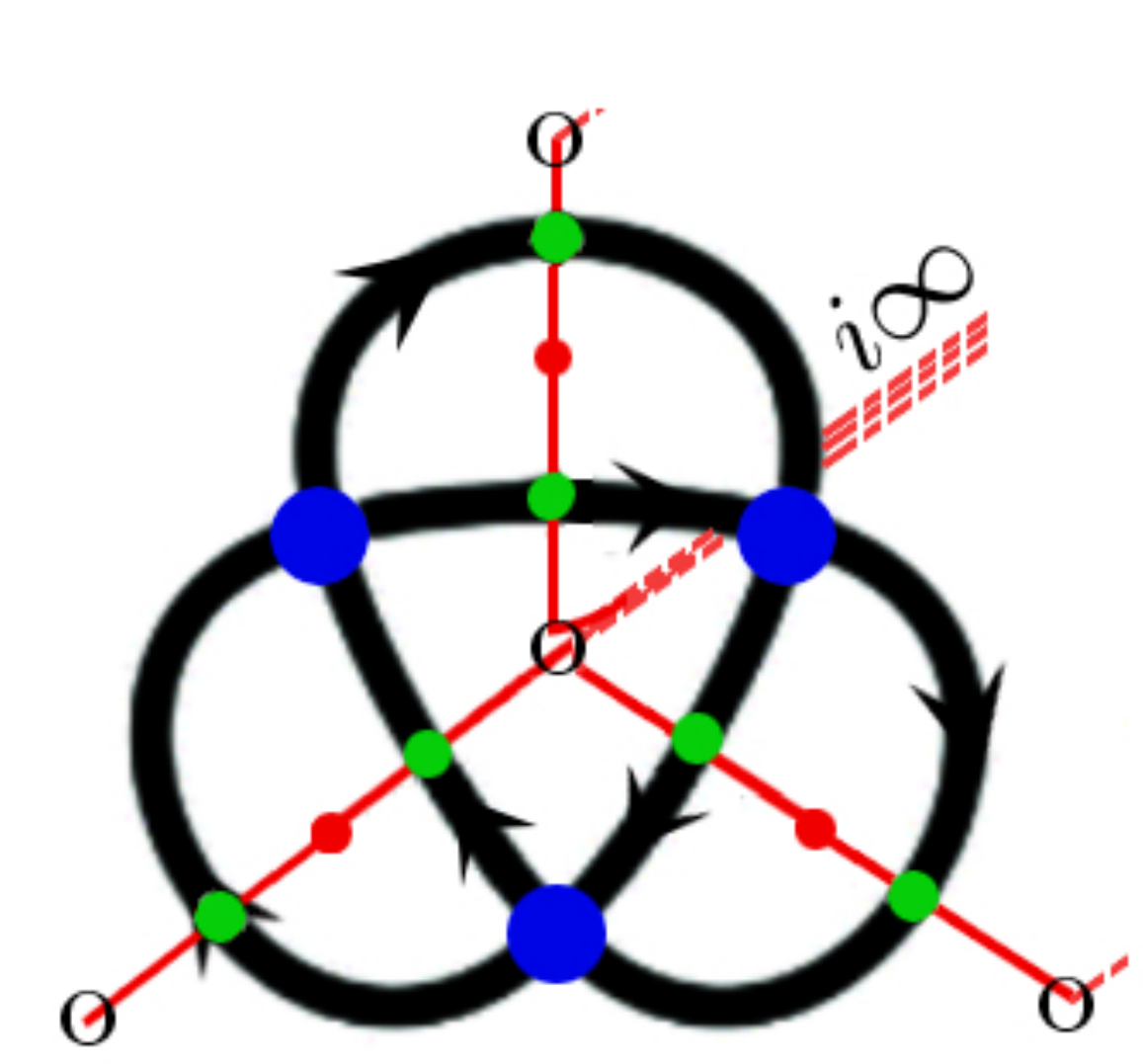}
		\caption{A possible planar thimble decomposition for trefoil knot in a regularized gauged quantum mechanics $\mathcal{A}$. Zero $\epsilon$-balls are shown to be at the crossings and $\mathcal J$ cycles are the directed curves, and red crosses are the singularities in the moduli space of some quiver theory. Red lines symbolize $\mathcal K$ cycles that intersect the Lefschetz thimbles orthogonally and flow to infinity along imaginary directions, consistent with the structure of thimbles in Fig.~\ref{fig:2-node_reg}. The green circles represent the saddle points. Therefore, $\mathcal{P}K=\sum_{i=1}^{6}(-1)^{t_{i-1}}\mathcal{J}_i,$ such that $\mathcal{J}_i$ is lifted up to the relative homology $H_{\tilde{\alpha}}(\Sigma^F,\Sigma^F_{\tau^\star})$ for $\Sigma$ being the $u$-space of $\mathcal{A}$. The knot formula (\ref{knot}) will not obviously detect trefoil knot so we must consider a different higher-rank gauge group $G$.}
		\label{fig:trefoil}
	 \end{figure}

	Identify zero $\epsilon$-balls in $\Sigma^F$ by the crossings of $K$ and Lefschetz thimbles described by 
	\begin{equation}
	\frac{d \bar{u}^j}{d \tau} = g^{i\bar{j}}\frac{\partial h_\mathbb{C}}{\partial u^i},\quad \frac{d u^i}{d \tau} = g^{i\bar{j}}\frac{\partial \overline{h}_\mathbb{C}}{\partial  \bar{u}^j}
	\end{equation}
	($g^{i\bar{j}}$ is a suitable metric on $u$-space), by the line segments connecting these crossings. A Lefschetz thimble contributing to the knot invariant always connects two zero $\epsilon$-balls which correspond to either $+-$ or $-+$ crossings for the connecting segment. Then we have defined a projection of the knot $K$
	onto the space $\Sigma^F$, called $\mathcal{P}K$. (See Fig.~\ref{fig:trefoil} to check what projected trefoil knot looks like in this setup). Then, inspired by (\ref{indexformula_v1}) the knot $\mathcal{P}K$ is conjectured to be given by 
	\begin{equation}
	\mathcal{P}K = \sum_{i=1}^{2^{\tilde{\alpha}}} (-1)^{t_{i-1}}\mathcal J_i. \label{knot}
	\end{equation}
	where there are $2^{\tilde{\alpha}}$ distinct types of saddle rims in ${\Sigma^F}$. 
	
	The HOMFLY polynomial corresponding to an unknot is given as
	\begin{equation}
	f(\lambda,y)(\text{unknot})= (\lambda^{-1}-\lambda)(y^{-1}-y)^{-1}.
	\end{equation}
	This leads to the $\mathfrak{sl}(k)$ knot invariant of the unknot if we put $\lambda=y^k$. But note that this exactly corresponds to the 
	refined Witten index of the Abelian 2-node quiver -$\mathcal{N} =4$ $ \mathbb{CP}^{k-1}$ model- discussed at full length in Sec.~\ref{sec:2nodequiver} and \ref{sec:2nodetheory} where unknot is
	topologically equivalent to the superposition of two Lefschetz thimbles:
	\begin{equation}
	\text{unknot} = \mathcal J_1 - \mathcal J_2.
	\end{equation}
	So one may 
	conjecture that the invariant (\ref{indexformula_v1}) may detect the alternating mirror symmetric knots upon the understanding of Reidemeister moves and crossing resolutions. The best candidate is Alexander polynomial $\Delta_K(t)$ that can be drawn from HOMFLY polynomial. It is a knot invariant that could be obtained from (\ref{knot}) once $y$ is identified with an appropriate function of $t$ due to the symmetry $\Delta_K(t)=\Delta_K(t^{-1})$. In a way, this sort of visualization of knots suggests what geometric sense we can draw from e.g. $\Delta_K(t)$ by relating every term in it directly to a specific cycle in the moduli space of a quantum system with a Hamiltonian that is conserved along every one of these cycles.
	\vspace{-.3cm}
	 \section*{Conclusions and Discussion}
     In this paper, we have considered the upgraded version of Morse theory, known Picard-Lefschetz theory in the
     context of supersymmetric quiver quantum mechanics. We first explained how localization principle may naturally arise
     by collapsing the path integral of bosonic harmonic oscillator onto the solution space of the holomorphizied Morse flow equations associated with its action in the Euclidean signature. The analytic continuation
     to real time then realized the Maslov index as the counting of Fourier modes appearing in the imaginary direction of the Lefschetz thimble.
  
     We then digressed to the computation of phases of Lefschetz thimbles in supersymmetric quantum theories from their singular algebraic curves 
     and discussed how the recently proposed hidden topological angles of the exact complex non-BPS solutions may be related to 
     the sheaf cohomology of holomorphic 1-forms, which aims to properly assemble a semi-classical means of understanding the topological
     invariants of singular spectral curves by replacing the cotangent bundle by cotangent sheaf. The phases of the non-perturbative solutions contributing to the ground states then indicate a close link to the topological phase of the path integral.
     
      A similar idea exploiting refined holomorphic anomaly equations around singular points of algebraic curves  borrowed from topological string theory has recently been proposed to also work for generic quantum mechanical systems \cite{Codesido:2016dld}. It would likewise be apt to understand how correction terms in complex phases discussed in this paper arise from anomaly equations. 
   
     These observations set the stage for building a passway to the study of integration cycles of the localized theory in the moduli space
     of BPS configurations in terms of Lefschetz thimbles that solve complex Morse equations in the presence of a gauge group action which in turn produces degeneracies in the critical points. We observed this in the Higgs branch moduli spaces of simple Abelian linear quivers and some oriented loops with bifundamental matter that could be decoupled. To lift up the flat directions and isolate the degenerate fixed points, one has to consider the full theory in the Coulomb branch. Then a proper perfect Morse function could be defined not only for the rank-one supersymmetric gauge theories but also generic non-Abelian ones, which is the subject of an ongoing work. In the class of quantum mechanical theories derived from $\mathcal{N}=(2,2)$ on a flat torus, after setting $D=0$ in the $g$ function \eqref{eq:g-func}, the supersymmetric vacua in the $u$-space become degenerate as shown in many examples before, leading to $\mathcal J$-cycles having to be Morsified to obtain a proper Picard-Lefschetz theory interpretation. Yet, it turns out that  the degeneracy would not be an issue and at least on a qualitative level, the cycles of the Morsified (regularized) theory
     mimic similar patterns seen in those of the original theory. 
     
     We used the path homotopy of degenerate Lefschetz thimbles to successfully build the basis of the relative homology in terms of $\mathcal J$- or similarly $\mathcal K$- cycles and calculated the of 2-node quiver aka $\mathcal{N}=4$ $\mathbb{CP}^{k-1}$ model by doing the integrals exactly along the thimbles. The two phases of the theory were found to be related to the 
    FI dependence of the intersection coefficients and different ways the Lefschetz thimbles may combine in the Morsified theory
    to piece together the integration cycle $\gamma$. We also discussed that the decoupling of nodes in the studied quivers together with the identification of boundaries of the bad cycles as the singularities in the $u$-space of Cartan directions, would make it possible to slice up the higher dimensional thimbles for higher-rank quivers for better understanding of Picard-Lefschetz decomposition. This is specially useful for treating the non-Abelian quivers but becomes harder as the number of Cartan directions increase and we do not have a clear understanding of this at the moment.
     
    The monodromies of the $R$-symmetry background field were proved to be the conserved charges of the Hamiltonian
    flow along the thimbles and the $R$-anomaly removal condition on $T^2$ was shown to force the $\mathcal{J}$-cycles to lie on Stokes lines that connect different vacua. Thus an integration over the Stokes line basically provides an index which is merely a difference between two chambers. Exactly for the values of the $U(1)_R$ parameter allowed by this condition, the index was interpreted as being on-the-wall and the analysis of generalized $XYZ$ model showed that the Stokes jumps in the Lefschetz thimbles and their coefficients could correctly recreate the index in every chamber. Thereby at least in 
    some easy cases, we combined  the usual wall crossing phenomenon via the discontinuities introduced by FI parameters,
    such as the Hori-Kim-Yi simple wall crossing formula of \cite{Hori:2014tda}, with the concept of Stokes phenomena. We showed that 
   in Picard-Lefschetz theory an on-the-wall index would come up as part of the non-invariant thimbles under Stokes phenomena 
   which are intrinsic to the wall, and integration over a cycle made out of them yields this index which in the context of quiver quantum mechanics, is alternatively called quiver invariant.
  
    We should reiterate that in this paper we kept $\zeta$ separate from the moduli space of BPS configurations, being a reason for why only intersection coefficients are affected by it. In the analysis of the index in Coulomb branch, FI parameters $\zeta$ appear in the effective potentials obtained by integrating out the matter chiral multiplets and  are thus intertwined with the moduli space, and therefore variations of $\mathcal{J}_i$ would explicitly depend on $\zeta$ in a piecewise manner. This is the subject of an ongoing work.
	\section*{Acknowledgments}
	We would like to thank Philip Argyres, Francesco Benini, Sung Bong Chun, Danielle Dorigoni, Sasha Gorsky, Sergei Gukov, Babak Haghighat, Masazumi Honda, Atsushi Kanazawa, Tatsuhiro Misumi, Thomas Sch\"afer, Shamil Shakirov, Shu-Heng Shao, Mithat \"Unsal, Masahito Yamazaki and Wenbin Yan for useful discussions. We are so indebted to Can Kozcaz for proposing this project and Yuya Tanizaki for his collaboration in the early stages of this work. We want to thank Shing-Tung Yau for his warm hospitality in Department of Mathematics and the Center for Mathematical Sciences and Applications (CMSA) at Harvard University,
	where most of this work was completed in the Fall of 2015. Special thanks to the organizers and participants of the “Resurgent Asymptotics in Physics and Mathematics” workshop at KITP in the Fall of 2017 where many great discussions were initiated that led to numerous improvements to an earlier version of the manuscript. This work was supported by the DOE grant DE-SC0013036. 
	
	\balance
	\bibliography{homsupersymmetricqmbib.bib}
	\bibliographystyle{apsrev4-1}
	
\end{document}